\documentclass[twocolumn]{aastex63}

\usepackage{fontawesome}
\usepackage[T1]{fontenc}
\usepackage{color}
\usepackage{graphicx}
\usepackage{amsmath, amssymb, bm}
\usepackage{CJK}
\usepackage[version=4]{mhchem} 
\usepackage{multirow}
\usepackage{comment}

\newcommand{\mub}{m_{\rm b}/M_\star}
\newcommand{\muc}{m_{\rm c}/M_\star}
\newcommand{\mud}{m_{\rm d}/M_\star}
\newcommand{\mue}{m_{\rm e}/M_\star}
\newcommand{\muunit}{M_\oplus/M_\odot}
\newcommand{\degree}{$^{\circ}$}
\renewcommand{\d}{$\mathrm{d}$}

\graphicspath{{./}{figures/}{transit_ground/}}

\shorttitle{A New Planet Revealed by JWST}
\shortauthors{Masuda, Libby-Roberts et al.}

\begin{document}
\turnoffeditone


\title{A Fourth Planet in the Kepler-51 System Revealed by Transit Timing Variations}

\correspondingauthor{Kento Masuda}
\email{kmasuda@ess.sci.osaka-u.ac.jp}

\author[0000-0003-1298-9699]{Kento Masuda}
\affiliation{Department of Earth and Space Science, Osaka University, Osaka 560-0043, Japan}
\author[0000-0002-2990-7613]{Jessica E. Libby-Roberts}
\affil{Department of Astronomy \& Astrophysics, 525 Davey Laboratory, The Pennsylvania State University, University Park, PA 16802, USA}
\affil{Center for Exoplanets and Habitable Worlds, 525 Davey Laboratory, The Pennsylvania State University, University Park, PA 16802, USA}
\author[0000-0002-4881-3620]{John H. Livingston}
\affiliation{Astrobiology Center, 2-21-1 Osawa, Mitaka, Tokyo 181-8588, Japan}
\affiliation{National Astronomical Observatory of Japan, 2-21-1 Osawa, Mitaka, Tokyo 181-8588, Japan}
\affiliation{Astronomical Science Program, Graduate University for Advanced Studies, SOKENDAI, 2-21-1, Osawa, Mitaka, Tokyo, 181-8588, Japan}
\author[0000-0002-7352-7941]{Kevin B. Stevenson}
\affiliation{JHU Applied Physics Laboratory, 11100 Johns Hopkins Rd, Laurel, MD 20723, USA}
\author[0000-0002-8518-9601]{Peter Gao}
\affiliation{Carnegie Science Earth and Planets Laboratory, 5241 Broad Branch Road, NW, Washington, DC 20015, USA}
\author[0000-0003-2527-1475]{Shreyas~Vissapragada}
\affiliation{Carnegie Science Observatories, 813 Santa Barbara Street, Pasadena, CA 91101, USA}
\author[0000-0002-3263-2251]{Guangwei Fu}
\affiliation{Department of Physics and Astronomy, Johns Hopkins University, Baltimore, MD, USA}
\author[0000-0002-7127-7643]{Te Han}
\affiliation{Department of Physics \& Astronomy, The University of California, Irvine, Irvine, CA 92697, USA}
\author[0000-0002-0371-1647]{Michael Greklek-McKeon}
\affil{Division of Geological and Planetary Sciences, California Institute of Technology, Pasadena, CA 91125, USA}
\author[0000-0001-9596-7983]{Suvrath Mahadevan}
\affil{Department of Astronomy \& Astrophysics, 525 Davey Laboratory, The Pennsylvania State University, University Park, PA, 16802, USA}
\affil{Center for Exoplanets and Habitable Worlds, 525 Davey Laboratory, The Pennsylvania State University, University Park, PA, 16802, USA}
\author[0000-0002-0802-9145]{Eric Agol}
\affiliation{Department of Astronomy, University of Washington, Seattle, WA 98195, USA}

\author[0000-0003-3355-1223]{Aaron Bello-Arufe}
\affiliation{Jet Propulsion Laboratory, California Institute of Technology, Pasadena, CA 91011, USA}
\author[0000-0002-3321-4924]{Zachory Berta-Thompson}\affiliation{Department of Astrophysical and Planetary Sciences, University of Colorado Boulder, Boulder, CO 80309, USA}
\author[0000-0003-4835-0619]{Caleb I. Ca\~nas}
\altaffiliation{NASA Postdoctoral Fellow}
\affiliation{NASA Goddard Space Flight Center, 8800 Greenbelt Road, Greenbelt, MD 20771, USA}
\author[0000-0002-0786-7307]{Yayaati Chachan}
\altaffiliation{CITA National Fellow}
\affiliation{Department of Physics and Trottier Space Institute, McGill University, 3600 rue University, H3A 2T8 Montr\'eal, QC, Canada}
\affiliation{Trottier Institute for Research on Exoplanets (iREx), Universit\'e de Montr\'eal, Quebec, Canada}
\author[0000-0003-1263-8637]{Leslie Hebb}
\affiliation{Hobart and William Smith Colleges, Geneva, NY 14456, USA}
\affiliation{Cornell University, Ithaca, NY, 14853, USA}
\author[0000-0003-2215-8485]{Renyu Hu}
\affiliation{Jet Propulsion Laboratory, California Institute of Technology, Pasadena, CA 91011, USA}
\affiliation{Division of Geological and Planetary Sciences, California Institute of Technology, Pasadena, CA 91125, USA}
\author[0000-0003-3800-7518]{Yui Kawashima}
\affil{Frontier Research Institute for Interdisciplinary Sciences, Tohoku University, 6-3 Aramaki aza Aoba, Aoba-ku, Sendai, Miyagi 980-8578, Japan}
\affil{Department of Geophysics, Graduate School of Science, Tohoku University, 6-3 Aramaki aza Aoba, Aoba-ku, Sendai, Miyagi 980-8578, Japan}
\affil{Cluster for Pioneering Research, RIKEN, 2-1 Hirosawa, Wako, Saitama 351-0198, Japan}
\author[0000-0002-5375-4725]{Heather A. Knutson}
\author[0000-0002-4404-0456]{Caroline V. Morley}\affiliation{Department of Astronomy, University of Texas at Austin, Austin, TX 78722, USA}
\author[0000-0001-8504-5862]{Catriona A. Murray}\affiliation{Department of Astrophysical and Planetary Sciences, University of Colorado Boulder, Boulder, CO 80309, USA}
\author[0000-0003-3290-6758]{Kazumasa Ohno}
\affiliation{Division of Science, National Astronomical Observatory of Japan, 2-21-1 Osawa, Mitaka-shi, Tokyo, Japan}
\author[0000-0002-4675-9069]{Armen Tokadjian}
\affiliation{Jet Propulsion Laboratory, California Institute of Technology, Pasadena, CA 91011, USA}

\author[0000-0002-8706-6963]{Xi Zhang}
\affiliation{Department of Earth and Planetary Sciences, University of California Santa Cruz, Santa Cruz, California, USA}
\author[0000-0003-0156-4564]{Luis Welbanks}
\affiliation{School of Earth and Space Exploration, Arizona State University, Tempe, AZ, USA}
\author[0000-0001-8236-5553]{Matthew C.\ Nixon}
\affiliation{Department of Astronomy, University of Maryland, College Park, MD, USA}
\author[0000-0001-9333-4306]{Richard Freedman}
\affiliation{SETI Institute, Mountain View, CA NASA Ames Research Center, USA}

\author[0000-0001-8511-2981]{Norio Narita}
\affiliation{Komaba Institute for Science, The University of Tokyo, 3-8-1 Komaba, Meguro, Tokyo 153-8902, Japan}
\affiliation{Astrobiology Center, 2-21-1 Osawa, Mitaka, Tokyo 181-8588, Japan}
\affiliation{Instituto de Astrof\'{i}sica de Canarias (IAC), 38205 La Laguna, Tenerife, Spain}
\author[0000-0002-4909-5763]{Akihiko Fukui}
\affiliation{Komaba Institute for Science, The University of Tokyo, 3-8-1 Komaba, Meguro, Tokyo 153-8902, Japan}
\affiliation{Instituto de Astrof\'{i}sica de Canarias (IAC), 38205 La Laguna, Tenerife, Spain}
\author[0000-0002-6424-3410]{Jerome P. de Leon}
\affiliation{Department of Multi-Disciplinary Sciences, Graduate School of Arts and Sciences, The University of Tokyo, 3-8-1 Komaba, Meguro, Tokyo 153-8902, Japan}
\author[0000-0003-1368-6593]{Mayuko Mori}
\affiliation{Astrobiology Center, 2-21-1 Osawa, Mitaka, Tokyo 181-8588, Japan
National Astronomical Observatory of Japan, 2-21-1 Osawa, Mitaka, Tokyo 181-8588, Japan}
\author[0000-0003-0987-1593]{Enric Palle}
\affiliation{Instituto de Astrof\'isica de Canarias (IAC), Calle V\'ia L\'actea s/n, 38200, La Laguna, Tenerife, Spain}
\author[0000-0001-9087-1245]{Felipe Murgas}
\author[0000-0001-5519-1391]{Hannu Parviainen}
\author[0000-0002-2341-3233]{Emma Esparza-Borges}
\affiliation{Instituto de Astrof\'isica de Canarias (IAC), Calle V\'ia L\'actea s/n, 38200, La Laguna, Tenerife, Spain}
\affiliation{Departamento de Astrof\'isica, Universidad de La Laguna, E-38206 La Laguna, Tenerife, Spain
}
\author[0000-0002-6227-7510]{Daniel Jontof-Hutter}
\affiliation{Department of Physics and Astronomy, University of the Pacific, 601 Pacific Avenue, Stockton, CA 95211, USA}
\author[0000-0001-6588-9574]{Karen A.\ Collins}
\affiliation{Center for Astrophysics \textbar \ Harvard \& Smithsonian, 60 Garden Street, Cambridge, MA 02138, USA}

\author[0000-0001-6981-8722]{Paul Benni}
\affiliation{Acton Sky Portal (private observatory), Acton, MA USA}
\author[0000-0003-1464-9276]{Khalid Barkaoui}
\affiliation{Astrobiology Research Unit, University of Li\`ege, All\'ee du 6 ao\^ut, 19, 4000 Li\`ege (Sart-Tilman), Belgium}
\affiliation{Department of Earth, Atmospheric and Planetary Sciences, MIT, 77 Massachusetts Avenue, Cambridge, MA 02139, USA}
\affiliation{Instituto de Astrof\'isica de Canarias (IAC), Calle V\'ia L\'actea s/n, 38200, La Laguna, Tenerife, Spain}
\author[0000-0003-1572-7707]{Francisco J. Pozuelos}
\affiliation{Instituto de Astrof\'isica de Andaluc\'ia (IAA-CSIC), Glorieta de la Astronom\'ia s/n, 18008 Granada, Spain}
\author[0000-0003-1462-7739]{Micha\"el Gillon}
\affiliation{Astrobiology Research Unit, University of Li\`ege, All\'ee du 6 ao\^ut, 19, 4000 Li\`ege (Sart-Tilman), Belgium}
\author[0000-0001-8923-488X]{Emmanu\"el Jehin}
\affiliation{STAR Institute, University of Li\`ege, All\'ee du 6 ao\^ut, 19, 4000 Li\`ege (Sart-Tilman), Belgium}
\author[0000-0001-6285-9847]{Zouhair Benkhaldoun}
\affiliation{Oukaimeden Observatory, High Energy Physics and Astrophysics Laboratory, Cadi Ayyad University, Marrakech, Morocco\label{oukaimeden}}

\author[0000-0002-6629-4182]{Suzanne Hawley}
\affil{Department of Astronomy, Box 351580, University of Washington, Seattle, WA, 98195, USA}
\author[0000-0002-9082-6337]{Andrea S.J.\ Lin}
\affil{Department of Astronomy \& Astrophysics, 525 Davey Laboratory, The Pennsylvania State University, University Park, PA, 16802, USA}
\affil{Center for Exoplanets and Habitable Worlds, 525 Davey Laboratory, The Pennsylvania State University, University Park, PA, 16802, USA} 
\author[0000-0001-7409-5688]{Guðmundur Stefánsson}  
\affil{Anton Pannekoek Institute for Astronomy, University of Amsterdam, Science Park 904, 1098 XH Amsterdam, The Netherlands} 

\author[0000-0001-6637-5401]{Allyson Bieryla} 
\affiliation{Center for Astrophysics \textbar \ Harvard \& Smithsonian, 60 Garden Street, Cambridge, MA 02138, USA}
\author[0000-0002-3276-0704]{Mesut Yilmaz}
\affiliation{Ankara University, Department of Astronomy and Space Sciences, TR-06100, Ankara, T\"{u}rkiye}
\affiliation{Ankara University, Astronomy and Space Sciences Research and Application Center (Kreiken Observatory), \.{I}ncek Blvd., TR-06837, Ahlatl{\i}bel, Ankara, T\"{u}rkiye}
\author[0000-0002-8961-277X]{Hakan Volkan Senavci}
\affiliation{Ankara University, Department of Astronomy and Space Sciences, TR-06100, Ankara, T\"{u}rkiye}
\affiliation{Ankara University, Astronomy and Space Sciences Research and Application Center (Kreiken Observatory), \.{I}ncek Blvd., TR-06837, Ahlatl{\i}bel, Ankara, T\"{u}rkiye}
\author[0000-0002-5443-3640]{Eric Girardin}
\affiliation{Grand-Pra Observatory, 1984 Les Hauderes, Switzerland}
\author[0000-0001-8134-0389]{Giuseppe Marino}
\affiliation{Wild Boar Remote Observatory, San Casciano in val di Pesa, Firenze, 50026 Italy}
\author[0000-0003-3092-4418]{Gavin Wang} 
\affiliation{Department of Physics \& Astronomy, Johns Hopkins University, 3400 N. Charles Street, Baltimore, MD 21218, USA}

\begin{abstract}

Kepler-51 is a $\lesssim 1\,\mathrm{Gyr}$-old Sun-like star hosting three transiting planets with radii $\approx 6$--$9\,R_\oplus$ and orbital periods $\approx 45$--$130\,\mathrm{days}$. Transit timing variations (TTVs) measured with past Kepler and Hubble Space Telescope (HST) observations have been successfully modeled by considering gravitational interactions between the three transiting planets, yielding low masses and low mean densities ($\lesssim 0.1\,\mathrm{g/cm^3}$) for all three planets. However, the transit time of the outermost transiting planet Kepler-51d recently measured by the James Webb Space Telescope (JWST) 10~years after the Kepler observations is significantly discrepant from the prediction made by the three-planet TTV model, which we confirmed with ground-based and follow-up HST observations.  
We show that the departure from the three-planet model is explained by including a fourth outer planet, Kepler-51e, in the TTV model. A wide range of masses ($\lesssim M_\mathrm{Jup}$) 
and orbital periods ($\lesssim 10\,\mathrm{yr}$) 
are possible for Kepler-51e. Nevertheless, all the coplanar solutions found from our brute-force search imply masses $\lesssim 10\,M_\oplus$ for the inner transiting planets. Thus their densities remain low, though with larger uncertainties than previously estimated. Unlike other possible solutions, the one in which Kepler-51e is around the $2:1$ mean motion resonance with Kepler-51d implies low orbital eccentricities ($\lesssim 0.05$) and comparable masses ($\sim 5\,M_\oplus$) for all four planets, as is seen in other compact multi-planet systems. This work demonstrates the importance of long-term follow-up of TTV systems for probing longer period planets in a system. 

\end{abstract}


\section{Introduction}\label{sec:intro}

Kepler-51 is a G-type star hosting unusually low-density planets. The star's $\approx8$-day rotation period as measured from its $\sim 1\%$ quasi-periodic flux modulation suggests youth \citep[$\sim0.5\,\mathrm{Gyr}$;][]{2013MNRAS.436.1883W, libbyroberts2020}. It has three Saturn-sized transiting planets  \citep{2013ApJS..204...24B,2013MNRAS.428.1077S} with orbital period ratios close to $1:2:3$ (45~days, 85~days, and 130~days for Kepler-51b, c, and d, respectively) detected with Kepler photometry. The inner two planets were confirmed by \citet{2013MNRAS.428.1077S} who demonstrated a significant anti-correlation in their transit timing variations (TTVs) and constrained their masses to be less than a few times that of Jupiter, considering dynamical stability. \citet{masuda2014} presented TTVs for the outermost transiting planet, Kepler-51d, and numerically modeled the TTVs of the three planets, finding that they all have masses of $<10\,M_\oplus$ and mean densities of $\sim 0.1\,\mathrm{g\,cm^{-3}}$. The TTV-based mass measurements have been confirmed in other works \citep{2016ApJ...818..177A, 2017AJ....154....5H, libbyroberts2020}. 

The nature of the Kepler-51 ``super-puff'' planets \citep{2016ApJ...817...90L} remains elusive. 
For example, their large inferred gas mass fractions \citep[$>$15\%;][]{lopez2014} challenge formation theories, with one proposed explanation positing rapid gas accretion in a dust-free, cold outer disk followed by inward migration \citep{2016ApJ...817...90L, 2021ApJ...919...63C}. Alternatively, their large sizes could be due to a hotter interior caused by obliquity tides inflating the observed radii \citep{2019ApJ...886...72M} instead of a high gas mass fraction. Meanwhile, their low gravity implies high atmospheric loss rates (assuming clear atmospheres) that may result in shorter gas envelope lifetimes than the age of the system \citep{wang2019}, making their low densities a mystery. 

Hubble Space Telescope (HST) Wide Field Camera 3 (WFC3) observations of the transmission spectra of Kepler-51b and d between 1.1 and 1.7 $\mu$m showed no detectable features \citep{libbyroberts2020}, which has led to the development of two hypotheses to explain both the flat spectra and the large sizes of the planets. One hypothesis involves optically thick, high-altitude dust and haze layers entrained in atmospheric outflows increasing the apparent planet radius by significantly reducing the pressure probed by transmission spectroscopy while simultaneously flattening the spectra \citep{wang2019, 2020ApJ...890...93G, 2021ApJ...920..124O}. The other hypothesis employs obliquely oriented planetary rings to block out more of the star than just the planet itself (thereby making the planetary shadow look larger) while also allowing the spectrally gray rings to dominate the transmission spectrum \citep{2020AJ....159..131P, 2020A&A...635L...8A, 2022ApJ...930...50O}. In both scenarios the planets themselves possess much lower gas mass fractions and higher gravity, thereby reducing their inferred atmospheric loss rates to more reasonable values.  

We observed a transit of Kepler-51d with the James Webb Space Telescope (JWST) NIRSpec-PRISM from 0.6--5 $\mu$m\footnote{The transmission spectrum will be presented in a subsequent follow-up publication (Libby-Roberts et al. In Prep)}, and an optical transit with the Apache Point Observatory (APO) simultaneously.
While the transmission spectrum will provide key information to characterize the chemical composition of Kepler-51d's atmosphere as well as the structure of a potential haze layer, the observation also revealed new aspects of the dynamics of the system:
the transit of Kepler-51d occurred two hours earlier compared to the prediction based on previous transit times and a three-planet TTV model with uncertainties of only 2.7~minutes.
The timing offset was confirmed by follow-up HST transit observations of both Kepler-51c and d that will be described herein. In addition, we checked the existing ground-based transit times of Kepler-51d and found that they also consistently deviated from the three-planet model. The simplest explanation is that there is a fourth planet, Kepler-51e, and thus the masses of the three transiting planets derived from the three-planet TTV modeling need to be revisited.

In this paper, we examine the evidence of the fourth planet in the system, Kepler-51e, by utilizing 
an extensive transit timing data set spanning over 14 years from various facilities including JWST, and update the mass constraints taking into account its presence.
In Section~\ref{sec:star}, we revisit the physical properties of the host star.
In Section~\ref{sec:obs}, we present relevant observations and analyses to derive transit times.
In Section~\ref{sec:spot}, we check the accuracy of transit time measurements in the presence of a starspot-crossing feature.
In Section~\ref{sec:model} we describe our TTV model. 
In Section~\ref{sec:3planet} we discuss evidence of the fourth planet in detail. In Section~\ref{sec:4planet} we explore four-planet TTV solutions and present updated physical parameters of the system. Section~\ref{sec:discussion} discusses the implications of our findings, and Section~\ref{sec:summary} summarizes the paper.

\section{Host Star Mass and Radius}\label{sec:star}

As the TTVs and transit light curves constrain only the planet masses and radii relative to those of the host star, we need external constraints on the stellar mass and radius to determine the physical properties of the planets. We fitted the MIST models \citep{2011ApJS..192....3P,2013ApJS..208....4P, 2015ApJS..220...15P, 2016ApJS..222....8D, 2016ApJ...823..102C} to the effective temperature $T_{\rm eff}$ and metallicity $\mathrm{[Fe/H]}$ of Kepler-51 inferred from high-resolution spectroscopy (see below), $K_s$ magnitude from the Two Micron All Sky Survey \citep[2MASS;][]{2006AJ....131.1163S} corrected for extinction using {\tt Bayestar17} \citep{2018MNRAS.478..651G}, and the parallax from Gaia DR3 \citep{2022arXiv220800211G} corrected for zero-point offset \citep{2021A&A...649A...4L} and underestimated error \citep{2021MNRAS.506.2269E}. 
We compute the likelihood function as a product of independent Gaussians for each measurement, and also incorporate the constraint on the stellar age from its rotation period of $8.2\,\mathrm{days}$ as measured from Kepler light curves \citep{2014ApJS..211...24M}, following the method described in \citet{2019AJ....158..173A} using a gyrochrone calibrated to the Praesepe cluster and the Sun.
We used the {\tt jaxstar} code \citep{2022ApJ...937...94M} to perform No-U-Turn sampling \citep{DUANE1987216, 2017arXiv170102434B} of the stellar parameters from the joint posterior probability density function (PDF), assuming a prior PDF uniform in the stellar mass--age--metallicity space. 

For $T_{\rm eff}=5674\,\mathrm{K}$ and $\mathrm{[Fe/H]}=0.047$ from the California Kepler Survey \citep{2017AJ....154..107P} and assuming 110~K and 0.1~dex uncertainties, we found the mass, radius, density, age to be $0.96\pm0.02\,M_\odot$, $0.87\pm0.02\,R_\odot$, $1.48\pm0.06\,\rho_\odot$, $0.7\pm0.5\,\mathrm{Gyr}$, respectively, where the quoted values are the means and standard deviations of the marginal posteriors. 
The assigned statistical errors of $\approx 2\%$ for the mass and radius are smaller than the systematic error floors set by the fundamental temperature scale and stellar-model dependence, which have been estimated as $\approx 4\%$ for radius and $\approx 5\%$ for mass \citep{2022ApJ...927...31T}. Thus we inflate the errors following that work and adopt the stellar mass and radius of $0.96\pm0.05\,M_\odot$ and $0.87\pm0.04\,R_\odot$, respectively, when we convert the planet-to-star mass/radius ratios to the planet masses/radii.
The values are consistent with those based on the Gaia DR2 parallax derived by \citet{2022AJ....163..179P}, whose method we essentially followed except for incorporating the gyrochronal information. They are also consistent with those found in \citet{libbyroberts2020} who adopted an age prior from gyrochronology and the parallax from Gaia DR2. Our results remain unchanged, within error bars, when we adopted $T_{\rm eff}=5574\,\mathrm{K}$ and $\mathrm{[Fe/H]}=-0.07$ from the Spectral Properties of Cool Stars project \citep{2018ApJS..237...38B}.

\section{Transit Observations and Data Analysis}\label{sec:obs}

Mid-transit times of all three Kepler-51 transiting planets were pulled from a wide range of observations spanning 14 years (2010--2024) of measurements made from both the ground and from space. We provide a full list of transit times for the three transiting planets in Table~\ref{tab:transit_times}. Below, we detail the observations and analysis of the individual data sets leading to the extraction of each reported mid-transit time.

\startlongtable
\begin{deluxetable}{ccccc}
\tablecaption{Transit Times of Kepler-51b, c, and d\label{tab:transit_times}}
\tablewidth{0.8\textwidth}
\tablehead{
\colhead{planet} & \colhead{transit time} 
& \colhead{$\sigma_{\rm lower}$} & \colhead{$\sigma_{\rm upper}$} & source\\
 & \colhead{(BJD$-$2454833)} & (days) & (days) &
}
\startdata
b & $159.1097$ & $0.0011$ & $0.0011$ & kepler \\
b & $204.2631$ & $0.0012$ & $0.0011$ & kepler \\
b & $249.4155$ & $0.0013$ & $0.0012$ & kepler \\
b & $294.5732$ & $0.0013$ & $0.0013$ & kepler \\
b & $339.7272$ & $0.0013$ & $0.0013$ & kepler \\
b & $384.8783$ & $0.0012$ & $0.0012$ & kepler \\
b & $430.0344$ & $0.0012$ & $0.0012$ & kepler \\
b & $520.3435$ & $0.0011$ & $0.0012$ & kepler \\
b & $565.5016$ & $0.0013$ & $0.0012$ & kepler \\
b & $610.6583$ & $0.0011$ & $0.0012$ & kepler \\
b & $655.8134$ & $0.0012$ & $0.0012$ & kepler \\
b & $700.9750$ & $0.0013$ & $0.0013$ & kepler \\
b & $746.1274$ & $0.0012$ & $0.0012$ & kepler \\
b & $791.2877$ & $0.0012$ & $0.0012$ & kepler \\
b & $836.4391$ & $0.0012$ & $0.0012$ & kepler \\
b & $881.5991$ & $0.0012$ & $0.0011$ & kepler \\
b & $926.7537$ & $0.0012$ & $0.0012$ & kepler \\
b & $971.9066$ & $0.0013$ & $0.0013$ & kepler \\
b & $1017.0599$ & $0.0013$ & $0.0013$ & kepler \\
b & $1062.2118$ & $0.0011$ & $0.0011$ & kepler \\
b & $1107.3669$ & $0.0011$ & $0.0011$ & kepler \\
b & $1152.52080$ & $0.00093$ & $0.00088$ & kepler \\
b & $1197.6764$ & $0.0011$ & $0.0011$ & kepler \\
b & $1242.83011$ & $0.00092$ & $0.00099$ & kepler \\
b & $1287.98472$ & $0.00091$ & $0.00098$ & kepler \\
b & $1333.1420$ & $0.0010$ & $0.0010$ & kepler \\
b & $1378.2980$ & $0.0010$ & $0.0010$ & kepler \\
b & $1423.45414$ & $0.00089$ & $0.00092$ & kepler \\
b & $1468.61325$ & $0.00093$ & $0.00098$ & kepler \\
b & $1513.7670$ & $0.0010$ & $0.0010$ & kepler \\
b & $2462.0321$ & $0.0020$ & $0.0015$ & hst \\
b & $2732.9527$ & $0.0030$ & $0.0029$ & hst \\
b & $3500.5934$ & $0.0015$ & $0.0015$ & m2-lco \\
b & $3816.6912$ & $0.0019$ & $0.0019$ & m2 \\
b & $5035.8767$ & $0.0011$ & $0.0010$ & m3 \\
b & $5261.6574$ & $0.0011$ & $0.0011$ & apo \\
c & $210.0118$ & $0.0057$ & $0.0057$ & kepler \\
c & $295.3159$ & $0.0043$ & $0.0045$ & kepler \\
c & $380.6394$ & $0.0037$ & $0.0039$ & kepler \\
c & $465.9534$ & $0.0036$ & $0.0034$ & kepler \\
c & $551.2621$ & $0.0037$ & $0.0038$ & kepler \\
c & $636.5701$ & $0.0038$ & $0.0038$ & kepler \\
c & $892.5190$ & $0.0042$ & $0.0040$ & kepler \\
c & $977.8359$ & $0.0056$ & $0.0056$ & kepler \\
c & $1148.4634$ & $0.0050$ & $0.0043$ & kepler \\
c & $1233.8052$ & $0.0039$ & $0.0042$ & kepler \\
c & $1319.1103$ & $0.0046$ & $0.0046$ & kepler \\
c & $1489.7539$ & $0.0043$ & $0.0044$ & kepler \\
c & $1575.0710$ & $0.0046$ & $0.0046$ & kepler \\
c & $3793.2836$ & $0.0078$ & $0.0086$ & m2 \\
c & $4646.4855$ & $0.0040$ & $0.0034$ & m2 \\
c & $5414.3352$ & $0.0084$ & $0.0043$ & hst \\
c & $5584.9712$ & $0.0047$ & $0.0045$ & palomar \\
d & $212.02406$ & $0.00076$ & $0.00076$ & kepler \\
d & $342.20782$ & $0.00095$ & $0.00095$ & kepler \\
d & $472.39081$ & $0.00079$ & $0.00073$ & kepler \\
d & $602.57379$ & $0.00079$ & $0.00073$ & kepler \\
d & $862.93213$ & $0.00079$ & $0.00079$ & kepler \\
d & $993.10431$ & $0.00073$ & $0.00079$ & kepler \\
d & $1123.28449$ & $0.00078$ & $0.00077$ & kepler \\
d & $1253.45026$ & $0.00070$ & $0.00070$ & kepler \\
d & $1383.62997$ & $0.00076$ & $0.00077$ & kepler \\
d & $1513.79021$ & $0.00087$ & $0.00087$ & kepler \\
d & $2555.2005$ & $0.0015$ & $0.0010$ & hst \\
d & $2945.7533$ & $0.0011$ & $0.0014$ & hst \\
d & $3466.4834$ & $0.0044$ & $0.0042$ & m2 \\
d & $3856.991$ & $0.021$ & $0.012$ & tess \\
d & $4247.4808$ & $0.0021$ & $0.0023$ & multiple \\
d & $5288.84734$ & $0.00006$ & $0.00006$ & jwst \\
d & $5419.0210$ & $0.0014$ & $0.0011$ & hst \\
\enddata
\tablecomments{Transit times are medians ($x_{50}$) of the marginal posterior; $\sigma_{\rm lower}$ and $\sigma_{\rm upper}$ are $x_{50}-x_{16}$ and $x_{84}-x_{50}$, where $x_{16}$ and $x_{84}$ are the 16th and 84th percentiles of the marginal posterior, respectively. In calculating the likelihood for the TTV analysis, these measurements were treated as ``$\mu\pm \sigma$,'' where $\mu=(x_{84}+x_{16})/2$ and $\sigma=(x_{84}-x_{16})/2$.
Abbreviations in the source column are defined as follows --- ``m2'': MuSCAT2, ``m3'': MuSCAT3, ``APO'': Apache Point Observatory, ``multiple'': TRAPPIST-North, Acton Sky Portal Observatory, AUKR, and KeplerCam.}
\end{deluxetable}

\subsection{JWST}\label{ssec:obs_jwst}

We observed a single transit of Kepler-51d with the JWST/NIRSpec-PRISM \citep{birkmann.nirpsec} on 26 June 2023 (Cycle 1, GO-2571, PI J. Libby-Roberts) using NIRSpec's Bright Object Time Series (BOTS). Given the relative faintness of Kepler-51 ($J$-mag: 13.56), we utilized 12 groups per integration, with an overall exposure time of 2.9 seconds and 18,082 total integrations across the $\sim$14 hours of observing time. We maintained a well-depth $<$ 75\% to avoid significant non-linearities. To maximize  efficiency, we used the NRSRAPID readout, with the S1600A1 slit and the SUB512 subarray. 

\begin{figure*}
    \epsscale{1}
    \plotone{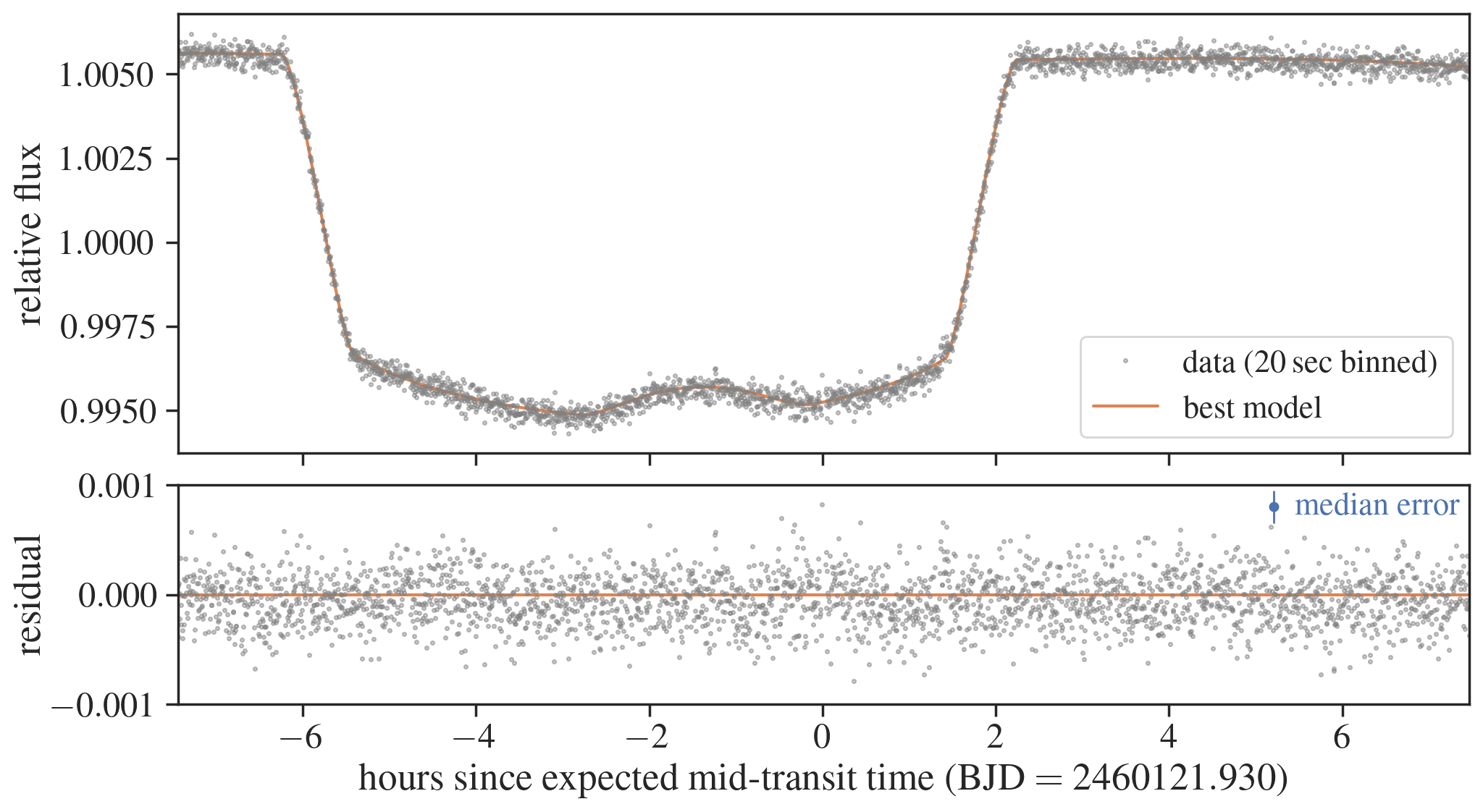}
    \caption{White light curve (0.519 -- 5.463 $\mu$m) of Kepler-51d observed with JWST NIRSpec/PRISM. The pronounced bump near the center of the transit is due to a star spot crossing event. Best fit model is included in orange, which is a combination of transit$+$starspot and systematics. The horizontal axis indicates the time (hours) since the expected mid-transit time based on the previous three-planet TTV model. The observed mid-transit time is significantly offset from the prediction.}
    \label{fig:jwst}
\end{figure*}

We reduced the data using the \texttt{Eureka!} package \citep{Bell2022} starting with Stage 0 uncalibrated files. We performed a Stage 1 reduction following the recommended steps and adopting a cosmic ray rejection threshold of 8$\sigma$ \citep{Bell2022}. We ignored a custom bias subtraction and performed a group-level column background subtraction by fitting a line to the top and bottom 7-pixels of the subarray. Stage 2 reduction followed the recommended steps as outlined in \citet{rustamkulov2022}. For Stage 3 we extracted the source position by fitting a Gaussian and extracting the 1D stellar spectrum by assuming a full-aperture width of 4-pixels. We also performed an overall background subtraction by calculating a column background value from 7-pixels away from the source position. Stage 4 combined the 1D stellar spectrum into a white-light curve by summing the flux from 0.519 -- 5.463 $\mu$m with outliers $>$3$\sigma$ beyond a rolling median flagged and removed. Figure~\ref{fig:jwst} plots the final broadband light curve of Kepler-51d.

The observations were originally planned to center on the expected mid-transit time leaving $\sim$3-hour baseline on either side of the $8.5$-hour transit. As the transit occurred earlier than expected, we were left with a 45-minute pre-transit baseline, the first 30-minutes of which demonstrated a slight slope due to instrument settling \citep{rustamkulov2022}. We also observed a pronounced star spot crossing event during the middle of the transit, which was unsurprising given the activity of this star.

We modeled the broadband light curve using the star spot package \texttt{fleck} \citep{fleck} which combines a star spot model with the \texttt{batman} transit package \citep{batman}. We fit for the spot latitude, longitude, size, and contrast as well as the transit depth, duration (scaled semi-major axis $a/R_\star$ and inclination), and mid-transit time. We also fitted the stellar quadratic limb darkening parameters and applied a second-order polynomial function for detrending. We used \texttt{emcee} assuming 50 walkers and 50,000 steps, and confirmed convergence by checking the autocorrelation lengths of each chain \citep{corelation_length}. The best-fit model is included in Figure~\ref{fig:jwst}. We determine a mid-transit time of 2460121.84734 $\pm$ 0.00006 BJD for Kepler-51d. Other planetary and stellar parameters will be reported and discussed in further detail in a separate planned publication (Libby-Roberts et al. In Prep).

\subsection{HST}\label{ssec:obs_hst}

Two full transits each of Kepler-51b and d were observed with HST between 2015 and 2017, and were analyzed by \citet{libbyroberts2020} (Cycle 23, GO 14218, PI Z. Berta-Thompson). For these transits, we adopted the transit times reported therein. More recently, HST observed a single transit each of Kepler-51c and d using the WFC3/G141 mode (Cycle 30, GO/DD 17585, PI P. Gao \& J. Libby-Roberts). Each observation lasted 6 HST orbits ($\sim$8.8 hours). The Kepler-51c observation was impacted by HST crossing the South Atlantic Anomaly (SAA) during orbits 3 -- 6.  This limited the number of frames to 10 -- 12 per HST orbit.  Otherwise, we obtained 23 frames per orbit.  The Kepler-51c observation was also impacted by a guide star acquisition failure in the first HST orbit.  Fortunately, HST was able to acquire the guide star in the subsequent orbit and the remainder of the observations proceeded normally. We purposefully obtained a partial transit of Kepler-51d due to its long duration.  Since the transit duration is well known, we were able to precisely constrain the mid-transit time by fitting the ingress.

The Kepler-51c and d visits were both observed and reduced using the following procedure. Each HST orbit started with a direct image using the F139M filter (GRISM256, NSAMP=15, SAMP-SEQ=RAPID, 4.167-sec exposures).  We then acquired spectroscopic data using the G141 grism (GRISM256, NSAMP=15, SAMP-SEQ=SPARS10, 103-sec exposures). The faintness of Kepler-51 permitted us to use WFC3's `stare' mode as opposed to the spatial scan mode, which is often done with time-series observations of bright targets.

We used the \texttt{Eureka!} data reduction pipeline \citep{Bell2022} to convert the FLT data files into a time series of 1D spectra.  Because WFC3 was operating in stare mode, there was no benefit to using the IMA files.  Starting with Stage 3, we selected a relatively small subarray region ($x=58$--$200$ pixels, y=$137$--$170$ pixels) to avoid nearby stars in this crowded field.  To search for outliers in the background region, we performed a double-iteration, 5$\sigma$ rejection threshold test for each pixel along the time axis.  We then performed column-by-column background subtraction by masking the region within 8 pixels of the trace and fitting for a constant term for each column.  We further rejected pixels above a 2.5$\sigma$ threshold in order to remove a visibly hot pixel from one of the columns.  We adopted a full-width aperture size of 7 pixels, centered on the trace, and used optimal spectral extraction \citep{Horne1986} to estimate the flux in each frame.

\begin{figure*}
    \epsscale{1.15}
	\plottwo{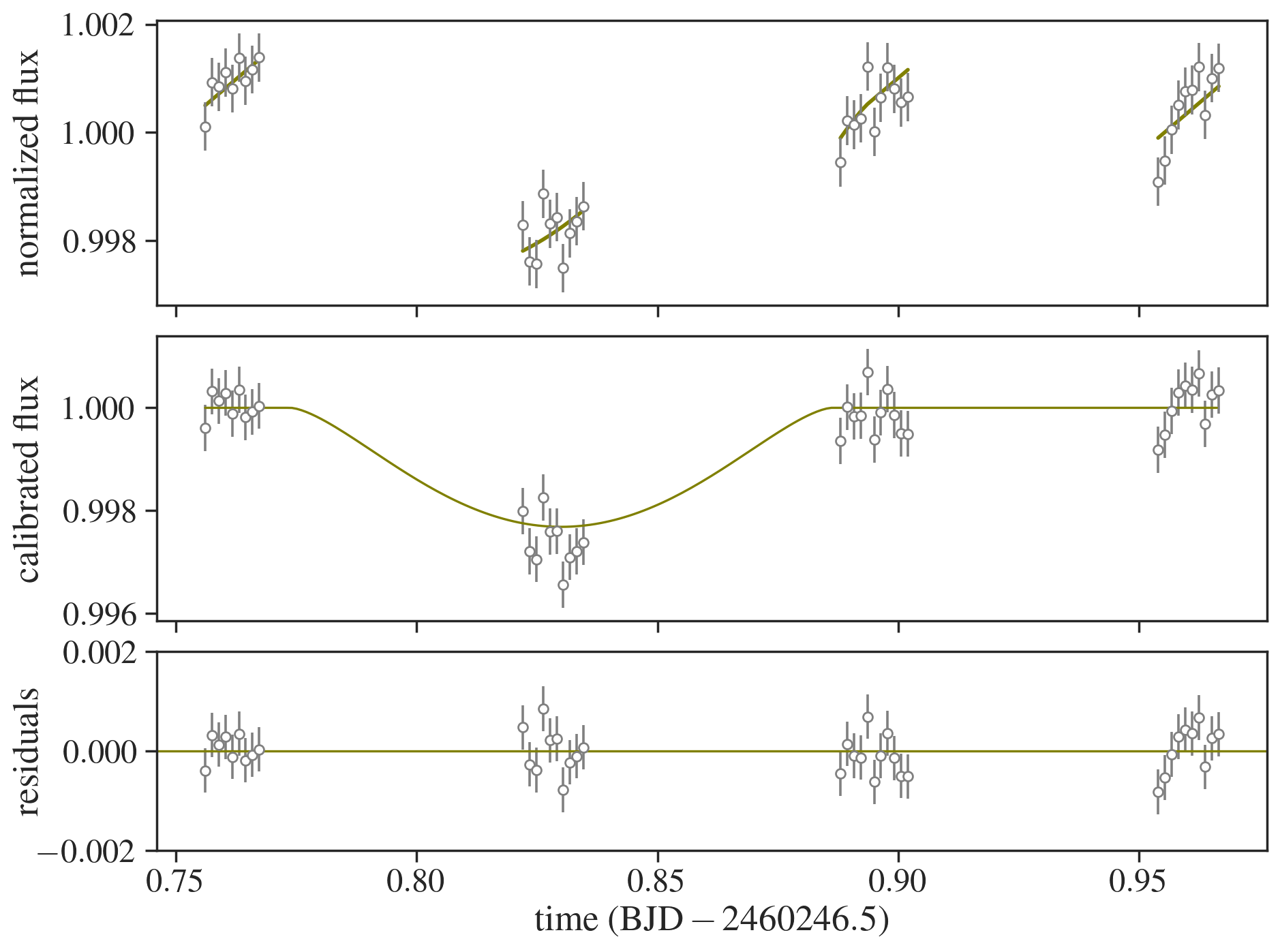}{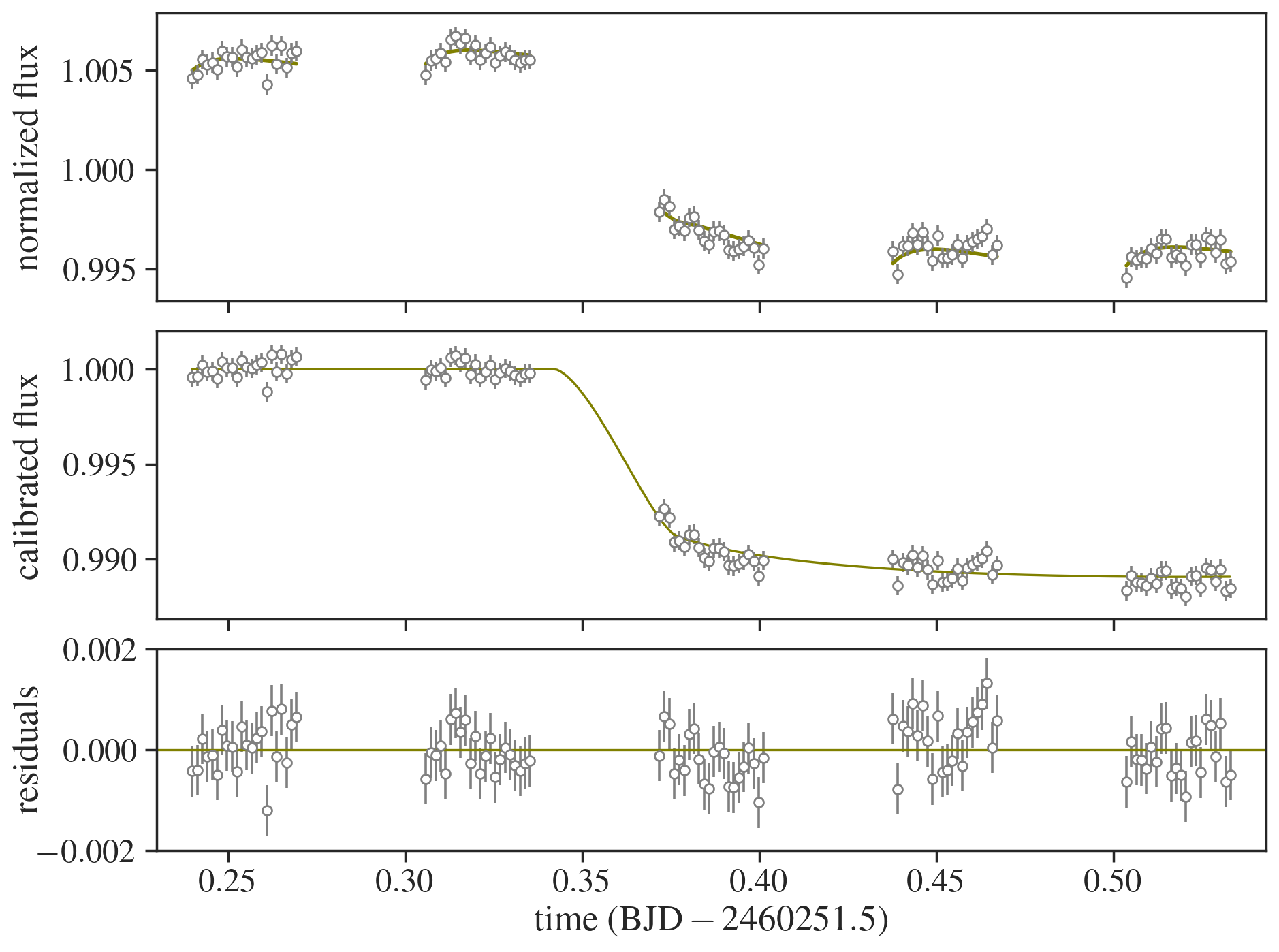}
	\caption{New HST transits of Kepler-51c (left) and d (right). The top panels show the normalized flux (circles) and the model including the transit, polynomial baseline, and the HST ramp (solid line). The middle panels show the data and the model after removing the systematic model (polynomial and ramp). The bottom panels show the residuals of the fits.}
	\label{fig:hst}
\end{figure*}

In Stage 4 of \texttt{Eureka!}, we measured and corrected for a slight drift in the position of the 1D spectra ($<$0.15 pixels) over the course of each observation.  In computing the white light curves, we summed flux from 1.125 -- 1.65$\,\mu\mathrm{m}$. Prior to fitting the white light curves (Stage 5), we manually clipped the first good HST orbit and the first frame within each orbit.  Because the guide star acquisition failed for the first orbit of the Kepler-51c observations, we effectively clipped the second HST orbit, leaving only four good orbits to work with.  We used an exponential plus linear function to fit the repeating ramp in each HST orbit. When fitting the transit shape using \texttt{batman} \citep{batman}, we fixed the orbital period, semi-major axis, and inclination to the values listed in Table~\ref{tab:orbitalParameters}.  We fixed the quadratic limb-darkening parameters to $u_1=0.235$ and $u_2=0.195$ computed by \citet{libbyroberts2020}. We found transit times of 
$2460247.33520^{+0.0043}_{-0.0084}$ for Kepler-51c and
$2460252.02100^{+0.0011}_{-0.0014}$ for Kepler-51d. Figure~\ref{fig:hst} plots the Kepler-51c and d HST light curves with the best-fit model derived from this analysis.

We also performed two additional, independent HST analyses and the mid-transit times agree to within one sigma. The choice of different reduction mid-transit times does not meaningfully change our TTV inference.

\begin{table}
    \centering
    \caption{Fixed orbital parameters for Kepler-51c and Kepler-51d from \citet{masuda2014}. These parameters were fixed for our HST white light curve fits.}
    \begin{tabular}{ccc}
        \hline\hline
        Orbital Parameter       & Kepler-51c    & Kepler-51d    \\
        \hline
        period (days)           & 85.31         & 130.18        \\
        inclination (\degree)   & 89.38         & 89.885        \\
        semi-major axis ($R_\star$) & 94.1          & 124.7         \\
        \hline
    \end{tabular}
    \label{tab:orbitalParameters}
\end{table}

\subsection{Kepler}\label{ssec:obs_kepler}

Transit times based on the Kepler observations have been reported in multiple works \citep[e.g.,][]{masuda2014, 2016ApJS..225....9H, libbyroberts2020}. Here we reanalyze the whole set of available Kepler light curves in a uniform manner, since some of the transits (including the simultaneous transit of b and d) were not incorporated in these analyses. 

All available Pre-search Data Conditioning (PDC) light curves for Kepler-51 were downloaded from the Mikulski Archive for Space Telescopes\footnote{\url{https://archive.stsci.edu}} using the {\tt Lightkurve} package \citep{2018ascl.soft12013L}. 
The time series show a $\sim 1\%$ quasi-periodic variation that has been attributed to starspots rotating in and out of view in sync with the 8.2-$\,\mathrm{day}$ rotation of the star \citep{2014ApJS..211...24M}. 
We first normalized the flux in each quarter by its median, and removed the quasi-periodic modulation by fitting the data from each quarter separately using a Gaussian process \citep[GP,][]{2022arXiv220908940A} and by subtracting the mean prediction from it. We excluded the data with non-zero quality flags assigned, and masked the data around known transits when computing the likelihood for fitting. The width of the transit mask was chosen to be 1.5 times the transit duration reported in the Kepler Object of Interest catalog. Here we assumed the following likelihood function given by a multivariate normal distribution $\mathcal{N}$:
\begin{equation}
    \label{eq:likelihood_kepler}
    \mathcal{L}(\mu, \alpha, \rho, \sigma_{\rm jit}) = \mathcal{N}(f; \mu, \Sigma),
\end{equation}
The flux $f$ is a vector of a length equal to the number of data points in a quarter. The mean flux $\mu$ is a vector of the same length, here taken to be constant and characterized by a single scalar. The $ij$ elements of the covariance matrix $\Sigma$ consists of the Mat\'ern-3/2 covariance function and the white-noise term:
\begin{equation}
    \label{eq:kernel}
    \Sigma_{ij}=(\sigma_i^2+\sigma_{\rm jit}^2)\delta_{ij}
    +\alpha^2 \left(1+{|t_i-t_j|\over 3\rho}\right)
    \exp\left(-{|t_i-t_j|\over 3\rho}\right),
\end{equation}
where $\delta_{ij}$ is the Kronecker delta, $t_i$ and $\sigma_i$ are the time and assigned flux error of the $i$th data point, and $\sigma_{\rm jit}$ accounts for the flux scatter in excess of the assigned error, if any.
The log-likelihood $\ln\mathcal{L}$ was evaluated using the {\tt JAX} \citep{jax2018github} interface of {\tt celerite2} \citep{celerite, celerite2}, and the scalar parameters $\mu$, $\ln \sigma_{\rm jit}$, $\alpha$, and $\rho$ were optimized using a truncated Newton algorithm as implemented in {\tt jaxopt} \citep{jaxopt_implicit_diff}. The mean of the predictive distribution conditioned on $f$ was computed for the optimal set of the parameters and was subtracted from $f$. The same procedure was applied to both long- and short-cadence data.
We then extracted the normalized and detrended flux within $0.5\,\mathrm{days}$ around all the known transits of the three planets for the subsequent analysis.

\begin{figure*}[h!]
	\epsscale{1.15}
	\plotone{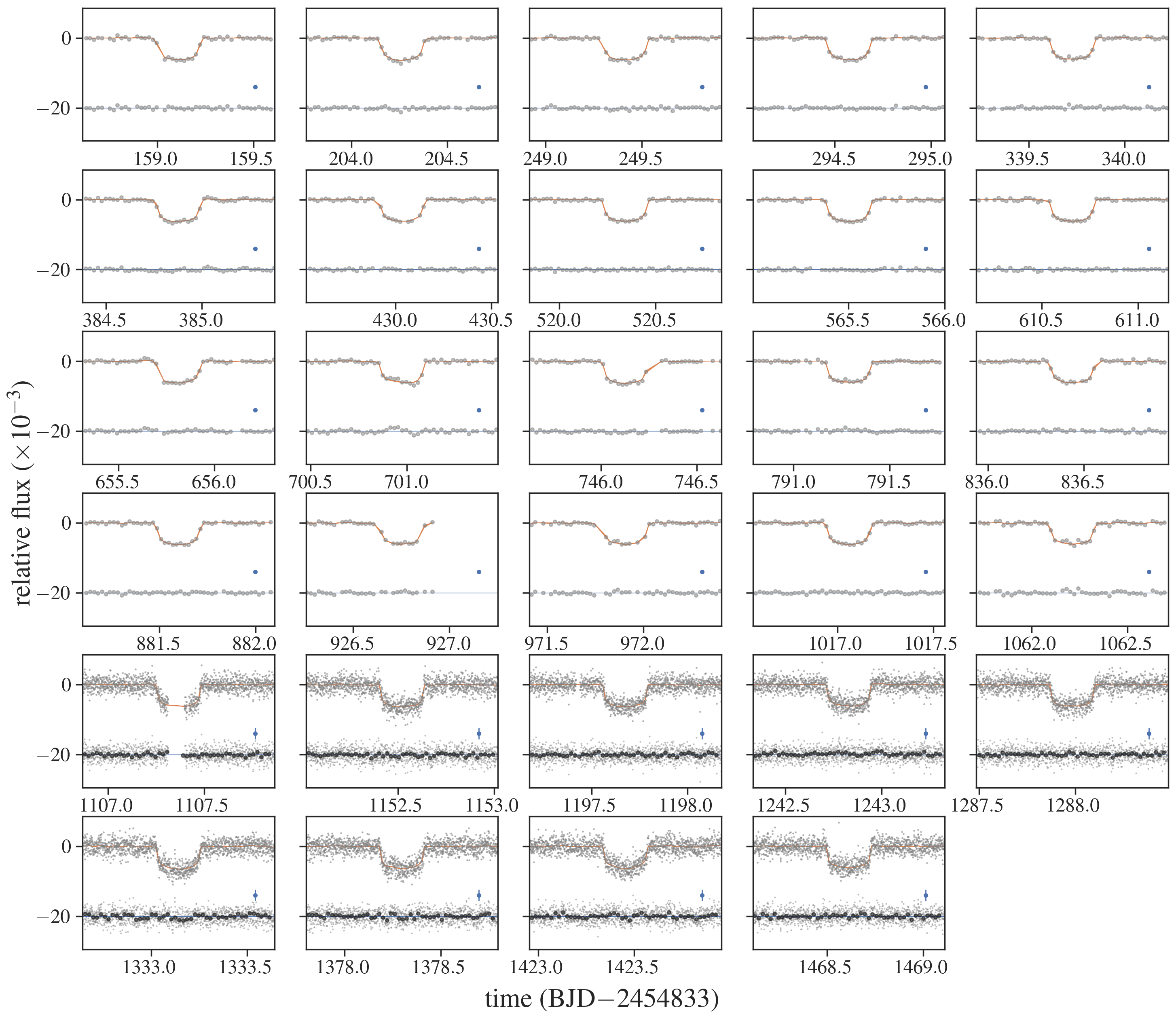}
	\caption{Transit light curves of Kepler-51b from Kepler. The orange lines show 20 GP-included models drawn from the posterior. Residuals from the mean model without GP are shown with offsets. The blue dots above the residuals show the median of the errors assigned to the fluxes. The data around the simultaneous transit of Kepler-51b and Kepler-51d are shown in Figure~\ref{fig:keptransit3}.}
	\label{fig:keptransit1}
\end{figure*}

\begin{figure*}[h!]
    \epsscale{1.15}
	\plotone{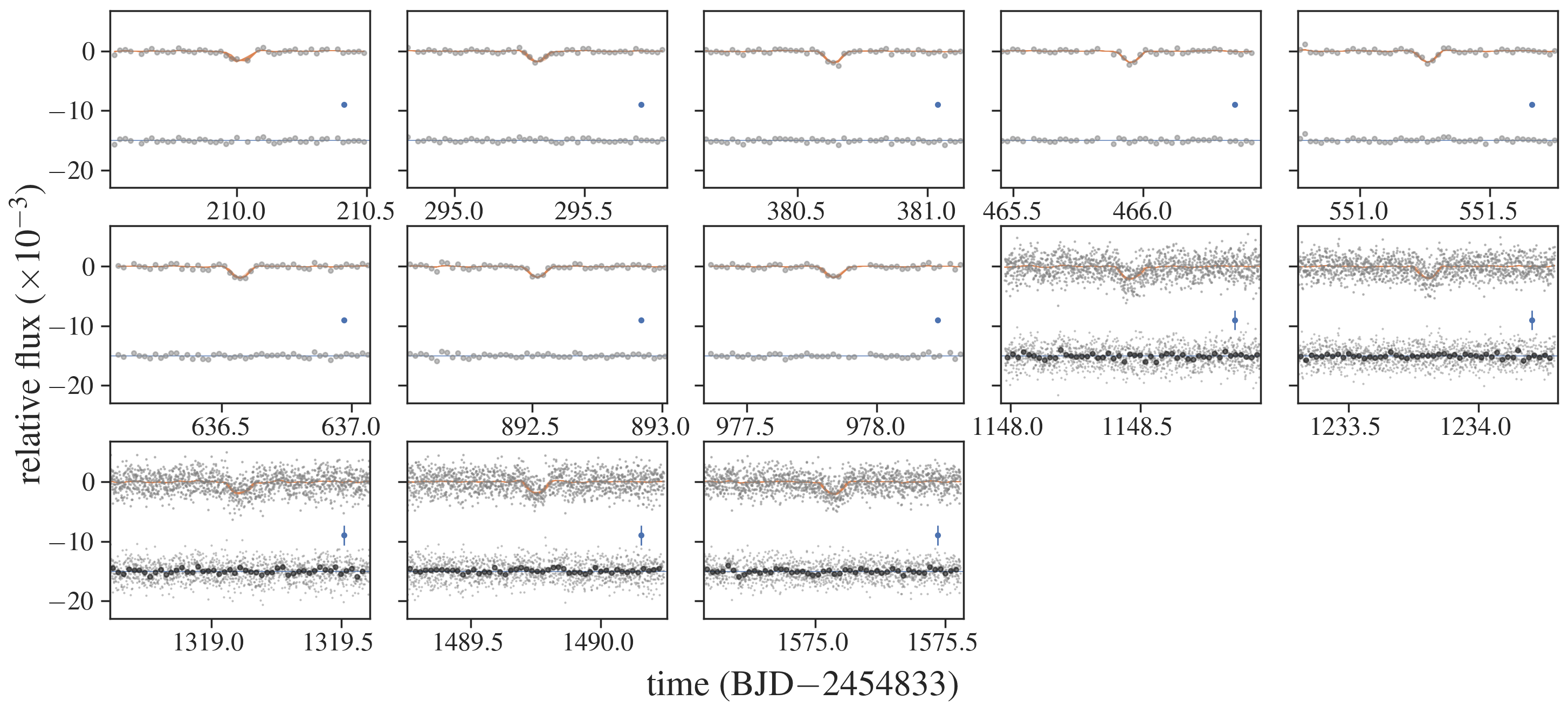}
    \caption{Same as Figure~\ref{fig:keptransit1}, but for Kepler-51c.}
    \label{fig:keptransit2}
    \vspace{0.5cm}
	\plotone{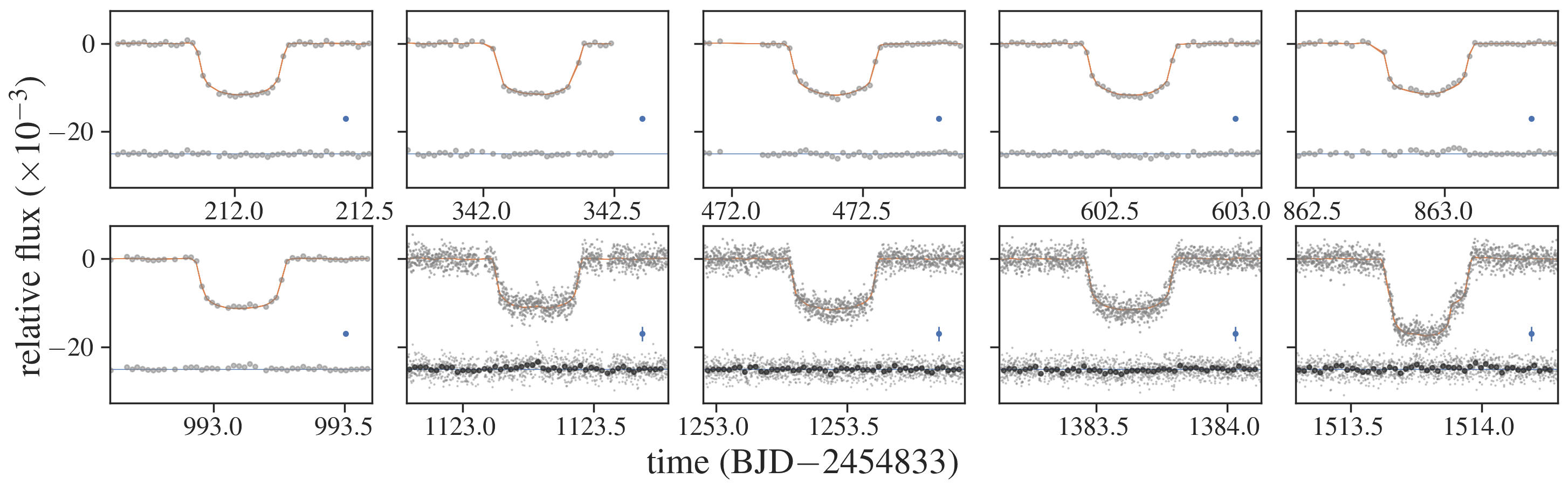}
	\caption{Same as Figure~\ref{fig:keptransit1}, but for Kepler-51d. In the last panel, both Kepler-51b and d are transiting simultaneously.}
	\label{fig:keptransit3}
\end{figure*}

We estimated transit times by fitting the transit light curves with the flux loss model $f_{\rm model}$ for a quadratically limb-darkened star \citep{2020AJ....159..123A}. The model was computed assuming circular Keplerian orbits, except that each transit was shifted in time to account for TTVs before being compared to the data. These timing shifts, $\Delta t_{ij}$ ($i$ is the index for planets and $j$ for transits), for all transits were optimized along with the impact parameters $b$ and radius ratios $r$ of the three transiting planets, and the mean density $\rho_\star$ and two limb-darkening coefficients $q_1$ and $q_2$ \citep{2013MNRAS.435.2152K} of the star. Here the linear ephemeris ($t_0$ and $P$) was fixed to the values obtained from fitting without TTVs, because they are completely degenerate with $\Delta t$; what matters is only the combination $t_{0,i} + P_i \cdot j + \Delta t_{ij}$, and this is the transit time reported in Table~\ref{tab:transit_times}. For the long-cadence data, we oversampled the model by a factor of 11 to account for the finite exposure time, and the simultaneous transit by planet b and d were modeled simply by summing the flux losses from the two planets (i.e., without considering the overlap). The noise (i.e., difference between the observed flux and the model light curve $f_{\rm model}$) was again modeled using a GP with the kernel in Eq.~\ref{eq:kernel}; the likelihood function is:
\begin{equation}
    \mathcal{L}(\Delta t, b, r, \rho_\star, q_1, q_2, \mu, \alpha, \rho, \sigma_{\rm jit}) 
    = \mathcal{N}(f-f_{\rm model}; \mu, \Sigma),
\end{equation}
We draw posterior samples for all the parameters including $\Delta t$ adopting uniform priors, where $|\Delta t|$ was restricted to be less than $100$~minutes. The sampling was performed using Hamiltonian Monte Carlo (HMC) and the No-U-Turn Sampler \citep{DUANE1987216, 2017arXiv170102434B} as implemented in {\tt NumPyro} \citep{bingham2018pyro, phan2019composable}. The long- and short-cadence data were modeled separately, and we adopted transit times from the short-cadence data when available. The results of this analysis are presented in Figures~\ref{fig:keptransit1}--\ref{fig:keptransit3} and Table~\ref{tab:transit_times}. The estimated mid-transit times and their errors agreed well with those reported in \citet{libbyroberts2020} for the transits that were also analyzed in that work: the difference in the best estimates normalized by the error bars is $0.05\pm0.45$, and the ratio of the error bars is $1.0\pm0.2$.

To obtain the prior information on the radius ratios, transit durations (here taken to be the total duration $T_{14}$ defined in \citet{2011exop.book...55W}), and impact parameters for the analyses of the ground-based transits (Section \ref{ssec:obs_ground}), we fit their marginal posteriors from the long-cadence data with normal distributions and truncated normal distributions with a lower bound of zero. The estimated means and standard deviations (location and scale parameters) are summarized in Table~\ref{tab:shape}. 
The mean stellar density inferred from the long-cadence and short-cadence data are $1.72\pm0.04\,\rho_\odot$ and $1.68\pm0.06\,\rho_\odot$ (mean and standard deviation), respectively. These values are consistent with the stellar mass and radius derived in Section~\ref{sec:star} accounting for the systematic errors, and with low orbital eccentricities ($\lesssim$ a few \%) inferred from the TTV modeling.

\begin{deluxetable}{l@{\hspace{.5cm}}cc}[!ht]
\tablecaption{Transit shape parameters from the Kepler long-cadence data.\label{tab:shape}}
\tablehead{
\colhead{Parameter} & \colhead{Location} & \colhead{Scale}
}
\startdata
\textit{(Kepler-51b)}\\
radius ratio   &  0.07225 &  0.00030\\
impact parameter   & 0.074&  0.072\\
total transit duration (days)  & 0.23975  & 0.00084\\
\textit{(Kepler-51c)}\\
radius ratio\tablenotemark{$\ast$}  & 0.068&   0.015\\
impact parameter\tablenotemark{$\ast$} & 0.988  &  0.020\\
total transit duration (days) & 0.1126 & 0.0035\\
\textit{(Kepler-51d)}\\
radius ratio   & 0.09857 & 0.00037\\
impact parameter  & 0.003  &  0.095\\
total transit duration (days) & 0.3501 & 0.0012\\
\enddata
\tablecomments{
For the impact parameters, the location and scale parameters are for the truncated normal distribution with the lower bound of zero. 
}
\tablenotetext{*}{The marginal posteriors for these parameters are not well approximated by (truncated) normal distributions due to the grazing nature of Kepler-51c.}
\end{deluxetable}


\subsection{Ground-Based Facilities}\label{ssec:obs_ground}

We observed 4 transits of Kepler-51b, 3 transits of c, and 3 transits of d with a variety of ground-based facilities between 2018 and 2023, which we detail below.

\subsubsection{Teide Observatory}\label{ssec:obs_ground_teide}

\begin{figure*}
    \epsscale{1.12}
    \plotone{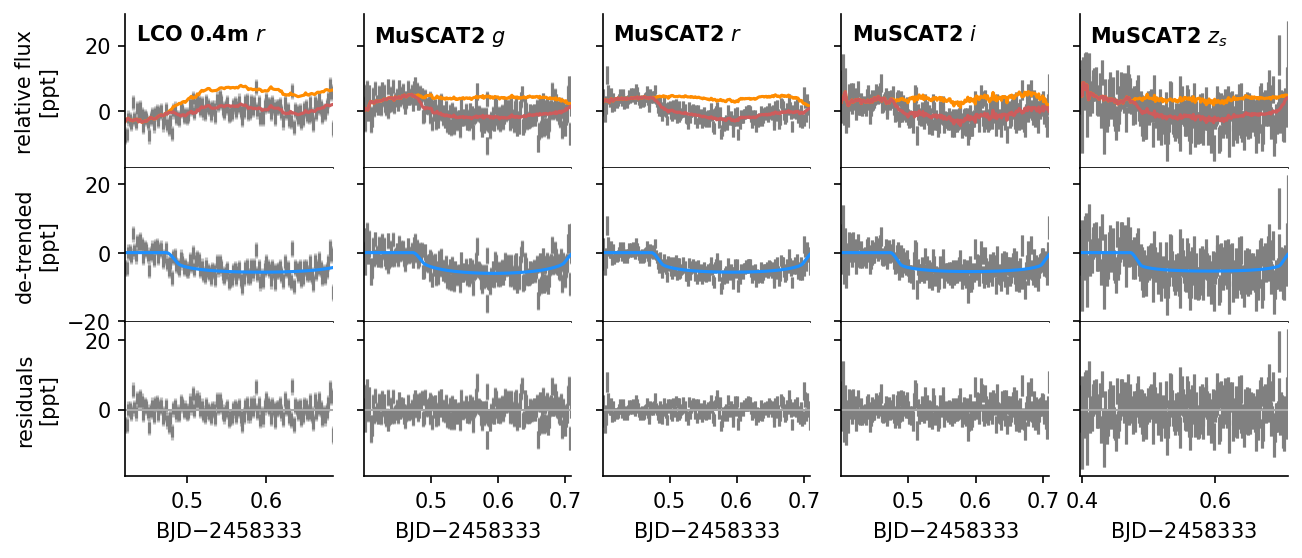}
    \caption{Transit of Kepler-51b observed with LCO ($r$-band) and MuSCAT2 ($g$, $r$, $i$, $z_\mathrm{s}$; from left to right) on UT 2018 August 2. 
    The gray lines show the measured flux values and their uncertainties, with the quadrature sum of the observed uncertainties and the best-fit jitter value shown in light gray. The top panels show the systematic model (orange) and the systematic + transit model (red). The middle panels compare the de-trended flux (measured flux minus the systematic model) with the transit model (blue). The bottom panels show the residuals.
    }
    \label{fig:180802-m2-lco}
\end{figure*}

\begin{figure*}
    \epsscale{0.71}
    \plotone{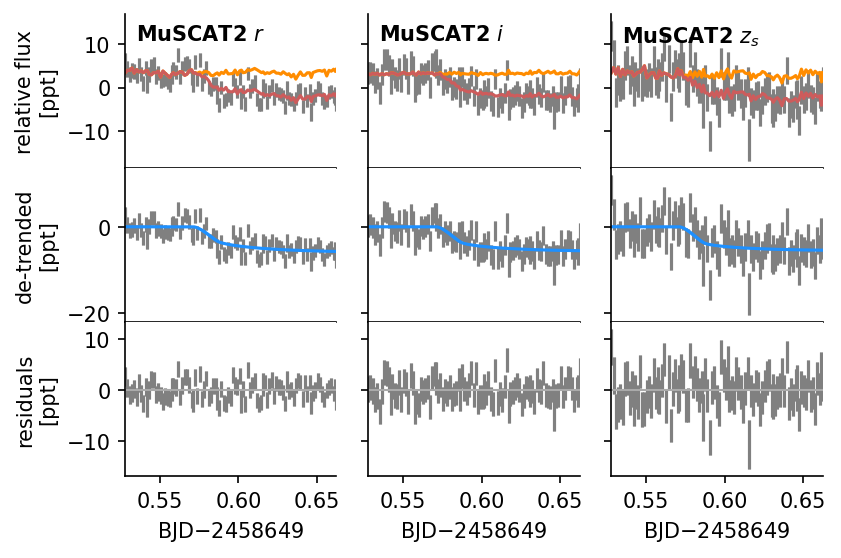}
    \caption{Transit of Kepler-51b observed with MuSCAT2 ($r$, $i$, $z_\mathrm{s}$ bands; from left to right) on UT 2019 June 14. The format of this figure is the same as Figure~\ref{fig:180802-m2-lco}.}
    \label{fig:190614-m2}
\end{figure*}

\begin{figure*}
    \epsscale{0.71}
    \plotone{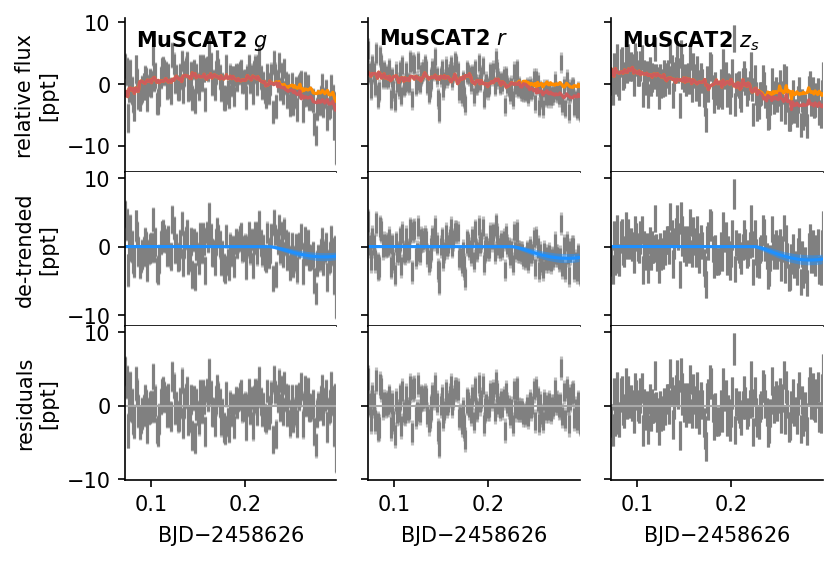}
    \caption{Transit of Kepler-51c observed with MuSCAT2 ($g$, $r$, $z_\mathrm{s}$ bands; from left to right) on UT 2019 May 22. The format of this figure is the same as Figure~\ref{fig:180802-m2-lco}.}
    \label{fig:190522-m2}
\end{figure*}

\begin{figure*}
    \epsscale{0.9}
    \plotone{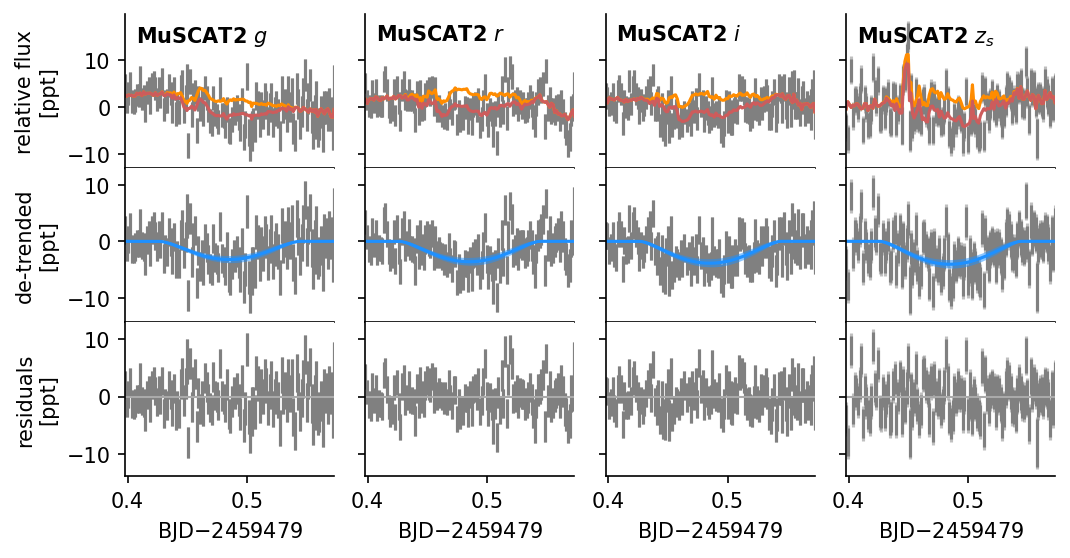}
    \caption{Transit of Kepler-51c observed with MuSCAT2 ($g$, $r$, $i$, $z_\mathrm{s}$ bands; from left to right) on UT 2019 September 21. The format of this figure is the same as Figure~\ref{fig:180802-m2-lco}.}
    \label{fig:210921-m2}
\end{figure*}

\begin{figure*}
    \epsscale{0.9}
    \plotone{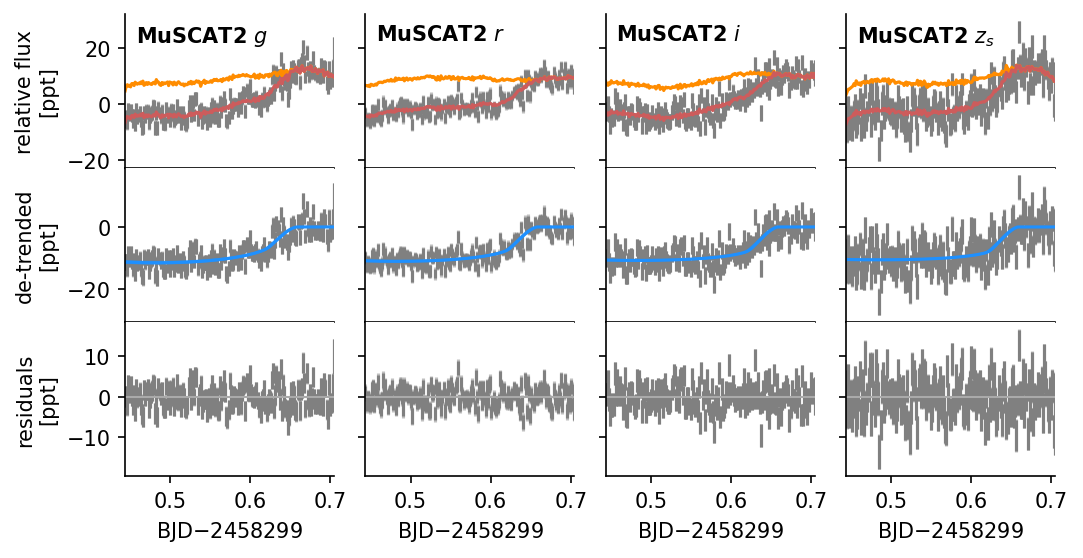}
    \caption{Transit of Kepler-51d observed with MuSCAT2 in $g$, $r$, $i$, and $z_\mathrm{s}$ bands (from left to right) on UT 2018 June 29. The format of this figure is the same as Figure~\ref{fig:180802-m2-lco}.}
    \label{fig:180629-m2}
\end{figure*}

We used the MuSCAT2 multiband imager \citep{Narita2018} installed at the 1.52~m Telescopio Carlos Sanchez (TCS) at the Teide Observatory, Spain, to observe transits of the Kepler-51 planets simultaneously in $g$, $r$, $i$, and $z_\mathrm{s}$. Transits of planet b were observed on the nights of UT 2018 August 2 (Figure~\ref{fig:180802-m2-lco}) and 2019 June 14 (Figure~\ref{fig:190614-m2}); planet c was observed on 2019 May 22 (Figure~\ref{fig:190522-m2}) and 2021 September 21 (Figure~\ref{fig:210921-m2}); and planet d was observed on 2018 June 29 (Figure~\ref{fig:180629-m2}). Exposure times (typically 30-60 seconds) were chosen to avoid saturation and maximize the duty cycle, while enabling precise auto-guiding of the telescope. However, they varied from night to night depending on the atmospheric seeing conditions and focus setting of the telescope. The images were calibrated and light curves were produced following \citet{2011Fukui}.

We fit the MuSCAT2 light curves using \texttt{PyMC3} \citep{pymc3}, \texttt{exoplanet}\footnote{\url{https://docs.exoplanet.codes/en/stable/}} \citep{exoplanet}, and \texttt{starry} \citep{luger18}. We obtained transit timing measurements by simultaneously fitting the light curves from all bandpasses of each observation/dataset. We used a wide uniform prior for the time of mid-transit, Gaussian priors for the transit duration and planet-to-star radius ratio, and a truncated Gaussian prior on the impact parameter ($X \sim \left [\max(0,N(\mu,\sigma),\min(1,N(\mu,\sigma) \right]$); see Table~\ref{tab:shape}. We computed quadratic limb darkening coefficients based on interpolation of the parameters tabulated by \citet{Claret2012}, using the effective temperature, metallicity, and surface gravity of Kepler-51 from \citet{libbyroberts2020}. We included separate white noise jitter parameters for each band to account for underestimated measurement errors and wavelength-dependent systematic noise. Systematics arising from variable atmospheric conditions and intrapixel gain non-uniformity were modeled as a linear combination of auxiliary variables (airmass, centroids, FWHM, and peak flux), as well as either a linear function of time or a third degree basis spline with five equally spaced knots. We used the gradient-based {\tt BFGS} algorithm \citep{NoceWrig06} implemented in {\tt scipy.optimize} to find initial maximum a posteriori (MAP) parameter estimates. We used these estimates to initialize an exploration of parameter space via ``no U-turn sampling'' \citep[NUTS,][]{HoffmanGelman2014}, an efficient gradient-based Hamiltonian Monte Carlo (HMC) sampler implemented in {\tt PyMC3}. The resulting chains had Gelman-Rubin statistic \citep{GelmanRubin1992} values of $<$1.01, indicating they were well-mixed.

\subsubsection{Las Cumbres Observatory Global Telescope}\label{ssec:obs_ground_lco}

 We used the SBIG camera mounted on a 0.4\,m LCOGT telescope of the Las Cumbres Observatory Global Telescope (LCOGT) network to observe a transit of Kepler-51b on 2018 August 2 (Figure~\ref{fig:180802-m2-lco}). The telescope we used was physically located at Teide Observatory in Spain. We observed in $r$ band with an exposure time of 150 seconds. The images were calibrated using the standard LCOGT BANZAI pipeline \citep{2018SPIE10707E..0KM}, and light curves were produced following \citet{2011Fukui}. As this dataset covered the same transit that was observed with MuSCAT2, we analyzed the data jointly with the MuSCAT2 light curves from that night, as described above.
 
\begin{figure*}
    \epsscale{0.9}
    \plotone{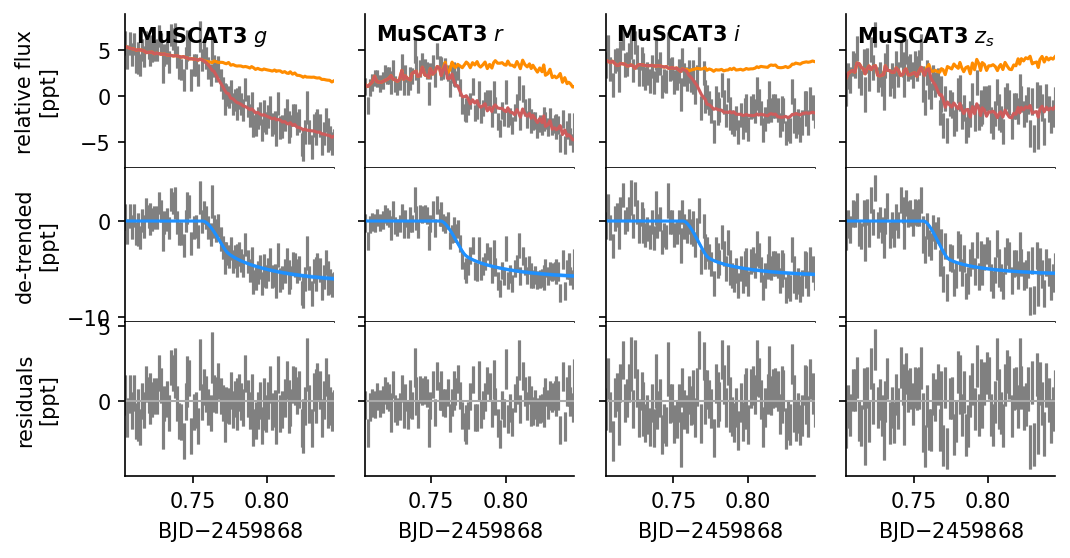}
    \caption{Transit of Kepler-51b observed with MuSCAT3 ($g$, $r$, $i$, $z_\mathrm{s}$ bands; from left to right) on UT 2022 October 15. The format of this figure is the same as Figure~\ref{fig:180802-m2-lco}.}
    \label{fig:221015-m3}
\end{figure*}

 We used the MuSCAT3 multi-band imager \citep{Narita2020} mounted on the LCOGT 2\,m Faulkes Telescope North at Haleakala Observatory on Maui, Hawai'i to observe a transit of Kepler-51b on UT 2022 October 15 simultaneously in the $g$, $r$, $i$, $z_\mathrm{s}$ bands, using exposure times of 60 seconds (Figure~\ref{fig:221015-m3}). The images were calibrated using the LCOGT BANZAI pipeline, and light curves were produced following \citet{2011Fukui}. We measured the mid-transit time by simultaneously fitting the four MuSCAT3 light curves, as was done for the MuSCAT2 light curves (see Section~\ref{ssec:obs_ground_teide}).

\subsubsection{Oukaïmeden Observatory}\label{ssec:obs_ground_tn}






We observed an egress of Kepler-51d on 2020 August 18 with the TRAPPIST-North (TRAnsiting Planets and PlanetesImals Small Telescope) 60-cm robotic telescope at the Oukaïmeden Observatory in Morocco \citep{Barkaoui2019_TN, Jehin2011, Gillon2011} as part of TESS Follow-up Observing Program (TFOP-SG1) photometric follow-up (Figure~\ref{fig:200818-trappistn-acton}). The transit was observed in the I+z filter with an exposure time of 120s. We reduced the photometry data and performed aperture photometry using the {\tt PROSE} pipeline \citep{garcia2021} assuming an 4.1 arcsecond aperture and using 5 comparison stars.
\edit1{
We measured the mid-transit time by simultaneously fitting the TRAPPIST-North light curve along with other light curves of the same transit described in Sections~\ref{ssec:obs_ground_acton}--\ref{ssec:obs_ground_keplercam}), as was done for the MuSCAT2 light curves (see Section~\ref{ssec:obs_ground_teide}).
}

\subsubsection{Acton Sky Portal Observatory}\label{ssec:obs_ground_acton}









\begin{figure*}
    \epsscale{1.1}
    \plotone{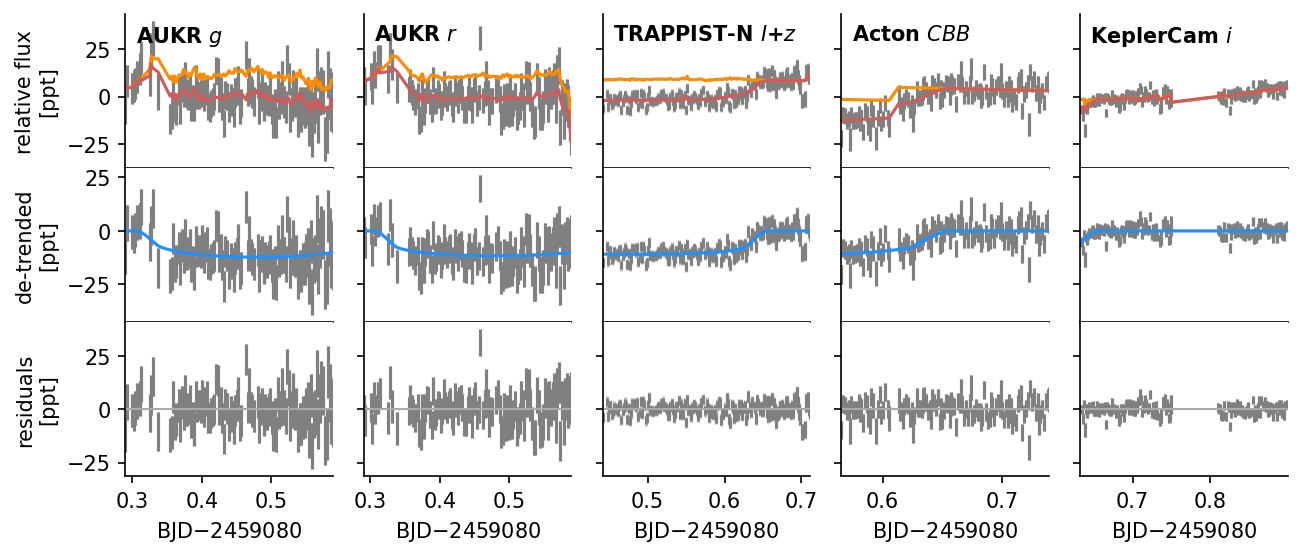}
    \caption{
    \edit1{Transit of Kepler-51d observed from the AUKR observatory (g' and r' filters), the TRAPPIST-North telescope (I+z filter), the Acton Sky Portal observatory (blue blocking clear Astrodon filter), and FLWO KeplerCam (Sloan i) on UT 2020 August 18. 
    }
    The format of this figure is the same as Figure~\ref{fig:180802-m2-lco}.}
    \label{fig:200818-trappistn-acton}
\end{figure*}

The same 2020 August 18 egress of Kepler-51d was observed using the 0.36-m telescope on the Acton Sky Portal private observatory in Acton, Massachusetts with a blue blocking clear Astrodon exoplanet filter (Figure~\ref{fig:200818-trappistn-acton}). We used 60-second exposure times and used \texttt{AstroImageJ} \citep{2017AJ....153...77C} to reduce and perform aperture photometry on the images. The egress ended $\sim$36 minutes early, and we experienced a meridian flip during the start of egress, which we accounted for during fitting.

\subsubsection{AUKR}

\edit1{
We observed the same 2020 August 18 transit of Kepler-51d in $g'$ and $r'$ filters with the Apogee Alta U47 CCD camera attached to the T80 Telescope located at the Ankara University Kreiken Observatory (AUKR). The exposure time for both $g'$ and $r'$ filters is 90 seconds. We used {\tt AstroImageJ} \citep{2017AJ....153...77C} to reduce and perform aperture photometry on the images. The reduced data includes an ingress and partial transit (Figure~\ref{fig:200818-trappistn-acton}). 
}

\subsubsection{KeplerCam} \label{ssec:obs_ground_keplercam}

\edit1{
We used the KeplerCam CCD mounted on the 1.2m telescope at the Fred Lawrence Whipple Observatory (FLWO) at Mount Hopkins, Arizona to observe the same 2020 August 18 transit of Kepler-51d. KeplerCam is a 4096$\times$4096 Fairchild CCD 486 detected which has a field-of-view of 23’$\times$23’ and an image scale of 0.672''/pixel when binned by two. A part of the transit egress along with long-baseline out-of-transit was observed (Figure~\ref{fig:200818-trappistn-acton}). Images were taken in the Sloan $i$ band with 100 second exposure time. Data were reduced using standard IDL routines and aperture photometry was performed using {\tt AstroImageJ} \citep{2017AJ....153...77C}.
}

\subsubsection{Apache Point Observatory}\label{ssec:obs_ground_apo}

\begin{figure*}
    \epsscale{1.15}
    \plottwo{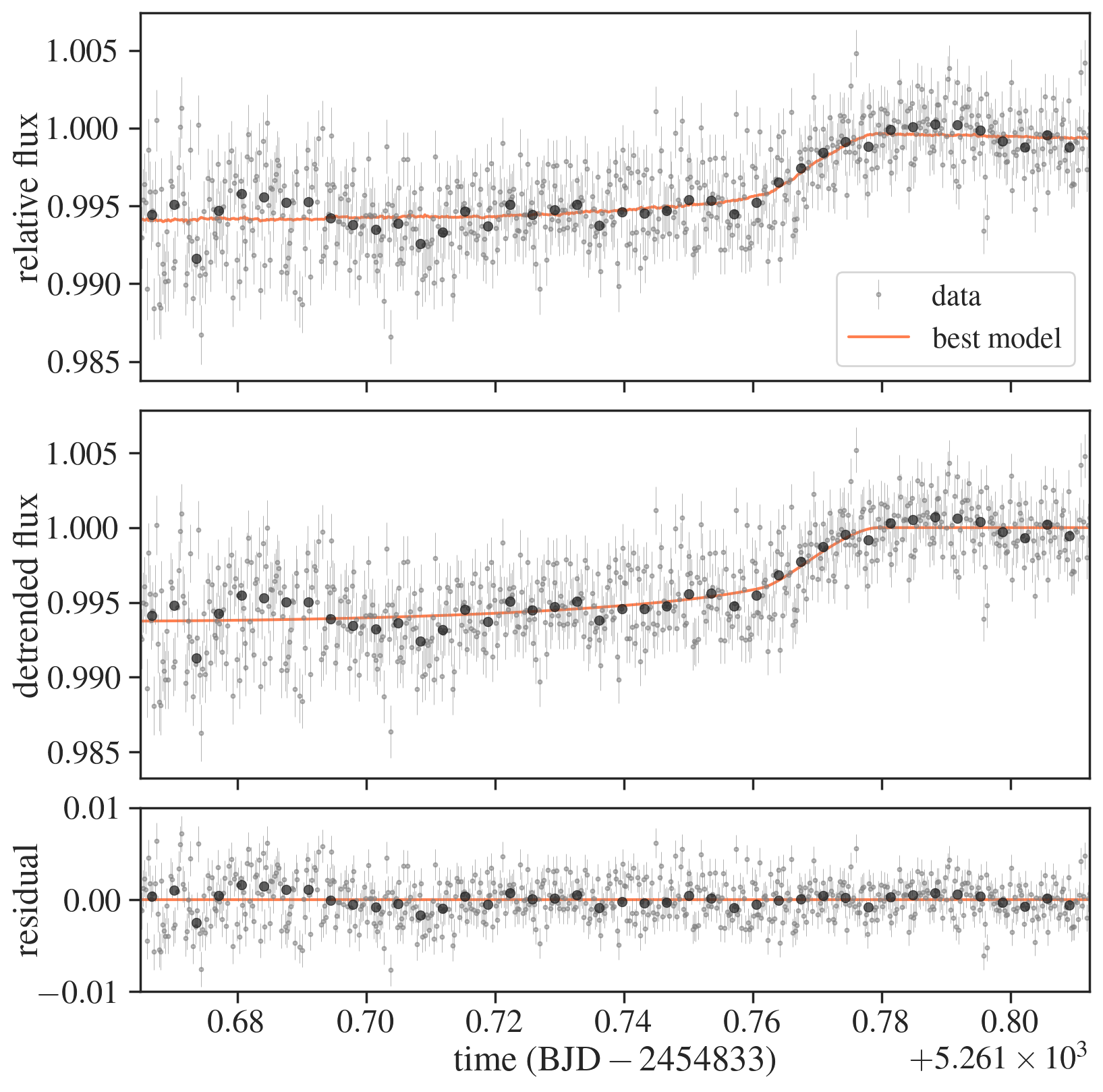}{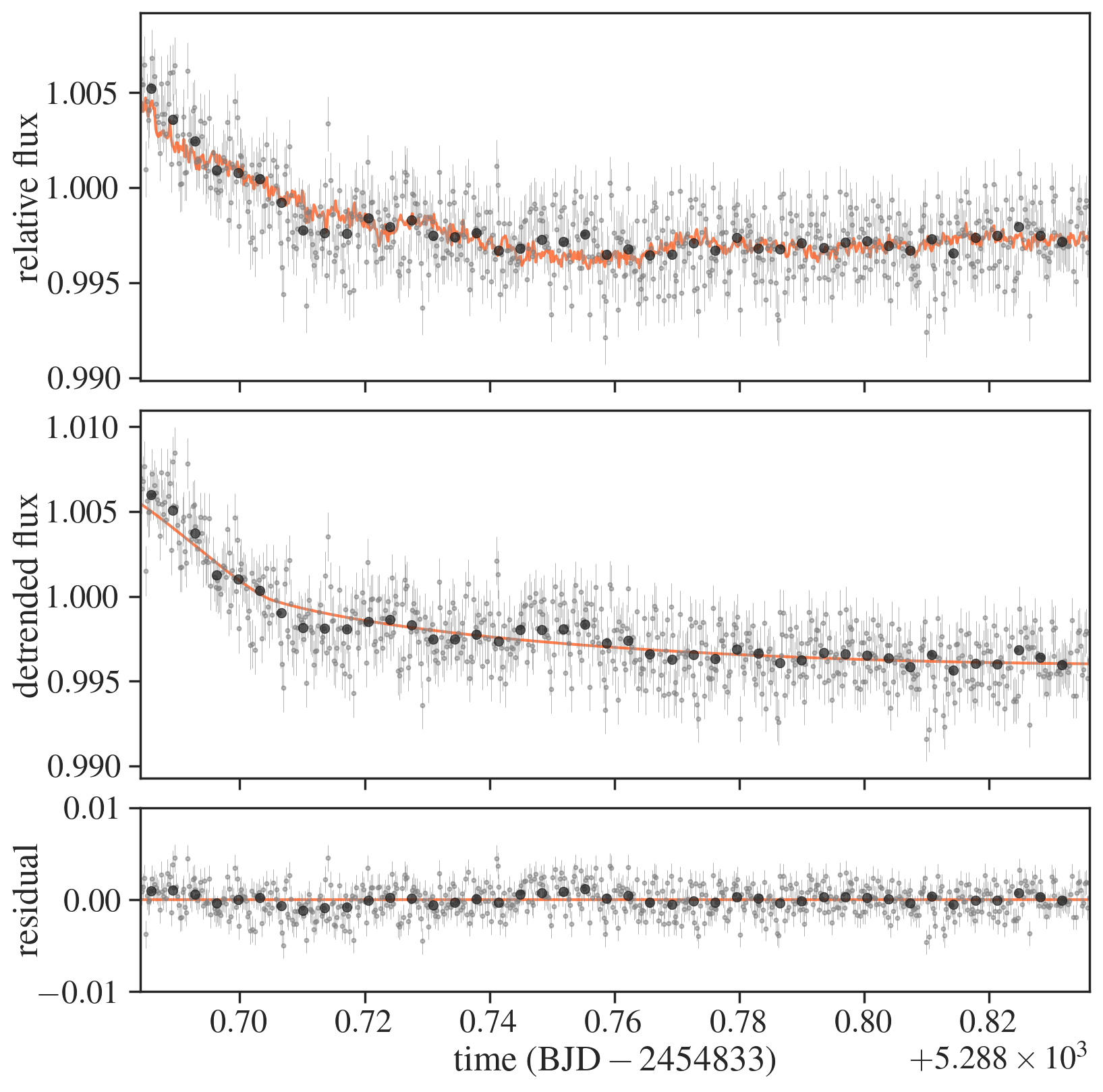}
    \caption{Partial transits of Kepler-51b (left) and -51d (right) observed with ARCTIC on the 3.5m APO Telescope. 
    The top panels show the normalized flux (gray dots) and the best-fit transit + systematics model (red solid line). The middle panels show the data and the model after removing the systematic model. The bottom panels show the residuals of the fits.
    The larger black points represent 5-minute binned data. Due to the transit time and duration, only the ingress and partial transit was observed for Kepler-51d, without any pre-transit baseline.}
    \label{fig:apo}
\end{figure*}

We observed partial transits of Kepler-51b and d with the ARCTIC optical CCD camera \citep{arctic.camera} attached to the 3.5-m ARC telescope at the Apache Point Observatory (APO) in Sunspot, New Mexico on 2023 May 30 and 2023 June 26 respectively -- simultaneous with the JWST observations. Due to scheduling we caught the transit-egress-baseline for Kepler-51b and ingress-transit for Kepler-51d (Figure~\ref{fig:apo}). We used the SDSS $r'$ filter with 15-second exposures and 4x4 detector binning. Due to a nearby companion, we kept the PSF's FWHM to $\sim$3 arcseconds over the night to minimize blending by nudging the focus to account for changes in the seeing. Images were bias-subtracted and flat-fielded\footnote{Dark current does not affect ARCTIC images with exposure times $<$60 seconds.} before we used \texttt{AstroImageJ} \citep{2017AJ....153...77C} to perform differential aperture photometry assuming a 4-arcsecond aperture radius. Flux errors were assumed to be a combination of photon noise from the star, background, and read-noise of the instrument. 

We modeled both light curves using the \texttt{batman} package, holding the transit duration constant to values reported in \citet{libbyroberts2020} and limb darkening to those reported by \texttt{ldtk} \citep{ldtk} for Kepler-51 with the SDSS $r'$ bandpass. We detrended the data using airmass, FWHM, sky, and centroid positions as a function of time and fit for the transit depth and mid-transit times for both transits. 
We determined mid-transit times of 2460094.6574 $\pm$ 0.0011 BJD for Kepler-51b and 2460121.8492$^{+0.0032}_{-0.0043}$ BJD for Kepler-51d. 
The latter value is consistent with that from the JWST transit reported in Section~\ref{ssec:obs_jwst}.
As JWST provides a tighter constraint, we opted to use the JWST time for Kepler-51d for the rest of the analysis.

\subsubsection{Palomar}\label{ssec:obs_ground_palomar}

We observed a transit of Kepler-51c in \textit{J} with the Wide-field Infrared Camera (WIRC) on the Hale Telescope at Palomar Observatory, California, USA. Palomar/WIRC is a 5.08-m telescope equipped with a 2048 x 2048 Rockwell Hawaii-II NIR detector, providing a field of view of 8.'7 x 8.'7 with a plate scale of 0''.25 per pixel \citep{wilson2003}. Our data were taken with a beam-shaping diffuser that increased our observing efficiency and improved the photometric precision and guiding stability \citep{stefansson2017,vissapragada2020}.

\begin{figure}
    \epsscale{1.15}
    \plotone{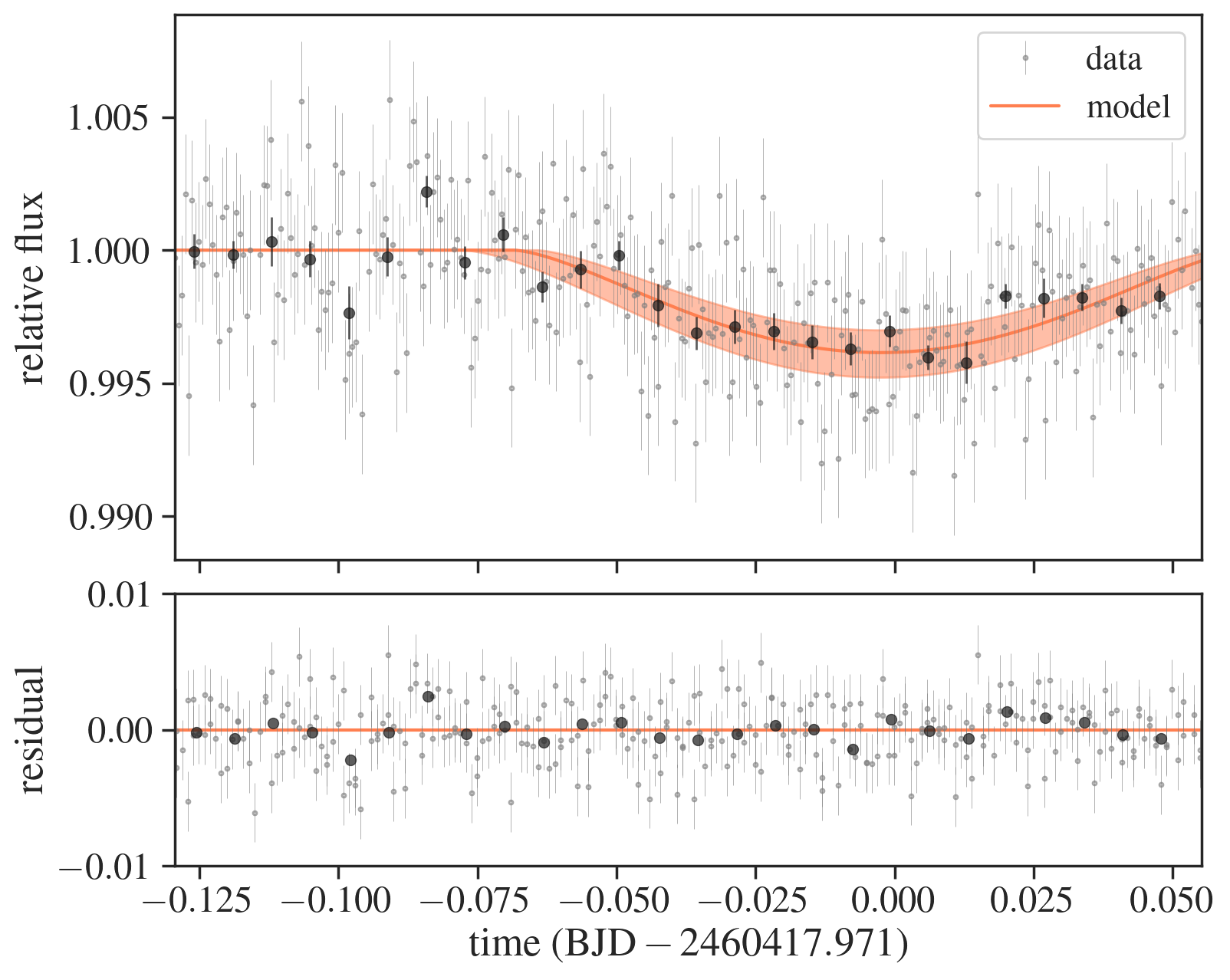}
    \caption{Full transit of Kepler-51c observed WIRC on the Hale Telescope at Palomar Observatory. The larger black points represent 10-minute binned data and the mean and standard deviation of the posterior models are plotted in red. Residuals are plotted in the bottom panels.}
    \label{fig:palomar}
\end{figure}

We observed the transit of Kepler-51c on UT 2024 April 17 from 08:10 to 12:38, between the airmasses of 2.11 – 1.07 (Figure~\ref{fig:palomar}). Our observing window covered $\sim$3.2 hours of pre-ingress baseline and nearly the full transit, ending shortly before transit egress was complete. We collected a total of 302 science images with an exposure time of 45 seconds per image. We initiated the observations with a five-point dither near the target to construct a background frame that we used to correct for the effects of a known detector systematic at short exposure times. We dark-subtracted, flat-fielded, and corrected for bad pixels following the methodology of \cite{vissapragada2020}. We scaled and subtracted the background frame from each science image to remove the full frame background structure. We then performed circular aperture photometry with the photutils package \citep{bradley2020} on our target, plus ten comparison stars. We tested aperture sizes between 5-25 pixels and selected the aperture that minimized the root-mean-square deviation of the final photometry, which was 10 pixels. We used uncontaminated sky annuli with inner and outer radii of 70 and 100 pixels, respectively, for local background subtraction. 

We fit the light curve using \texttt{exoplanet} \citep{exoplanet} with a combined systematics and transit model. Our systematics model includes a linear combination of comparison star light curves, an error inflation jitter term, and a linear trend. We also tested systematics models with linear combinations of weights for the target centroid offset, PSF width, airmass, and local background. We compared the Bayesian Information Criterion \citep[BIC]{schwarz1978} for all possible combinations of these systematic noise parameters and found that the model with the optimized BIC value included none of these four parameters. We fit a transit model with the framework of \cite{greklekmckeon2023}, including a wide uniform prior of $\pm$10 hours on the transit time, normal priors on the radius ratio $r$, semi-major axis $a/R_\star$, impact parameter $b$, and orbital period from \cite{libbyroberts2020}. We modeled the data with two different limb darkening frameworks to ensure the results were consistent despite the grazing transit configuration. First we fixed the limb darkening parameters $u_1$ and $u_2$ to 0.27 and 0.15, which are the predictions from \texttt{ldtk} \citep{ldtk} in the \textit{J} bandpass using the stellar effective temperature, metallicity, and surface gravity values from \cite{libbyroberts2020}. Then we ran the same model with free quadratic limb darkening parameters. The posterior results were consistent across the both cases, so we report the results from the model with free limb darkening here. We explore the parameter space with the NUTS sampler in \texttt{PyMC3} \citep{pymc3} and ensure that the chains have evolved until the Gelman-Rubin statistic \citep{BB13945229} values are <1.01 for all parameters. We measure a transit time of $2460417.9711\pm0.0047$ BJD.

\subsubsection{Other Attempts}

\edit1{
We attempted to observe transits of Kepler-51b 
on UT 2020 July 25 with a 0.4\,m LCOGT telescope, and
on UT 2020 September 8 with
the Grand-Pra Observatory 0.4\,m and Wild Boar Remote Observatory 0.24\,m telescopes.
The light curves from these observations yielded inconclusive transit detections due to the faintness of the target star, poor seeing conditions, and/or limited in-transit coverage; we did not consider these data further in our analyses. These data are publicly available on the ExoFOP website.\footnote{\url{https://exofop.ipac.caltech.edu/tess/}}
}

\subsection{TESS}\label{ssec:obs_tess}

The TESS photometry is analyzed using the \texttt{tglc} package \citep{han.tglc}. \texttt{tglc} removes contamination from nearby stars in the Full-Frame Images (FFI) using locally-fitted effective Point-Spread Functions derived using Gaia DR3 as priors of star positions and magnitude. A 3x3 aperture is then applied to the decontaminated FFI image to produce the \texttt{cal\_aper\_flux} light curve that we use in this work, which is also detrended with the bi-weight method from \texttt{wotan} \citep{wotan}.

\begin{figure}
    \epsscale{1.15}
    \plotone{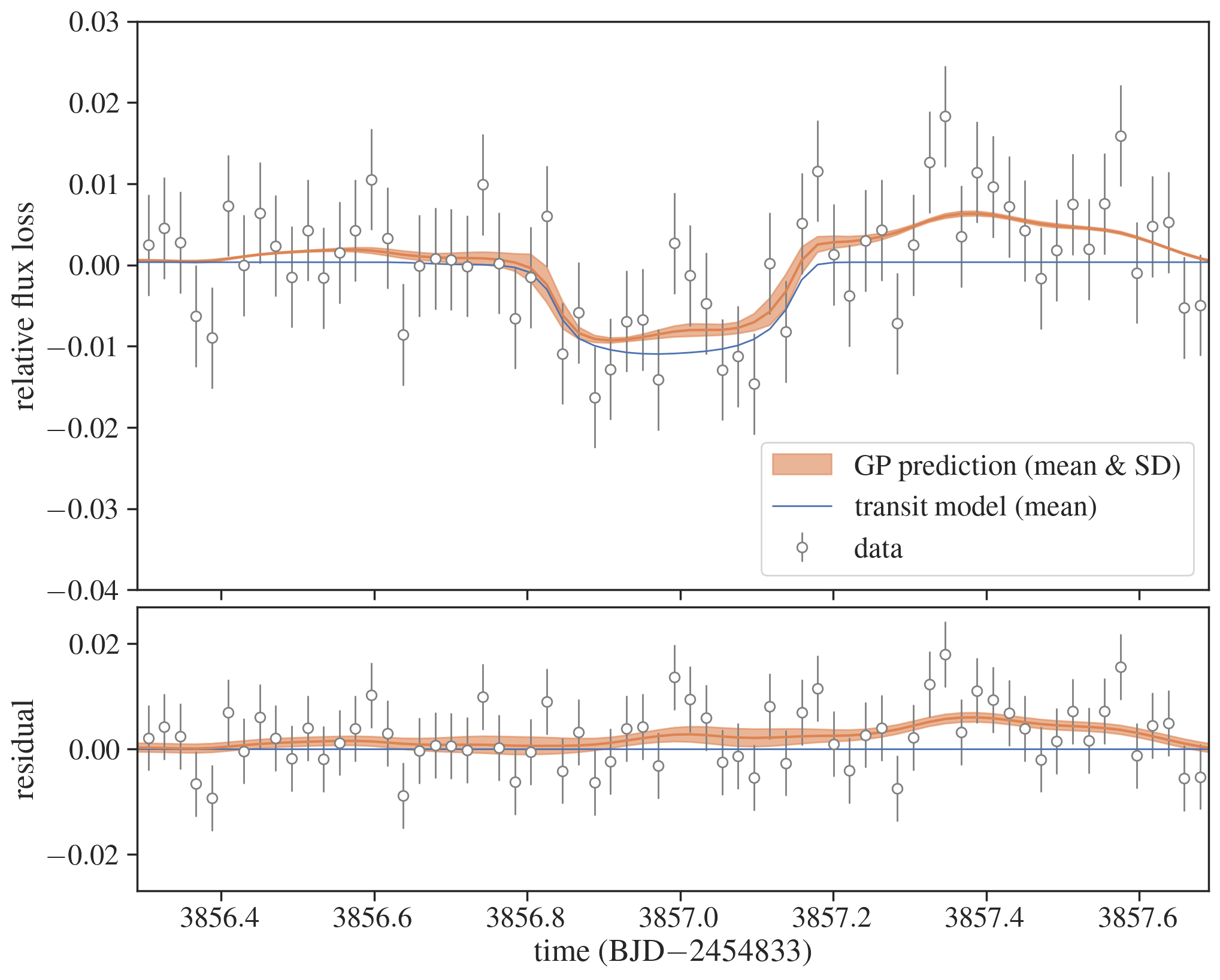}
    \caption{TESS sector 14 transit of Kepler-51d. (Top) The white circles with error bars show the relative flux loss. The orange line and shade show the mean and standard deviation of the predictions of the GP-included transit models computed from posterior samples. The blue line shows the mean of the transit models computed from posterior samples. (Bottom) Residuals of the fit relative to the mean transit model (blue line in the top panel).}
    \label{fig:tess_planet3}
\end{figure}

TESS observed Kepler-51 in six different Sectors. The transits of Kepler-51b were caught in Sectors 14, 54, 55 and Kepler-51d in Sector 14. 
We first fit the Sector 14 data with a similar GP model as described in Section~\ref{ssec:obs_kepler}, adopting the (truncated) normal priors in Table~\ref{tab:shape} as well as the priors on \edit1{$u_1+(u_2/2)$ and $u_1-(u_2/2)$} derived from the Kepler data in a similar manner.
For Kepler-51d, we obtained the transit time of $\mathrm{BJD}-2457000 = 1689.991_{-0.021}^{+0.012}$ (median and symmetric 68\% interval of the marginal posterior),\footnote{We found a slightly bimodal marginal posterior for the transit time. This summary based on percentiles is associated with the higher mode by construction.} which is consistent with the measurement by \citet{2022AJ....164...42J}.
The former measurement is adopted in Table~\ref{tab:transit_times}, and the light curve data and model are shown in Figure~\ref{fig:tess_planet3}.
For Kepler-51b, we obtained $\mathrm{BJD}-2457000 = 1694.875^{+0.08}_{-0.02}$. This is again consistent with \citet{2022AJ....164...42J}, but the uncertainty we derived is larger ($\sim100\,\mathrm{min}$) as the planet's shallower transit is not clearly distinguished from the correlated noise in our light curve model. Since we are not confident in the accuracy of this timing measurement, we discard this transit of Kepler-51b from the subsequent TTV modeling, although adding this transit does not affect the result due to its large uncertainty in any case. Similarly, we could not clearly identify the expected transits of Kepler-51b in Sectors 54 and 55, so they are not included in our timing analysis either.   

\section{Impact of Spot Crossing Events on Transit Time Measurements}\label{sec:spot}

The JWST transit light curve of Kepler-51d shows clear evidence of a spot-crossing event (Figure~\ref{fig:jwst}). The hints of similar events are also seen in the residuals of the Kepler transits of planet b and d (Figures~\ref{fig:keptransit1} and \ref{fig:keptransit3}).\footnote{This is a plausible explanation for a ``bump'' during the simultaneous transit of the two planets, as discussed in \citet{masuda2014} and \citet{2022AJ....164..111G}.}
It has been pointed out that such events, if unaccounted for, can introduce systematic errors in transit times up to a few minutes for transit parameters similar to Kepler-51b and d \citep[e.g.,][]{2013MNRAS.430.3032B,2013A&A...556A..19O, 2015ApJ...800..142M, 2016A&A...585A..72I}. 
In most of the timing measurements in Section~\ref{sec:obs}, we explicitly modeled the correlated noise using a Gaussian process, and expect that the impact of the wiggles caused by spot-crossing events is mitigated. That said, the GP model is not necessarily the correct model for the correlated noise introduced by such events, and so it is a priori unclear how accurate the resulting transit times and their uncertainties are. 

\begin{figure}
    \epsscale{1.15}
    \plotone{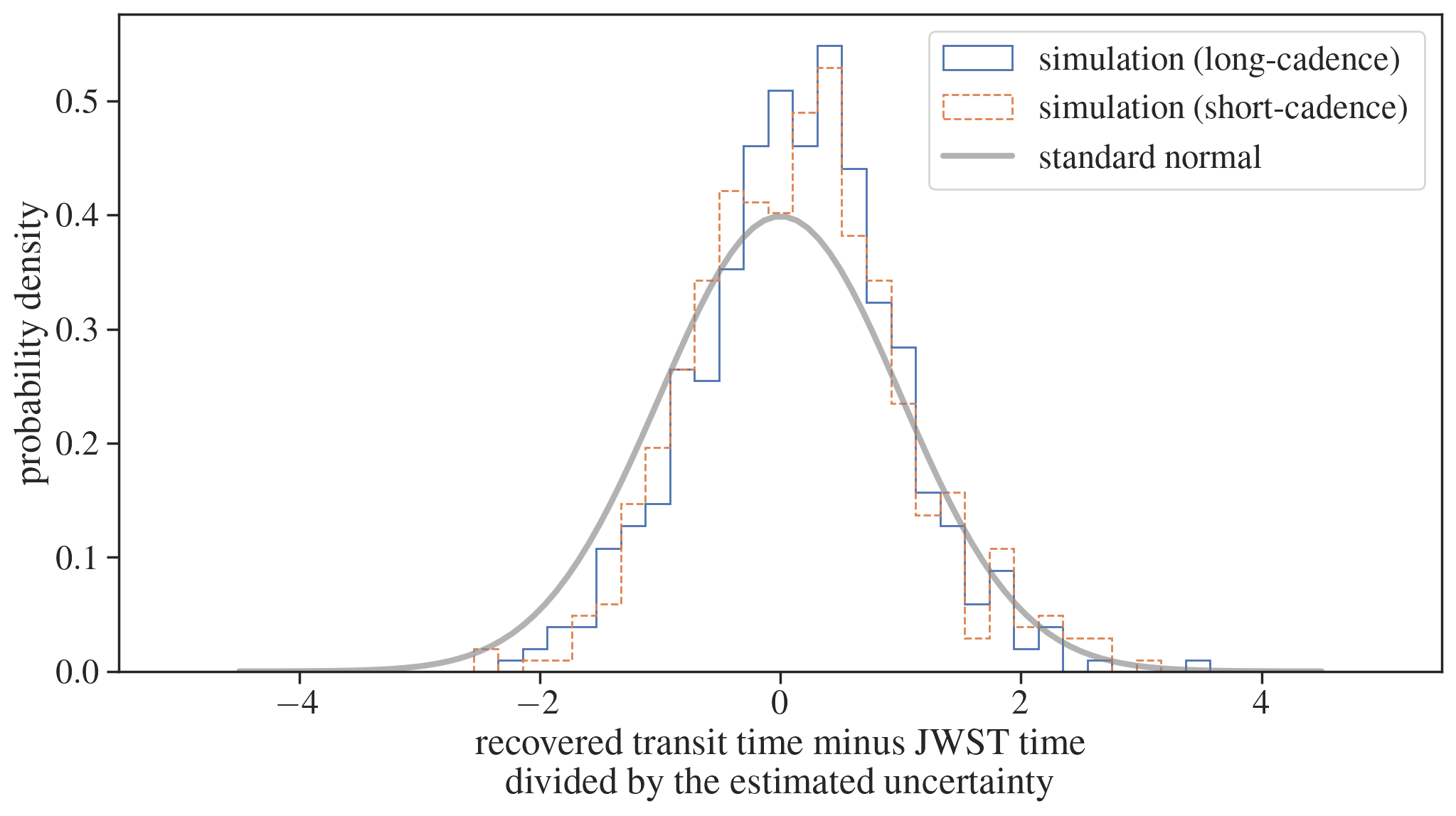}
    \caption{Results of injection-and-recovery tests for the timing bias using the simulated Kepler-like light curves including a spot-crossing event.
    The blue solid and orange dashed histograms show the distributions of the bias of the recovered transit times normalized by the estimated uncertainty. The gray solid line shows the standard normal distribution (a normal distribution with a mean of zero and a variance of one).}
    \label{fig:tc_sim}
\end{figure}

This motivates us to explore the accuracy of our timing measurements using the JWST transit light curve as a ``real template'' of a transit including a significant spot-crossing event. Because the noise in the JWST data is much smaller and the cadence is much shorter than in the Kepler observations, the data can be used to simulate how this transit would have been observed by Kepler, by adding extra noise and then by appropriately downsampling the JWST light curve. We simulated 500 realizations of the Kepler long-cadence observations of this transit, changing the random seed for the extra white noise injected to the JWST data. Then for each realization, we performed the same analysis as described in Section~\ref{ssec:obs_tess} to obtain posterior samples for the mid-transit time, derived their mean $\mu_{\rm sim}$ and standard deviation $\sigma_{\rm sim}$, and checked the distribution of $\delta=(\mu_{\rm sim}-\mu_{\rm JWST})/\sigma_{\rm sim}$, where $\mu_{\rm JWST}$ is the transit time derived in Section~\ref{ssec:obs_jwst} incorporating the spot crossing in the model and is considered to be the ground truth in this simulation. If our timing estimate is not biased and its uncertainty is correctly estimated, $\delta$ would follow the standard normal distribution with the mean of zero and the standard deviation of unity. We found the mean and standard deviation of $\delta$ to be $0.14$ and $0.83$, respectively (blue solid line in Figure~\ref{fig:tc_sim}). This suggests that the bias is small, $\sim 0.1\,\mathrm{min}$ for typical timing error of $\sim 1\,\mathrm{min}$, and that the estimated uncertainty is reasonably accurate, though it could be overestimated by $\sim10\%$. We repeated the same analysis simulating the Kepler short cadence observations too (see orange dashed line), and found similar values (0.16 for the mean and 0.90 for the standard deviation). Of course, this result applies to only Kepler-51d, and only for spot-crossing events near the middle of the transit just as observed by JWST, and the impact is likely larger for similar events happening in the ingress/egress. However, such events should be rarer by the fraction of the ingress/egress duration to the total transit duration. 
We have not simulated transit observations with other facilities either, but we suspect other noise sources will be more dominant in the lower signal-to-noise light curves: the relative flux change due to the observed spot crossing ($\sim 10^{-3}$) is comparable to or smaller than the measurement errors in the ground-based light curves analyzed in Section~\ref{ssec:obs_ground}.
Thus we do not believe that a majority of the measured transit times and their uncertainties are systematically in error due to spot-crossing events that we did not explicitly model.

\section{The TTV Model}\label{sec:model}

We model the TTVs by numerically computing the transit times of the three transiting planets considering gravitational interactions between the planets and the star. The interaction was assumed to be Newtonian, and the light travel time was ignored. The model parameters are the planet-to-star mass ratio $m$ and the osculating orbital elements at the start of integration, here chosen to be $t_{\rm epoch}(\mathrm{BJD}_{\rm TDB})=2454833+155$.
These quantities are defined for each planet and are specified by the subscript b, c, d, and e when necessary. 
The mass ratios (rather than masses) are adopted as fitting parameters, because the TTVs
(in which the light-travel time effect is not apparent) only constrain the mass ratios between the star and the planets. Throughout the paper, we will frequently quote $m$ in units of $M_\oplus/M_\odot$. This value will be the planet mass in units of Earth mass for a solar-mass host star, and the actual planet mass in units of $M_\oplus$ is $m$ multiplied by the stellar mass in units of $M_\odot$: if the stellar mass is $2\,M_\odot$, for example, the planet mass is $2m$ Earth masses. For Kepler-51 with $\approx 0.96\,M_\odot$ (see Section~\ref{sec:star}), the mass ratio quoted in this unit is close to the planet mass in units of $M_\oplus$.
The osculating orbital elements are: orbital period $P$, eccentricity $e$, argument of periastron $\omega$, and time $T$ of inferior conjunction nearest to the start of integration, which was converted to the time of periastron passage $\tau$ via 
$2\pi (T-\tau)/P = E_0 - e\sin E_0$ with $E_0=2\arctan\left[\sqrt{{1-e}\over{1+e}}\tan\left({\pi \over 4}-{\omega \over 2}\right)\right]$.\footnote{Here $f=\pi/2-\omega$ at inferior conjunction is converted to $E$ via $\tan (E/2)=\sqrt{(1-e)/(1+e)}\tan(f/2)$. This is why the argument of $\arctan$ includes $\pi/4-\omega/2$.} 
We assume coplanarity and fix the orbital inclination $i$ to be $\pi/2$, and the longitude of the ascending node $\Omega$ to be $0$. 
We choose the sky plane to be the reference plane, and adopt the $+Z$-axis pointing toward the observer.\footnote{Some authors choose the opposite \citep[e.g.,][]{2021MNRAS.507.1582A}, in which case the ``ascending'' node moves to the opposite side of the line of nodes, and $\omega$ changes by $\pi$. Our choice is the same as that adopted in {\tt TTVFast} \citep{2014ApJ...787..132D}.}

The input masses and osculating orbital elements are converted to Jacobi coordinates. Here we use the total interior mass as the mass entering in the conversion from the orbital periods to coordinates, as in {\tt REBOUND} \citep[][see Section~2.2]{2015MNRAS.452..376R}.\footnote{We note that this convention is different from {\tt TTVFast}. This results in the difference in the ``orbital period'' corresponding to the same position and velocity, and so appropriate conversion is needed to specify the same initial state using Jacobi elements. See also footnote 51 of \citet{2023AJ....165...33D}.}
We then perform an $N$-body integration using a symplectic integrator \citep{1991AJ....102.1528W, 2006AJ....131.2294W, 2014ApJ...787..132D}. The resulting orbits are used to derive mid-transit times of each planet where the planet-star distance in the sky plane is minimized, following the iterative scheme described in \citet{2010arXiv1006.3834F}.\footnote{The difference between the ``transit times'' computed this way and the times of inferior conjunctions as constrained in Section~\ref{sec:obs} is negligible for Kepler-51b, c, and d since their orbital periods are sufficiently long and eccentricities are sufficiently small \citep[cf.][]{2019AJ....158..133H}.} During this iteration, a fourth-order Hermite integrator \citep{2004PASJ...56..861K} is used. 
The $N$-body code is implemented in {\tt JAX} \citep{jax2018github} to enable automatic differentiation with respect to the input mass ratios and orbital elements \citep[see also][]{2021MNRAS.507.1582A}, and is available through GitHub as a part of the {\tt jnkepler} package;\footnote{\url{https://github.com/kemasuda/jnkepler}} see Masuda (2024, in prep.) for details of the implementation, and \citet{2023AJ....165...33D} for an application of the code. In this work, the time step of the symplectic integrator was fixed to be one day. For this step size and for typical system parameters as presented in Section~\ref{sec:4planet}, the fractional energy change during integration was $\sim 10^{-9}$,
and the resulting transit times agreed with those computed using {\tt TTVFast} \citep{2014ApJ...787..132D} within $\sim 1\,\mathrm{sec}$.

\section{Evidence for a Fourth Planet}\label{sec:3planet} 

\begin{figure*}
	\epsscale{1.05}
	\plotone{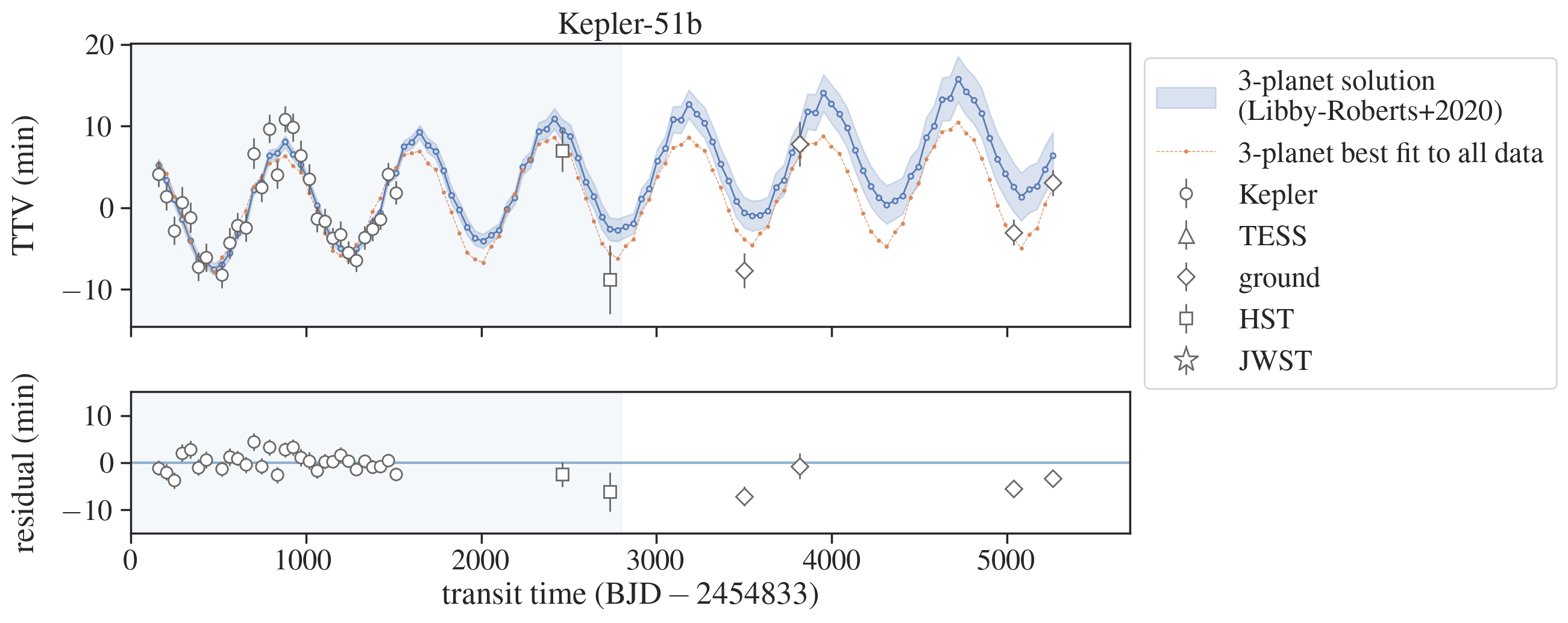}
    \plotone{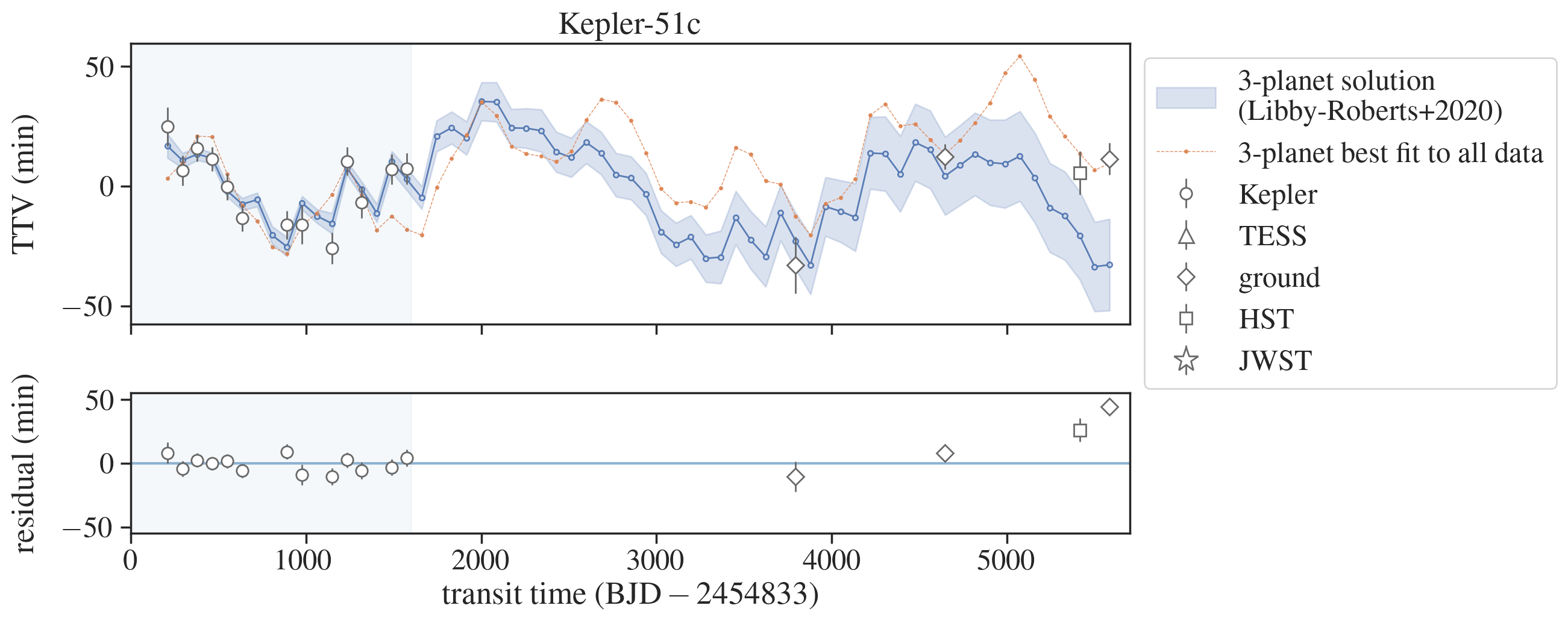}
    \plotone{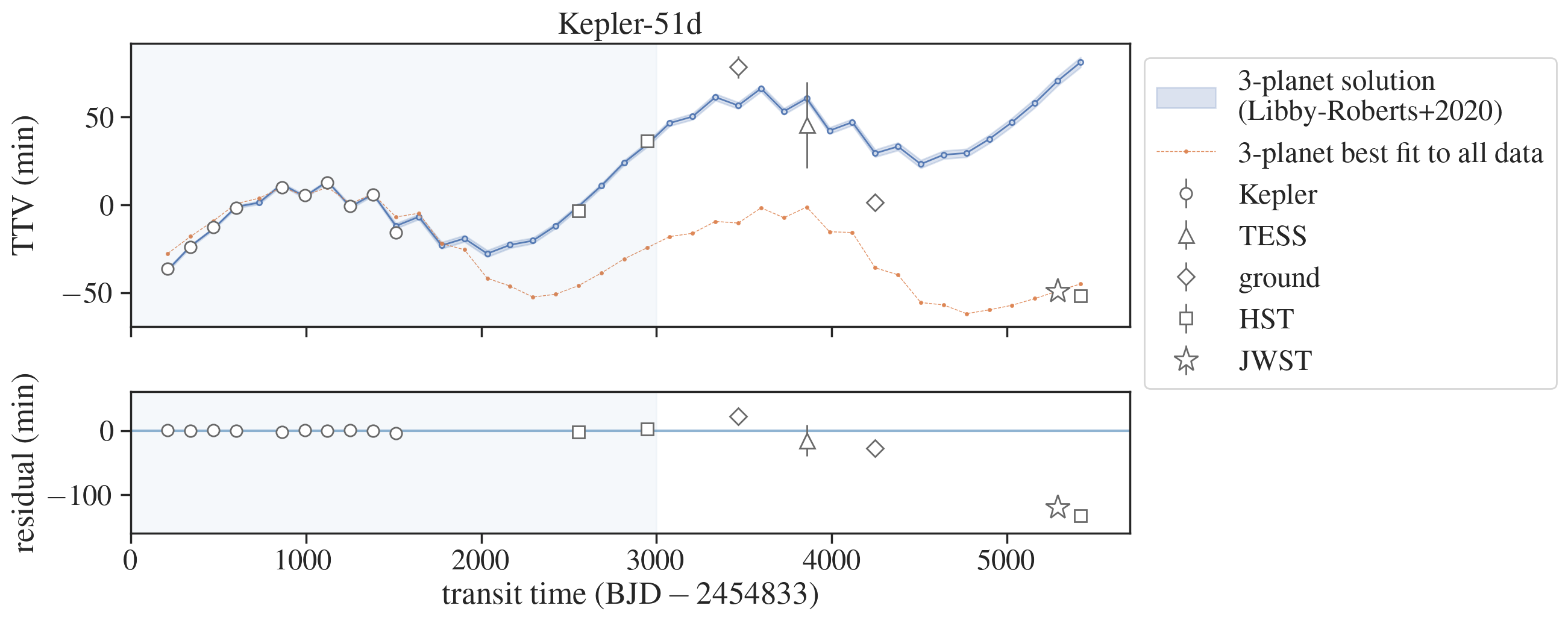}
	\caption{Comparison between the three-planet model prediction and all the timing data presented in this paper (Table~\ref{tab:transit_times}).
    Here the TTVs in the vertical axes are shown with respect to the linear ephemerides given by $t_0(\mathrm{BJD}-2454833)=159.1068, 209.9946, 212.0493$ and $P(\mathrm{days})=45.155296,  85.316963, 130.175183$ for planet b, c, and d respectively; note that the appearance of the plot depends on these arbitrary choices.
    The three-planet model based on the Kepler and HST data presented in \citet{libbyroberts2020} is shown with the blue circles (mean) and shade (standard deviation); the solution is {\it not} conditioned on the other data at time $>3000$.
    The JWST data for Kepler-51d (star symbol in the bottom panel) was $\approx 2\,\mathrm{hr}$ off from the predicted transit time. The discrepancy was confirmed with the ground-based data (diamonds) and the later HST data (rightmost squares for Kepler-51c and d).
    \edit1{The best three-planet model based on {\it all} transit times (Section~\ref{ssec:3planet_all}) is shown with orange dashed lines.}}
	\label{fig:ttv_3planet}
\end{figure*}

To validate the model in Section~\ref{sec:model}, we repeated the TTV analysis in \citet{libbyroberts2020} using the same timing data and including no planets other than the three transiting planets. We ran four Hamiltonian Monte Carlo (HMC) chains in parallel for 500 warm-up steps and for 1,500 sampling steps (i.e., 6,000 samples in total). We found split Gelman--Rubin statistics \citep{BB13945229} of $<1.01$ and effective number of samples $>1,000$ for all parameters. The distribution of the posterior samples shows an excellent agreement with that in \citet{libbyroberts2020}, as shown in Figure~\ref{fig:corner_3planets} in Appendix~\ref{sec:app_corner}.
The mean and standard deviations of the marginal posteriors for the mass ratios are $\mub(\muunit)=3.9\pm1.8$, $\muc(\muunit)=4.5\pm0.5$, $\mud(\muunit)=5.8\pm1.1$; the 95\% highest-density intervals (HDIs) are $[0.7, 7.3]$, $[3.5, 5.5]$, and $[3.7, 7.9]$ for planet b, c, and d, respectively.\footnote{The value of $\omega$ for Kepler-51b reported in Table 10 of \citet{libbyroberts2020} is erroneously shifted by $\pi$. Posterior samples in their Table~9 are correct.} 

In Figure~\ref{fig:ttv_3planet}, we show TTVs of the three transiting planets computed from the above three-planet fit (blue circles and shades). 
This model predicts that the mid-transit time of Kepler-51d we aimed to observe with JWST should have been $\mathrm{BJD}=2460121.9304 \pm 0.0018$ (mean and standard deviation of the posterior distribution). 
However, the JWST transit of Kepler-51d (Section~\ref{ssec:obs_jwst}) happened earlier by 120~minutes compared to this prediction (which exhibits a $1\sigma$ uncertainty of 2.6~minutes), as shown in Figure~\ref{fig:jwst} and by the star symbol in the bottom panel of Figure~\ref{fig:ttv_3planet}. We see evidence of a spot-crossing event in the light curve, but obviously it does not explain the offset, 
and the JWST timing measurement was also confirmed independently by the APO observation from the ground (Section~\ref{ssec:obs_ground_apo}).
This indicates that the three-planet TTV model is incorrect.

The timing discrepancy was further supported by the HST observation of the subsequent transit of Kepler-51d (Section~\ref{ssec:obs_hst}). The transit times from the ground-based facilities obtained before these measurements (Section~\ref{ssec:obs_ground}) also show (smaller) deviations from the three-planet prediction. These data are shown in Figure~\ref{fig:ttv_3planet} with square and diamonds, respectively.\footnote{On the other hand, the TESS transit time for Kepler-51d happened to be consistent with the three-planet model, as shown in the bottom panel. This explains why the deviation from the three-planet model was not noted by \citet{2022AJ....164...42J}.} Thus the data from different space and ground-based observatories all show that the model involving only the three known transiting planets does not replicate the observed TTVs. The simplest explanation is that there is a fourth planet in the system. We explore this scenario in Section~\ref{sec:4planet}.

\subsection{Light Travel Time Effect}\label{ssec:3planet_ltte}

\edit1{
We modeled TTVs only due to the physical variations of the orbits and ignored the light travel time effect associated with the reflex motion of the central star around the common center-of-mass. This is because the expected amplitude of this signal \citep{2005MNRAS.359..567A}, 
\begin{align}
    \delta t_{\rm LTTE}={a\over c}{M_{\rm p}\over M_\star}\approx 0.5\,\mathrm{s}\left(M_{\rm p}/M_{\rm Jup} \over M_\star/M_\odot\right)\left(a \over 1\,\mathrm{au}\right),
\end{align}
is small for Kepler-51b, c, and d. This equation shows that a stellar-mass perturber on an au-scale orbit could generate a signal comparable to the excess TTVs of $\sim 100\,\mathrm{min}$ observed for planet d. However, we do not consider this to be a plausible explanation, as the light-travel time effect should similarly impact Kepler-51b and c for which such TTVs are not observed.
}

\subsection{Three-planet Fit to All Timing Data}\label{ssec:3planet_all}

\edit1{
Above we demonstrated the discrepancy between the three-planet prediction based on the Kepler and previous HST data, and the new timing measurement from JWST. This implies that both data sets cannot be consistently modeled. To check this, we also fitted all available transit times with a three-planet model, minimizing the chi-squared
\begin{align}
\label{eq:chi2}
    \chi^2 = \sum_{i \in \mathrm{all\ data}} \left(t_i - m_i \over \sigma_i\right)^2,
\end{align}
using the Trust Region Reflective method implemented in {\tt scipy.optimize.curve\_fit}: here $t_i$ and $\sigma_i$ are the observed transit time and error (see the note in Table~\ref{tab:transit_times}), $m_i$ is the transit time from the $N$-body model, and the summation is over all transit times in Table~\ref{tab:transit_times}.
This optimization was repeated by fixing the mass ratio of each planet to be $1, 2, \dots, 20\,\muunit$; the search ranges for the other parameters are listed in Table~\ref{tab:grid}.
The best model we found is shown with the orange dashed lines in Figure~\ref{fig:ttv_3planet}, where the model for Kepler-51d now fits the JWST data with a small error but fails to explain the others. The model gives $\chi^2\approx 2500$ for 70 data points, which again argues against the three-planet model.
}

\subsection{Impact of Non-zero Mutual Orbital Inclinations}\label{ssec:3planet_photod}

\edit1{
To explore the potential impact of non-zero mutual inclinations, we also performed a ``photodynamical'' modeling of the Kepler light curves, in which models of transit light curves were computed based on numerically-integrated orbits and the mutual inclinations of the planetary orbits were allowed to vary; see Appendix~\ref{sec:app_photod} for details of the analysis setup and results. This modeling incorporates the information on the transit shapes and is therefore more sensitive to the mutual orbital inclinations than the analysis of TTVs alone. We found the inclinations of the orbits of Kepler-51b and Kepler-51c relative to that of Kepler-51d to be $0$--$13\,\mathrm{deg}$ and $0$--$3\,\mathrm{deg}$ (95\% highest density intervals), and the mass ratios for all three transiting planets consistent with the coplanar TTV modeling as described above. The TTV predictions extrapolated after Kepler observations also remained similar. Thus the three-planet model fails to explain the data even considering non-zero mutual inclinations. This analysis also suggests that the planetary masses from coplanar TTV models are largely unaffected by mutual orbital inclinations of $\lesssim 10\,\mathrm{deg}$.
}

\begin{deluxetable*}{lccc}[!ht]
\tablecaption{Parameter ranges of the brute-force searches in Section~\ref{ssec:4planet_brute}.\label{tab:grid}}
\tablehead{
\colhead{} & \multicolumn{1}{c}{Search1 (Kepler-like, inside d)} & \multicolumn{1}{c}{Search2 (Kepler-like, outside d)} & \multicolumn{1}{c}{Search3 (outer giant)}
}
\startdata
\multicolumn{3}{l}{\textbf{Fourth planet}}\\
period $P_{\rm e}$ &  $[1.1\,P_{\rm b}, 1.7\,P_{\rm b}]$, $N=50$  &  $[1.1\,P_{\rm d}, 5\,P_{\rm d}]$, $N=100$ & $[5\,P_{\rm d}, 50\,P_{\rm d}]$, $N=200$\\
 &  $[1.1\,P_{\rm c}, 1.4\,P_{\rm c}]$, $N=50$  &  & \\
orbital phase $(T_{\rm e}-T_{\rm epoch})/P_{\rm e}$ & $[0, 1]$, $N=50$ & $[0, 1]$, $N=50$ & $[0, 1]$, $N=250$ \\
$e\cos\omega$, $e\sin\omega$ & $[-0.1, 0.1]$, $N=1$ & $[-0.2, 0.2]$, $N=1$ & $[-0.7, 0.7]$, $N=1$\\
mass ratio $\mue$ & $[10^{-6}, 10^{-4}]$, $N=1$ & $[3\times10^{-7}, 3\times10^{-4}]$, $N=4$ & $[3\times10^{-5}, 3\times 10^{-3}]$, $N=2$ \\
\multicolumn{3}{l}{\textbf{Inner three transiting planets}}\\
period $P$ & \multicolumn{3}{c}{$[P_0 - 0.1, P_0 + 0.1]$, $N=1$}\\
orbital phase $T$ & \multicolumn{3}{c}{$[T_0 -0.01, T_0 + 0.01]$, $N=1$}\\
$e\cos\omega$, $e\sin\omega$ & \multicolumn{3}{c}{$[-0.1, 0.1]$, $N=1$}\\
mass ratio & \multicolumn{3}{c}{$[10^{-6}, 10^{-4}]$, $N=1$}\\
\enddata 
\tablecomments{$[a, b]$ shows the edges of the bins, and $N$ shows the number of bins between $a$ and $b$ ($N=1$ means the interval was not divided). We adopt log-uniform bins for the period and mass ratio, and uniform bins for the orbital phase. For the inner transiting planets, $P_0$ and $T_0$ denote the linear ephemeris for each planet obtained by linear fitting to the transit times against the number of transit.}
\end{deluxetable*}

\section{Four-planet TTV Modeling Assuming Coplanarity}\label{sec:4planet}

Here we explore possible four-planet solutions to the observed TTVs and assess how adding this planet may or may not impact other planets' parameters, especially their masses. 
We find no clear evidence of transits of Kepler-51e in the Kepler data, leaving its orbital period and inclination unknown. 
Any planet outside Kepler-51b, whose semi-major axis is about 60 times larger than the stellar radius, will not transit the star from our perspective if its orbital inclination is less than $\arccos(1/60)=89\,\mathrm{degrees}$. 
As such, the lack of transit detection is not informative, and does not necessarily imply significant deviation from coplanarity (see also Section~\ref{ssec:disc_tdv}).

Deriving the mass and orbit of a planet without detected transits using TTV data alone has been possible only in a few cases where strong gravitational interactions and good data coverage enabled detection of TTV signals with multiple frequencies \citep[e.g.,][]{2012Sci...336.1133N}. This is not the case here as the TTV residuals of Kepler-51d only show a part of an excess long-term modulation, and we will assume that the orbital planes of the four planets are aligned in the following analyses. 
We expect that the mass constraints will not change significantly for deviations of $\lesssim 10\,\mathrm{deg}$ from perfect coplanarity (see also Section~\ref{ssec:3planet_photod}), but other qualitatively different solutions involving larger mutual inclinations for Kepler-51e could exist. That said, the assumption of near coplanarity is not without foundation: the mutual inclinations of the orbits of non-transiting planets in systems with multiple transiting planets have been inferred to be small ($\lesssim 10\,\mathrm{deg}$) in general, for those that are both near and far from the transiting planets \citep{2018ApJ...860..101Z, 2020AJ....159...38M}.

\subsection{Brute-force Search for the Solutions}\label{ssec:4planet_brute}

TTV inversions involving a non-transiting planet typically result in multiple solutions with different orbital periods and phases for the non-transiting planet \citep[e.g.,][]{2023arXiv231011775J, 2024A&A...683A..96A}. This motivates us to perform a brute-force search over a ``grid'' of orbital periods and phases for Kepler-51e. 

\subsubsection{Grid Optimization and Stability Analysis}

We set up bins in the parameter space as shown in Table~\ref{tab:grid}, and performed bounded optimization in each of the 
\edit1{$100 \times 50 \times 1$ (Search1), $100 \times 50 \times 4$ (Search2), or $200 \times 250 \times 2$ (Search3) bins
minimizing the chi-squared in Equation~\ref{eq:chi2}.
Search1 and Search2 are to probe ``Kepler-like'' systems where the fourth planet is near the inner three and its orbit has a modest eccentricity. The fourth planet is assumed to be inside and outside of Kepler-51d's orbit in Search1 and Search2, respectively. 
}
Here the lower boundary of the period ratio (1.1) is smaller than the period ratios of known Kepler multis \citep{2015ARA&A..53..409W}, and the corresponding orbits are likely dynamically unstable \citep[][also see below]{2015ApJ...807...44P}. For Search2, the upper bounds of
$|e\cos\omega|$ and $|e\sin\omega|$ were set to $0.2$ for most parameter combinations, but they were truncated for small $P_{\rm e}/P_{\rm d}$ so that the orbits do not cross.
Search3 is to explore another common situation where (not-so-metal-poor) stars with Kepler-like planets have an outer giant planet whose orbit is not necessarily circular \citep{2018AJ....156...92Z, 2019AJ....157...52B}.

\edit1{
Search1, Search2, and Search3 result in optimal solutions in each of the $100 \times 50 \times 1=5000$, $100 \times 50 \times 4=20000$, and $200 \times 250 \times 2=100000$ bins, respectively.} For those that give reasonably good fits (see below), we used {\tt FeatureClassifier} in the {\tt SPOCK} \citep{2020PNAS..11718194T} software to compute the probability that the system is stable over $10^9$ orbits, which is comparable to our rough age estimate of $\sim 0.7\,\mathrm{Gyr}$ (Section \ref{sec:star}).

\subsubsection{Results}\label{sssec:4planet_results}

\begin{figure*}
	\epsscale{1.2}
	\plotone{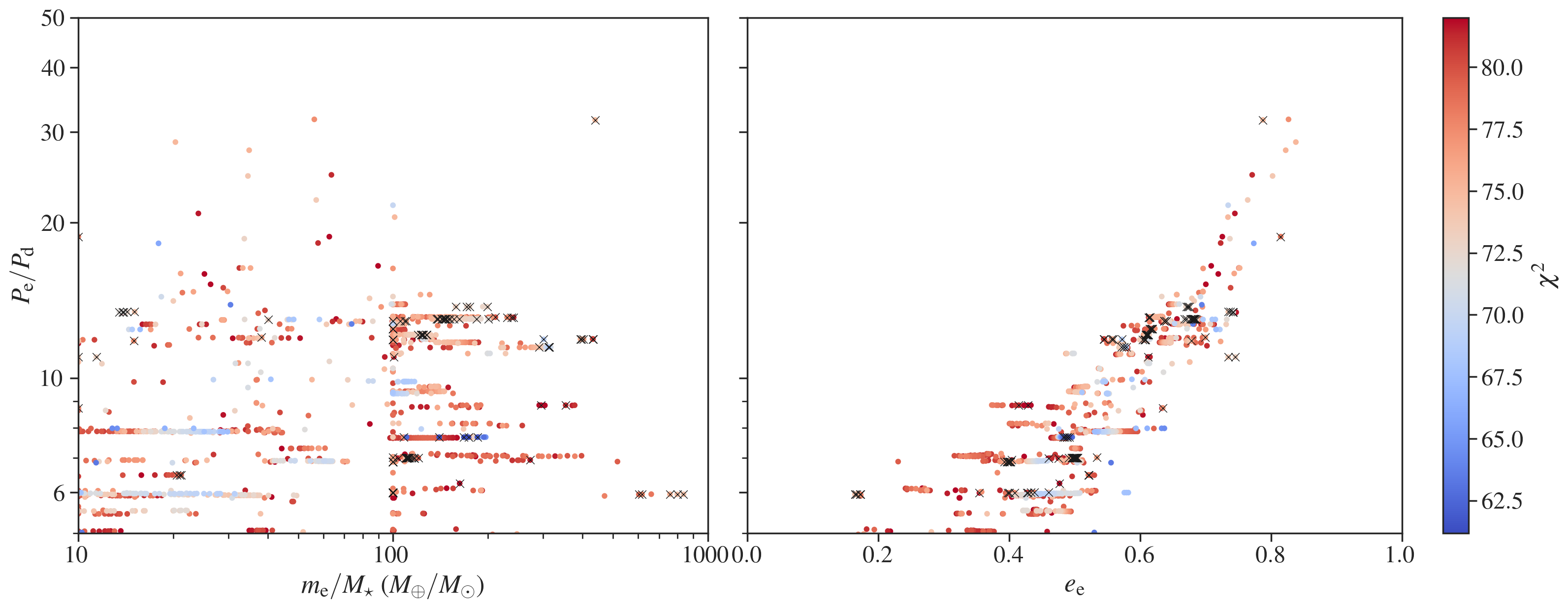}
    \plotone{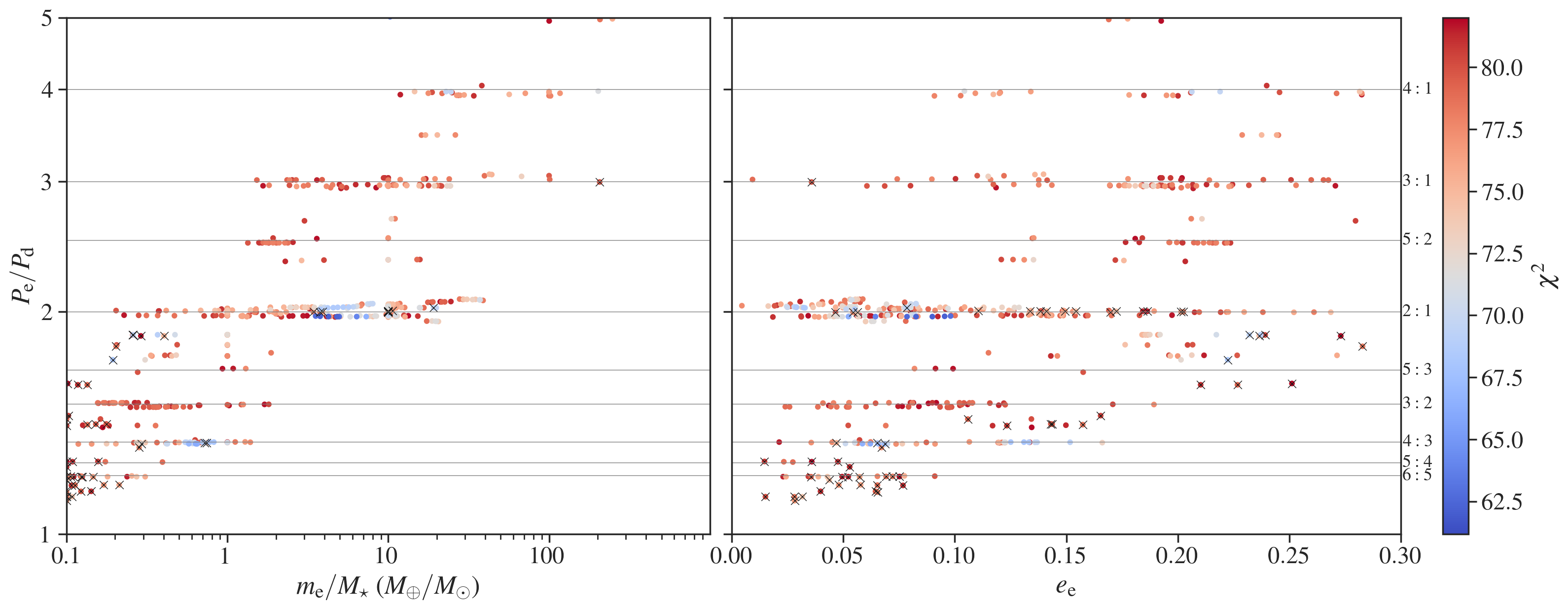}
    \plotone{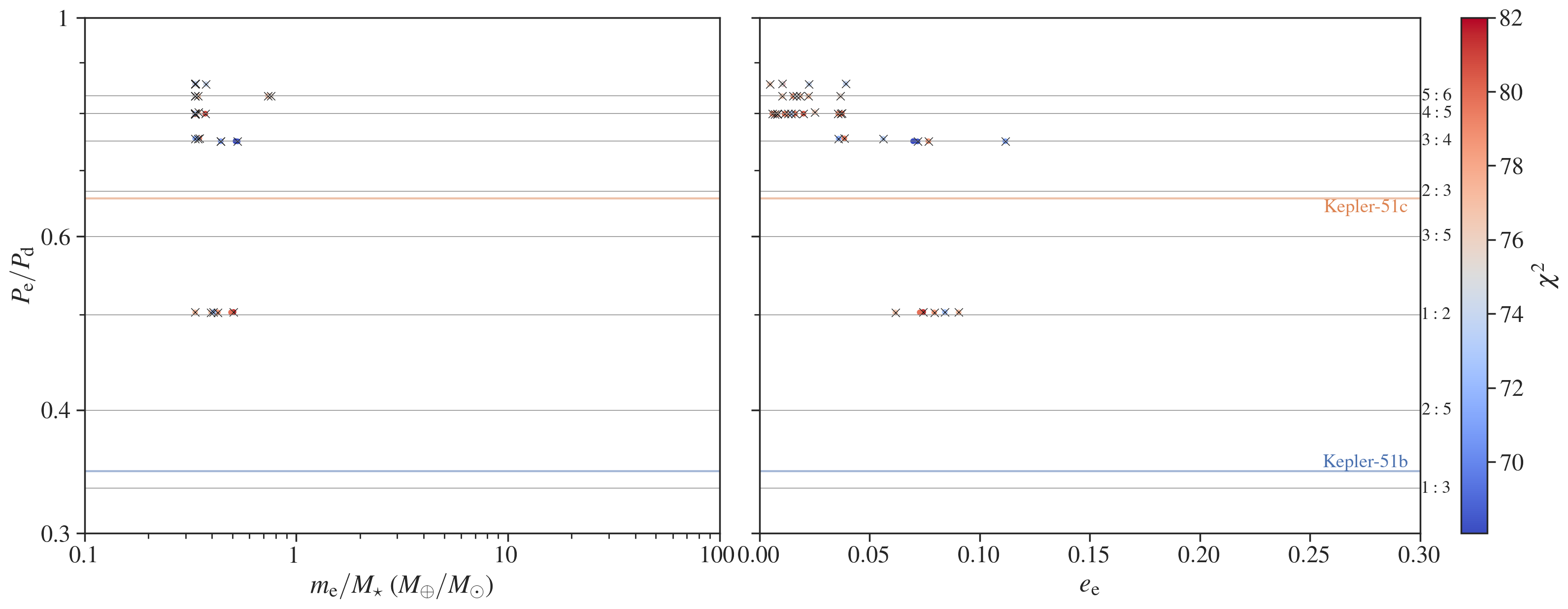}
	\caption{Results of the brute-force search and stability analysis for 
    Search3 (Top), Search2 (Middle), and Search1 (Bottom)
    in Table~\ref{tab:grid}. Note the different scales of the $x$-axes. 
    The colored points show the period ratio of planet e and planet d
    against the mass ratio (left column) and orbital eccentricity (right column) of planet e. Their colors correspond to the $\chi^2$ values shown in the right color bar. The crosses show the solutions for which the probability that the system is long-term stable is lower than 50\% (see Section~\ref{ssec:4planet_brute}).} 
	\label{fig:p4_m4_e4}
\end{figure*}

\begin{figure*}
	\epsscale{1.14}
	\plotone{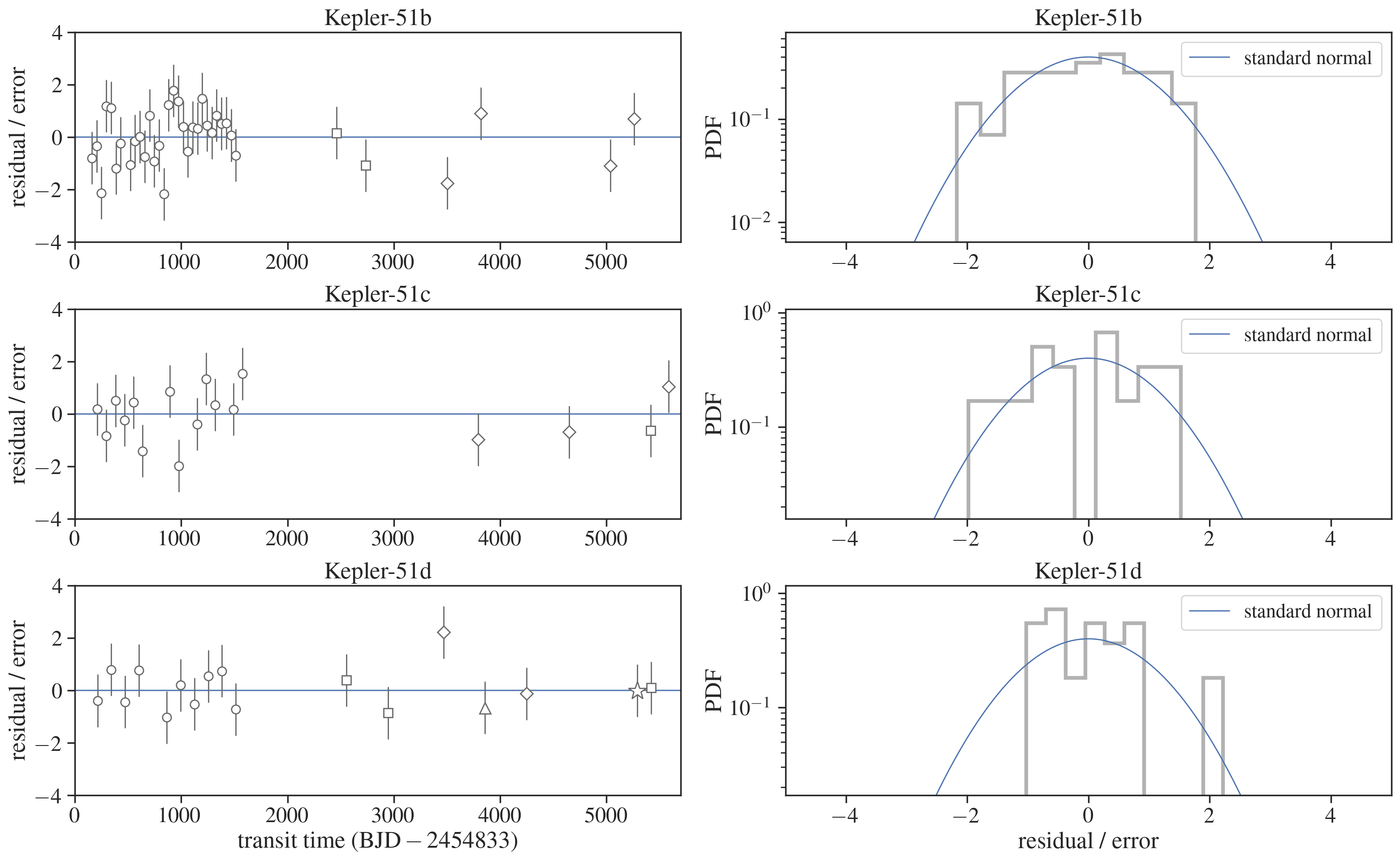}
    \epsscale{1.15}
    \plotone{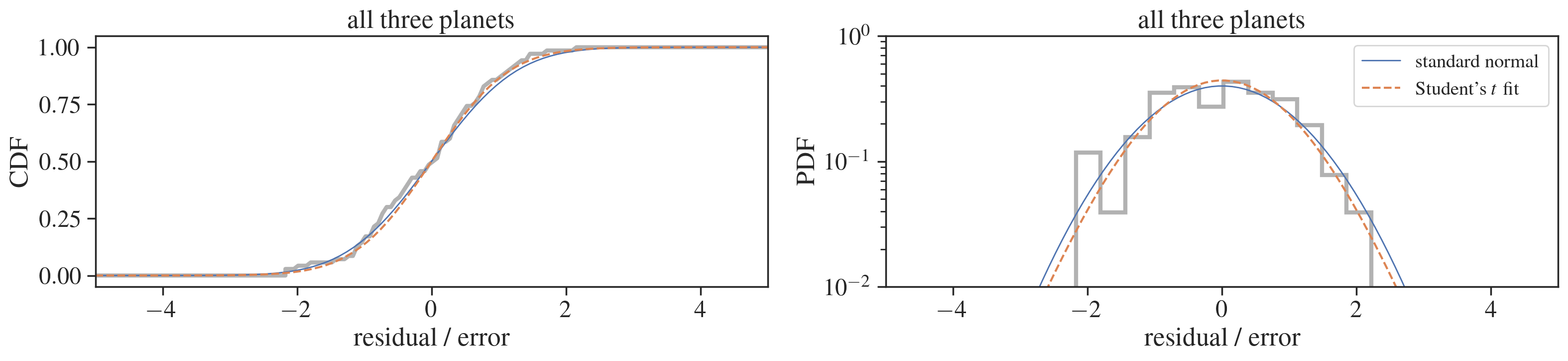}
	\caption{Normalized fitting residuals versus time (top three panels on the left) and their distributions (the other panels) for the best $2:1$ solution found from the brute-force search described in Section~\ref{ssec:4planet_brute}. The length of the error bars is unity for all points by construction. In the residuals versus time plots, the symbols represent observations as in Figure~\ref{fig:ttv_3planet}. In the distribution plots, the (cumulative) distributions of the normalized residuals are shown with the thick gray lines. The blue solid lines show the standard normal distribution. The orange dashed lines in the bottom two panels show the Student's $t$ distribution fitted to the normalized residuals (see Section~\ref{sssec:4planet_results}).}
	\label{fig:residual_2to1}
\end{figure*}

Figure~\ref{fig:p4_m4_e4} shows a subset of the optimal solutions from the brute-force search in the period--mass--eccentricity plane of planet e. The color of each point corresponds to the value of $\chi^2$, and the solutions for which the stability probability computed with {\tt SPOCK} is less than 50\% are shown with crosses. Here we show the solutions with $\chi^2<82$, which corresponds to a $p$-value of $\approx 0.3\%$ for 50 degrees of freedom, i.e., 70 (number of data points) minus 20 (number of fitted parameters). We set a rather generous threshold, because the effective number of degrees of freedom in this non-linear problem may be larger than calculated above \citep{2010arXiv1012.3754A}. Indeed, we performed a Kolmogorov-Smirnov test for the null hypothesis that the distribution of the fitting residuals normalized by the assigned errors are drawn from the standard normal distribution, and found $p$-values larger than 0.1 for all of these solutions with $\chi^2<82$. Thus, none of these solutions are completely excluded by the available data, although the solutions in which planet e is inside Kepler-51d (Search1, bottom panels) generally seem disfavored from a stability point of view. The list of parameters corresponding to these solutions are available through the author's GitHub.\footnote{\url{https://github.com/kemasuda/kep51_jwst}\label{footnote:github}}
The minimum $\chi^2$ value of $\approx 61$ was found for $P_{\rm e}/P_{\rm d}\approx 2$ (Search2) and $P_{\rm e}/P_{\rm d}\approx 8$ (Search3).

In Figure~\ref{fig:residual_2to1}, we show the distribution of the normalized residuals (i.e., residuals divided by the assigned errors) for the minimum $\chi^2$ solution around $P_{\rm e}/P_{\rm d}=2$ as an example. The TTV models corresponding to this class of solutions are shown in Figure~\ref{fig:ttv_2to1} (see also Section~\ref{ssec:4planet_2to1}). As stated above, we see no clear evidence that the normalized residuals deviate from the standard normal distribution. This is consistent with our assessment of the accuracy of timing errors in Section~\ref{sec:spot}.
We also fit the normalized residuals with a Student's $t$ distribution to estimate the number of degrees of freedom, $\nu$, and the scale of the distribution squared, $V_1$, considering that more heavy-tailed residual distributions have been reported in previous transit timing analyses \citep[e.g.,][]{2016ApJ...820...39J, 2021PSJ.....2....1A}. We find $\ln \nu=3.3\pm0.9$ (or $\nu\approx 27$) and $\ln V_1 \approx -0.21\pm0.22$, which support the consistency with the standard normal distribution (see orange dashed lines in the bottom panels of Figure~\ref{fig:residual_2to1}).

In short, we have found that the four-planet model works well to explain the observed TTVs. However, the property of the fourth planet remains unconstrained by the available data: it could either be a relatively low-mass planet on a near-circular orbit outside of and close to Kepler-51d, or a relatively massive planet on a distant, eccentric orbit. A general tendency, as shown in Figure~\ref{fig:p4_m4_e4}, is that the required mass and eccentricity of the fourth planet increases as the period ratio $P_{\rm e}/P_{\rm d}$ increases. More timing data are needed to pin down the properties of Kepler-51e. 
They will also be important to better characterize the noise distribution including its tails. While we see no clear evidence for the non-Gaussianity in the current timing residuals, if more timing data reveal that it is more appropriate to treat part of the data as outliers, the parameter estimates and their uncertainties may be quantitatively affected.

\begin{figure*}
	\epsscale{1.15}
	\plotone{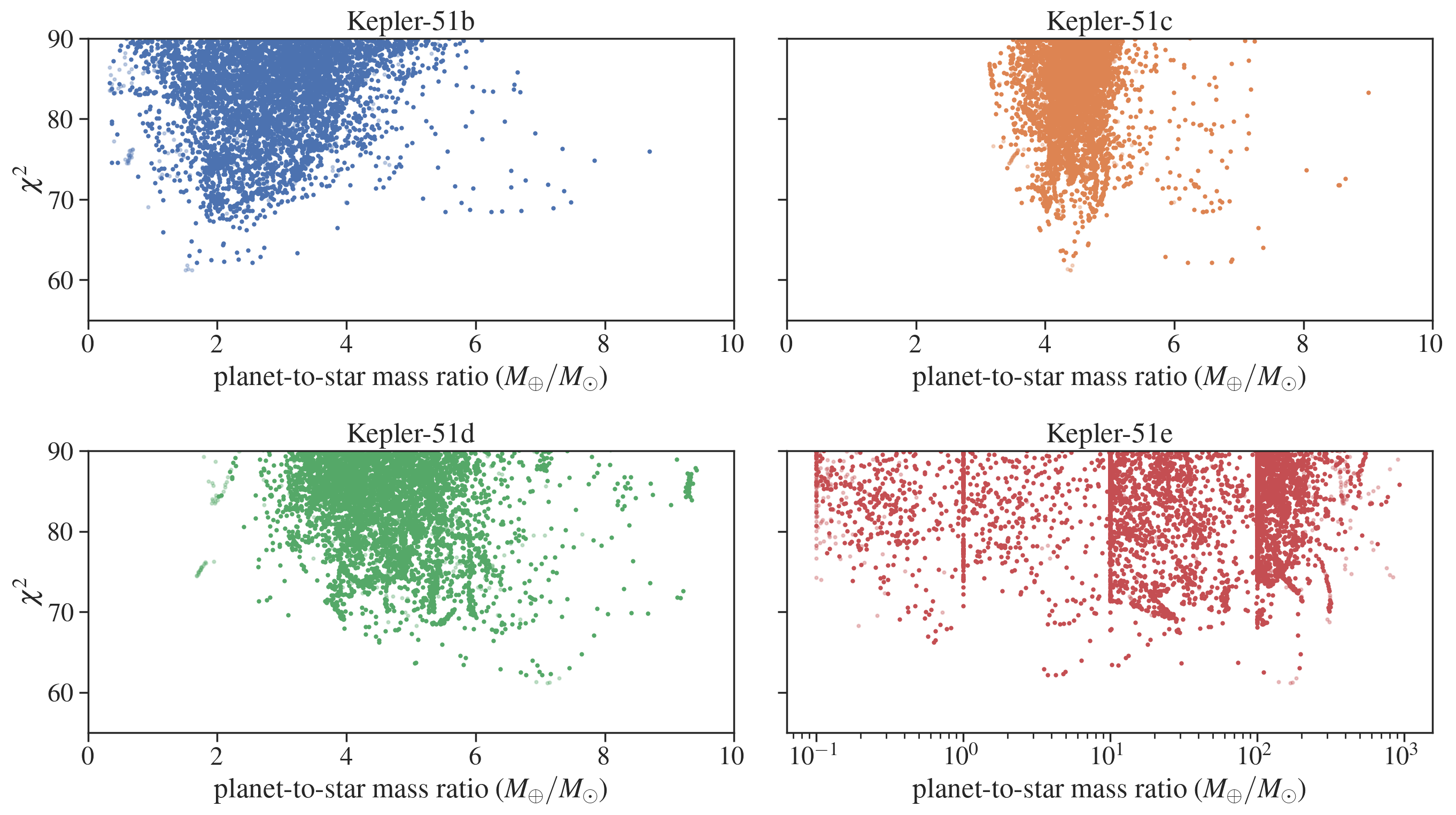}
	\caption{Mass ratios of the four planets from the brute-force search. Note the different axis scale in the bottom-right panel for planet e; the vertical stripes correspond to the edges of the mass grids (see Table~\ref{tab:grid}). This plot shows only the best masses corresponding to each set of the fourth planet's parameters (but see also Figure~\ref{fig:profile}).}
	\label{fig:masses}
\end{figure*}

In contrast, we found that all of these solutions consistently imply mass ratios $\lesssim 10\,M_\oplus/M_\odot$ for the inner three transiting planets (Figure~\ref{fig:masses}). The values are comparable to those found from the previous three-planet model (see Section~\ref{sec:3planet}). We interpret this result as a reasonable consequence of the fact that the masses of planet c and d are mainly determined by the chopping signal in the TTVs of planets d and c from Kepler, respectively, and that the mass of planet b is constrained by the absence of its gravitational impact on the TTVs of planet c. The amplitude of the chopping signal is mainly determined by the mass ratios and the period ratios of the planets \citep{2015ApJ...802..116D}, the latter of which does not change much by adding a fourth planet. 

That said, Figure~\ref{fig:masses} shows only the best-fit masses for given properties of the fourth planet, 
and the above searches do not quantify the uncertainty of the inner planets' masses associated with each given set of the fourth planet's parameters. For example, if the planet mass was constrained to be $10\pm5\,M_\oplus$ (with $10\,M_\oplus$ giving the highest likelihood values) for all possible values of $P_{\rm e}$, Figure~\ref{fig:masses} would have shown points only at $10\,M_\oplus$ without conveying any information about the $5\,M_\oplus$ uncertainty.
More generally, the above searches could have missed solutions with larger masses for the inner planets that are qualitatively different from the three-planet solutions, given that the search involves non-linear optimization in a high-dimensional parameter space. We thus further explore the constraints on the masses of the inner three transiting planets in the next subsection.

\subsection{Constraints on the Mass Ratios of the Inner Three Transiting Planets}\label{ssec:4planet_profile}

Here we attempt to quantify the uncertainties of the mass ratios of the inner three transiting planets by evaluating the profile likelihood for these parameters. For each of planets b, c, and d, we fix the mass ratio to be 1, 2, 3, $\dots$, 15$\,\muunit$, perform the brute-force search as described in the previous subsection for each fixed mass ratio, and determine the minimum value of $\chi^2$ as a function of the fixed mass ratio after optimizing all other parameters. To save computational time, we limited the search range of $P_{\rm e}/P_{\rm d}$ around the values for which good solutions were found (see Figure~\ref{fig:p4_m4_e4}). The search ranges for the other parameters are the same as in Table~\ref{tab:grid}. 

\begin{figure}
	\epsscale{1.15}
	\plotone{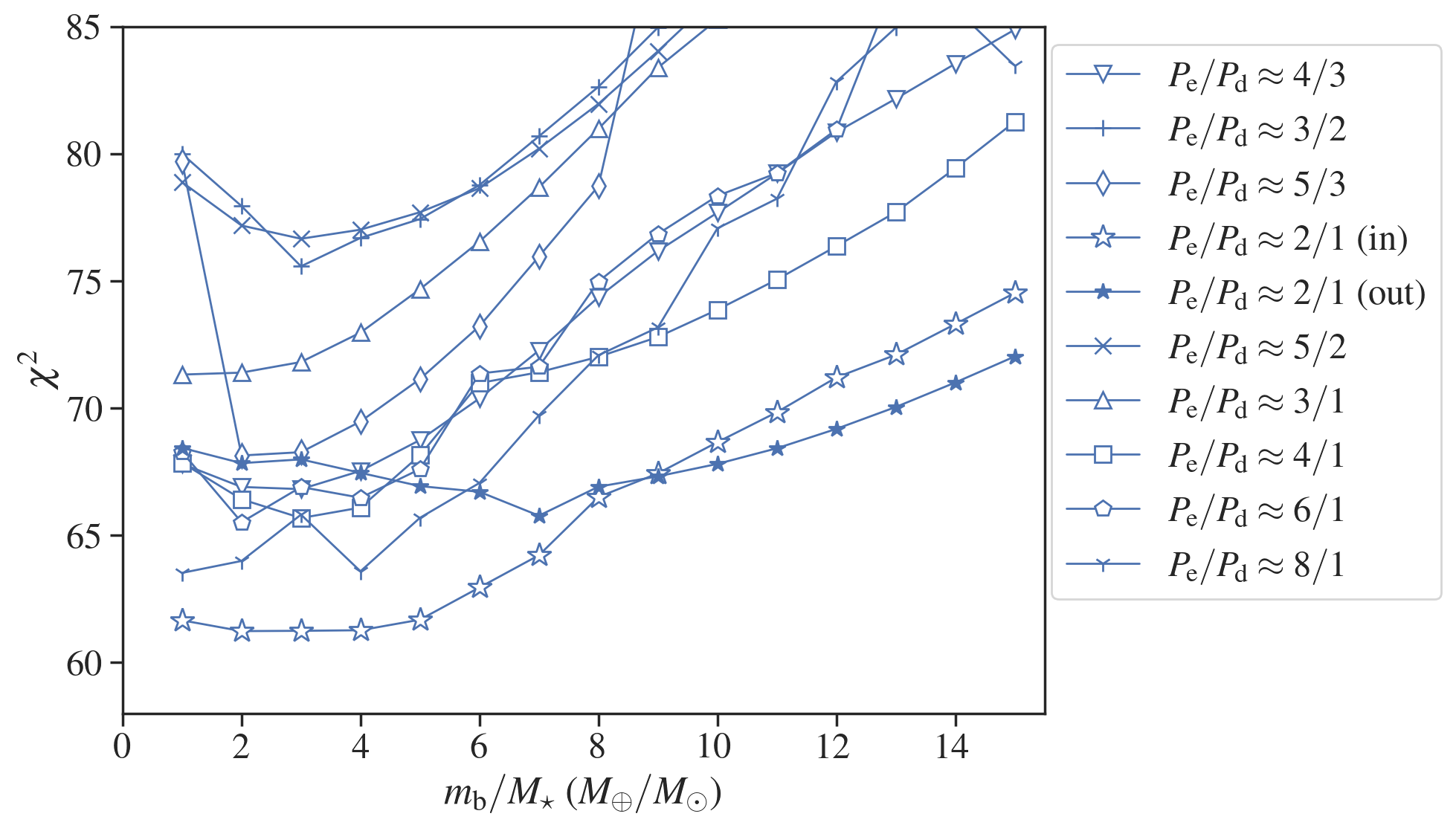}
    \plotone{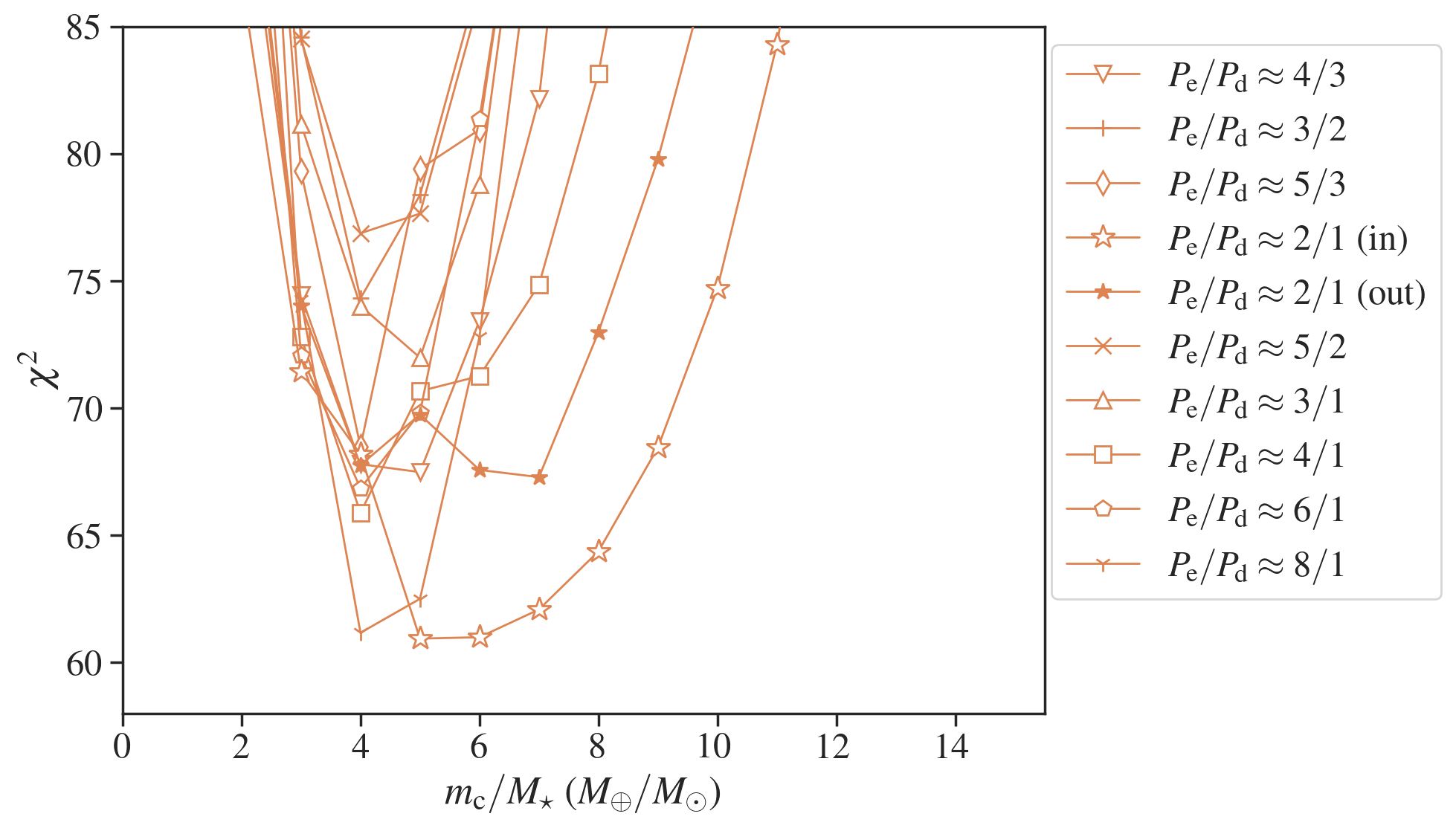}
    \plotone{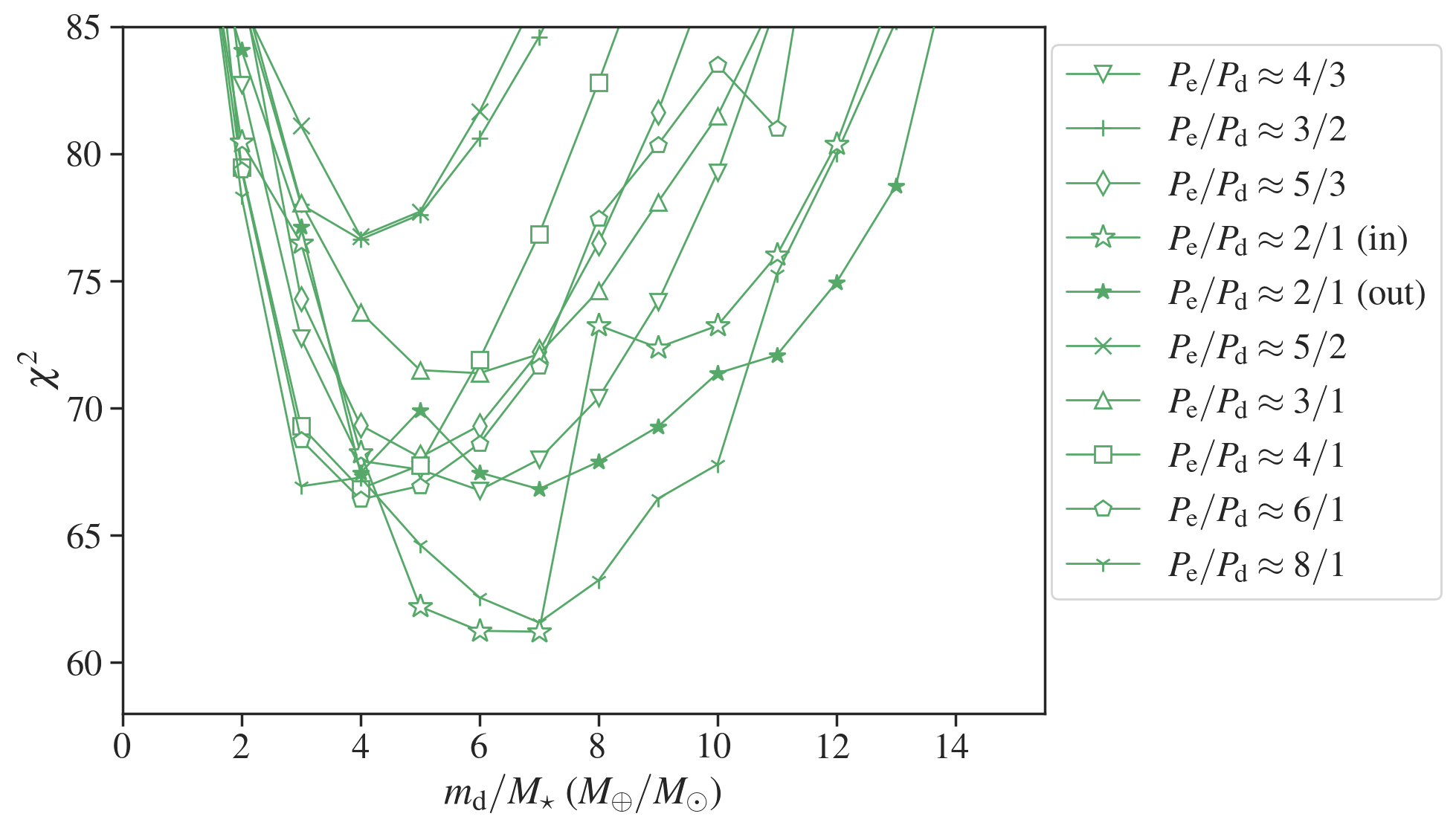}
	\caption{The minimum $\chi^2$ values as a function of the mass ratio of planet b (top), c (middle), and d (bottom), for selected orbital periods of planet e shown in the right.} 
	\label{fig:profile}
\end{figure}

The results are shown in Figure~\ref{fig:profile} for the period ratios of $P_{\rm e}/P_{\rm d} \approx 4/3,\ 3/2,\ 5/3,\ 2/1,\ 5/2,\ 3/1, 4/1, 6/1$, and $8/1$. 
For the $2:1$ ratio, solutions with $P_{\rm e}/P_{\rm d}<2$ (in) and $P_{\rm e}/P_{\rm d}>2$ (out) are shown separately (see also Section~\ref{ssec:4planet_2to1}).
For the period ratios examined here, we confirm that the best-fit mass ratios for planets b--d are low, as in Figure~\ref{fig:masses}, and also that the models involving larger mass ratios do yield poorer fits to the data and lower likelihood values. 
Quantitatively, the largest mass ratios are allowed for either $2:1$ (in) or $2:1$ (out) solutions depending on the planet. In these cases, the ranges of mass ratios corresponding to $\Delta\chi^2<4$ are: $<13\,\muunit$ for Kepler-51b ($2:1$ out), $4.5$--$8.1\,\muunit$ for Kepler-51c ($2:1$ in), and $3.8$--$9.2\,\muunit$ for Kepler-51d ($2:1$ out).
This analysis therefore further supports the conclusion that the inner three transiting planets have masses $\lesssim 10\,M_\oplus$ (given $M_\star\approx 0.96\,M_\odot$; Section~\ref{sec:star}) regardless of the properties of the fourth planet, although the mass uncertainties in the four-planet model have become larger than estimated from three-planet models \citep{libbyroberts2020}.

\subsection{Posterior Sampling for the 2:1 Solution}\label{ssec:4planet_2to1}

\begin{figure*}
	\epsscale{1.05}
	\plotone{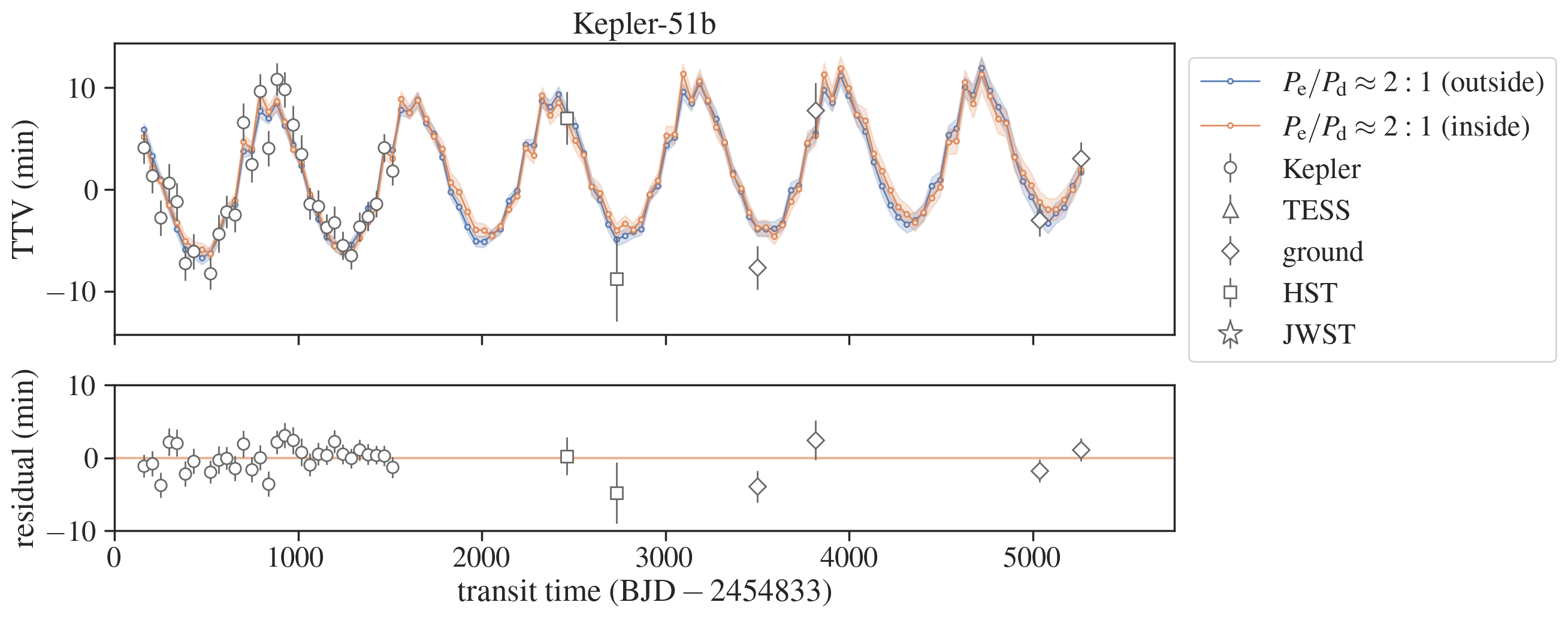}
    \plotone{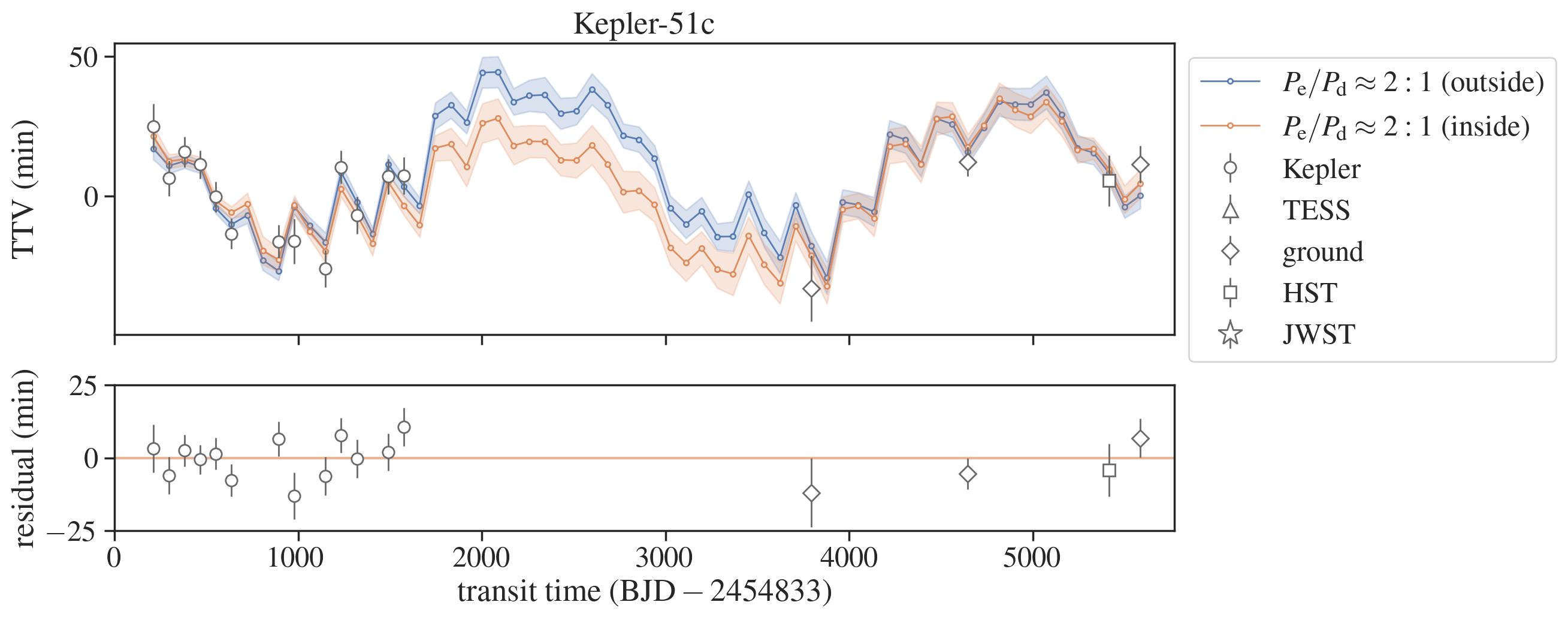}
    \plotone{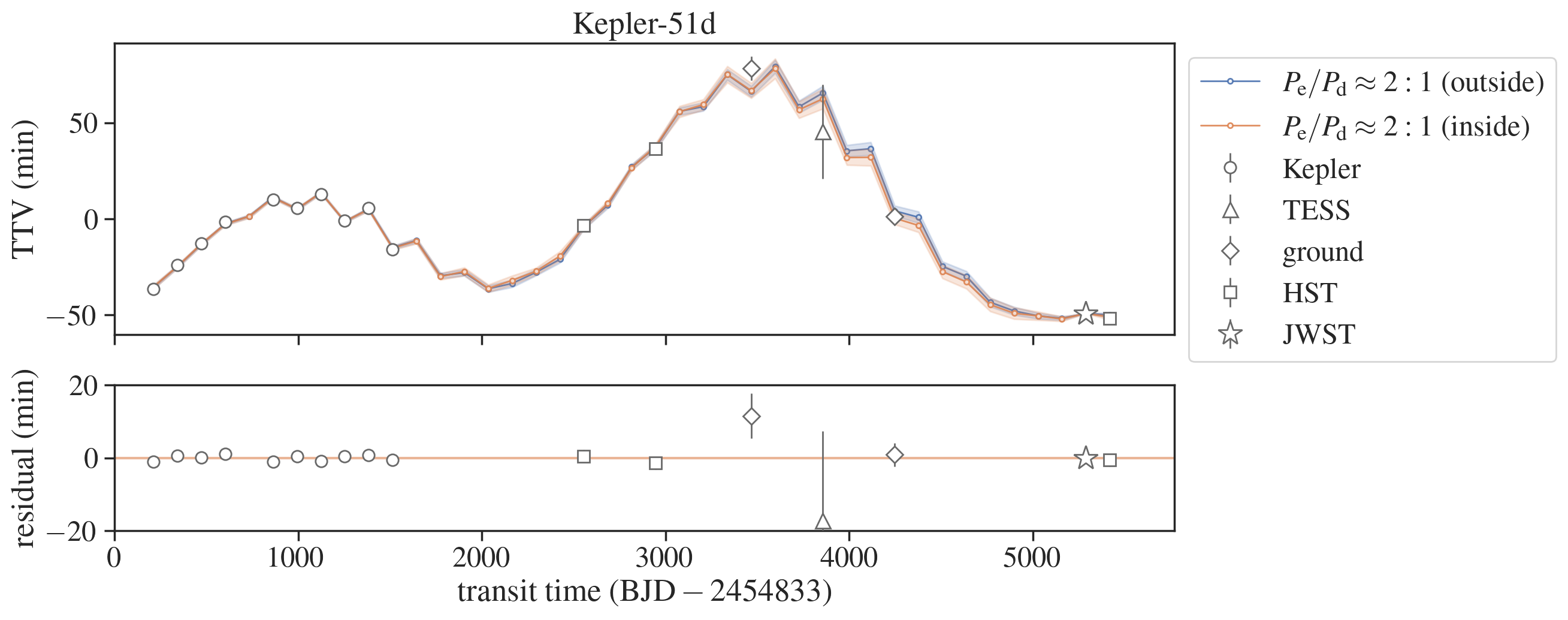}
	\caption{Comparison between the solutions in which Kepler-51e is around the $2:1$ resonance with Kepler-51d, 
    and all the timing data presented in this paper (Table~\ref{tab:transit_times}).
    Here the TTVs in the vertical axes are shown with respect to the linear ephemerides given by $t_0(\mathrm{BJD}-2454833)=159.1068, 209.9946, 212.0493$ and $P(\mathrm{days})=45.155296,  85.316963, 130.175183$ for planet b, c, and d respectively; note that the appearance of the plot depends on these arbitrary choices.
    The solid line and the shaded region show the mean and standard deviation of the TTV models at each time computed for the posterior samples obtained in Section~\ref{ssec:4planet_2to1}.}
	\label{fig:ttv_2to1}
\end{figure*}

\begin{figure*}
	\epsscale{1.1}
	\plotone{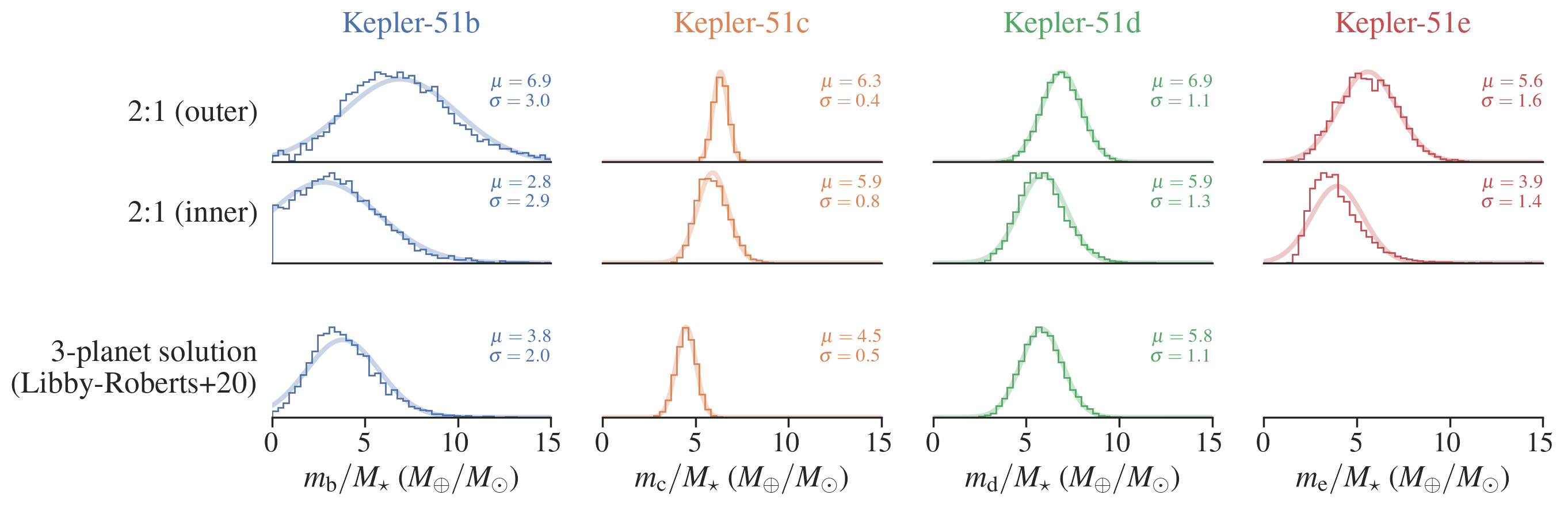}
	\caption{Comparison of the mass ratio posteriors from the four-planet model assuming the near $2:1$ period ratio for the outermost pair (top two rows) and from the previous three-planet model (bottom). The thick solid lines show the truncated normal distributions with lower bounds of zero fitted to the posterior samples, whose location ($\mu$) and scale ($\sigma$) parameters are given in each panel. These values agree with the means and standard deviation of the posteriors for planet c and d, but not for planet b because the distribution is truncated at zero. As discussed in Section~\ref{sec:4planet}, the TTV data do not exclude other solutions with different orbital periods for planet e, in which the masses of the inner three planets vary only slightly, while that of Kepler-51e can vary by orders of magnitudes (see Figures~\ref{fig:p4_m4_e4} and \ref{fig:masses}--\ref{fig:profile} and the discussion in Sections~\ref{sssec:4planet_results}--\ref{ssec:4planet_profile}).}
	\label{fig:mass_ratios}
\end{figure*}

We further examine the mass constraints from Section~\ref{ssec:4planet_profile} with a Markov Chain Monte Carlo (MCMC) analysis. Since it is not computationally feasible to perform the analysis for all possible period ratios, here we focus on the $2:1$ solution, which yields the largest mass upper limit for planet b among the solutions searched (Figure~\ref{fig:profile}). Such a solution deserves attention in light of the puzzlingly low inferred densities of the Kepler-51 planets, as it can accommodate higher densities. 
Therefore it is useful to examine the accuracy of the mass limits for this solution with an independent analysis.

There are several additional properties of the $2:1$ solution that are worth noting. 
First, this is one of the best solutions in terms of the $\chi^2$ values (Figure~\ref{fig:profile}).
Second, at this period ratio the largest number of ``good'' solutions are found among the $T_4$ bins, i.e., the points in the middle panels of Figure~\ref{fig:p4_m4_e4} are the most densely clustered around this period ratio. This solution is therefore less fine-tuned in terms of the unknown orbital phase of planet e than the others.
Third, the $2:1$ solution favors $e_{\rm e}\lesssim 0.1$ and $\mue\sim 1$--$10\,M_\oplus$, as shown in the middle panels of Figure~\ref{fig:p4_m4_e4}, and this is essentially the only solution from the brute-force search that satisfies both. The inferred low eccentricity is reminiscent of planets in the compact multi-transiting systems from Kepler \citep{2014ApJ...787...80H, 2015ApJ...808..126V}, while the inferred mass range is comparable to the masses of the inner three transiting planets (as will be further quantified below). Mass uniformity has been suggested for other known compact multi-planet systems \citep{2017ApJ...849L..33M}. The $2:1$ period ratio is also fairly common among Kepler multi-planet systems \citep{2015ARA&A..53..409W}.
Thus the properties of Kepler-51e from the $2:1$ solution appear to be a priori plausible, although it is difficult to quantify the plausibility relative to the other solutions. This provides further motivation to examine this solution in detail.

We draw samples from the joint posterior PDF for the set of system parameters $\theta$ conditioned on the timing data $d$:
\begin{align}
    p(\theta|d) = {p(d|\theta)\,p(\theta) \over p(d)}.
\end{align}
The parameter $\theta$ consists of the mass ratios and orbital elements of the four planets (see Section~\ref{sec:model}). 
Here we adopt the likelihood function given by
\begin{align}
\label{eq:loglike_ttv}
    \ln p(\theta|d) = -{1\over 2} \sum_{i \in \mathrm{all\ data}} \left[{(t_i - m_i)^2 \over {\sigma_i^2}} + \ln (2\pi\sigma_i^2)\right].
\end{align}
The prior PDF $p(\theta)$ is assumed to be separable for each model parameter, as summarized in Table~\ref{tab:ttvpriors}. Because we found it difficult to sample simultaneously from solutions inside (i.e., $P_{\rm e}/P_{\rm d}<2$) and outside ($P_{\rm e}/P_{\rm d}>2$) the $2:1$ resonance due to the presence of a low-likelihood region near $P_{\rm e}/P_{\rm d}=2$ (see the middle-left panel of Figure~\ref{fig:p4_m4_e4}), we split the prior for $P_{\rm e}$ at $2P_{\rm d}$ and sampled separately. 
The sampling was performed using the No-U-Turn Sampler \citep{DUANE1987216, 2017arXiv170102434B} as implemented in {\tt NumPyro} \citep{bingham2018pyro, phan2019composable}, turning on the dense mass matrix option that adjusts the non-diagonal elements of the mass matrix in addition to the diagonal ones. We ran four independent chains in parallel for 2,000 steps, after which we found the effective number of samples $n_{\rm eff}$ to be a few 100--1,000 and the split Gelman--Rubin statistic $\hat{R}$ to be $1.00$--1.04 for each parameter \citep{BB13945229}. 

\begin{deluxetable}{lc}
\caption{The prior PDFs adopted when sampling from the posterior distributions for the $2:1$ solutions in Section~\ref{ssec:4planet_2to1}.}
\label{tab:ttvpriors}
\tablehead{
	\colhead{Parameter} & \colhead{Prior}
} 
\startdata
\textit{(inner three transiting planets)}\\
planet-to-star mass ratio & $\mathcal{U}(0,5\times 10^{-4})$\\
orbital period (days) & $\mathcal{U}(P_0-0.05, P_0+0.05)$\\
orbital eccentricity & $\mathcal{U}(0,0.2)$\\
argument of periastron & $\mathcal{U}(0,2\pi)$\tablenotemark{$\dagger$}\\
time of inferior conjunction (days) & $\mathcal{U}(T_0-0.05, T_0+0.05)$\\
\textit{(Kepler-51e)}\\
planet-to-star mass ratio & $\mathcal{U}(0,5\times 10^{-4})$\\
orbital period (days) & $\mathcal{U}(250.35, 260.35)$ (inside)\\
 & or $\mathcal{U}(260.35, 270.35)$ (outside)\\
orbital eccentricity & $\mathcal{U}(0,0.2)$\\
argument of periastron & $\mathcal{U}(0,2\pi)$\tablenotemark{$\dagger$}\\
time of inferior conjunction  & $\mathcal{U}(200,400)$\\
($\mathrm{BJD}-2454833$) & \\
\enddata
\tablecomments{$\mathcal{U}(a,b)$ denotes the uniform distribution between $a$ and $b$. $\mathcal{LU}(a,b)$ denotes the log-uniform distribution between $a$ and $b$. The symbols $P_0$ and $T_0$ denote the linear ephemeris computed from the observed transit times for each planet.} 
\tablenotemark{$\dagger$}{These parameters were treated as angles and the distributions were wrapped at the edges.}
\end{deluxetable}

Our results are summarized in Table~\ref{tab:2to1}, which also includes the absolute masses of the four planets as well as the radii and mean densities of the three transiting planets, computed using the posterior samples for the mass ratios from the TTVs, the posterior samples for the radius ratios from the Kepler long-cadence light curves (Section~\ref{ssec:obs_kepler}), 
and the stellar mass and radius from Section~\ref{sec:star}. As mentioned in 
Section~\ref{sec:model},
the planetary masses in units of $M_\oplus$ are mostly the mass ratios in units of $\muunit$, but their uncertainties include that of the stellar mass and are therefore larger. 
The full posterior samples for each solution are available through the author's GitHub (see footnote~\ref{footnote:github}).
The corresponding TTV models and corner plot are shown in Figures~\ref{fig:ttv_2to1} and \ref{fig:corner_2to1}, respectively.

The 95\% highest density intervals for the mass ratios 
are comparable to those from the $\chi^2$ analysis in Section~\ref{ssec:4planet_profile} and Figure~\ref{fig:profile}, and confirm those mass constraints from a Bayesian point of view. We note that exact agreement is not necessarily expected, because the two constraints have different meanings even for the uniform priors on the mass ratios adopted here. Figure~\ref{fig:profile} shows the {\it projection} of the (negative log) likelihood profiles onto the mass ratio axes, while Figure~\ref{fig:mass_ratios} shows the probability distributions for the mass ratios after {\it integrating} over the other parameters. Thus the latter, in principle, depends on how the mass ratios are degenerate with the other parameters, as well as on the prior PDFs assumed for all the model parameters. The two constraints could therefore differ significantly if the posterior distribution is sensitive to the prior. This does not appear to be the case here.

As shown in Figure~\ref{fig:mass_ratios}, the mass of planet e is inferred to be $\approx 5\,M_\oplus$ for solutions both outside and inside the $2:1$ period ratio, 
consistent with the masses of the three transiting planets within the uncertainties. 
The latter masses did not change drastically from the previous three-planet model (bottom row of Figure~\ref{fig:mass_ratios}), although the mass ratio posterior for Kepler-51c shifted upwards by $\sim 1\,\muunit$. This is likely due to the fact that Kepler-51d's TTVs now include contributions from both Kepler-51c and e.

\begin{deluxetable*}{l@{\hspace{.1cm}}ccc@{\hspace{.4cm}}ccc}
\tablecaption{Planetary and orbital parameters of the $2:1$ solution computed from the posterior samples obtained in Section~\ref{ssec:4planet_2to1}.\label{tab:2to1}}
\tablehead{
\colhead{} & \multicolumn{3}{c}{outside $2:1$} 
& \multicolumn{3}{c}{inside $2:1$}\\
\colhead{} & \colhead{mean} & \colhead{SD} & \colhead{$95\%$ HDI} & \colhead{mean} & \colhead{SD} & \colhead{$95\%$ HDI} 
}
\startdata
\textit{\textbf{(Kepler-51b)}}\\
mass ratio $m$ ($M_\oplus/M_\odot$)&$6.9$&$2.9$&$[1.53,12.84]$&$3.7$&$2.2$&$[0.0,7.6]$\\
time of inferior conjunction (BKJD)&$159.11087$&$0.00039$&$[159.1101,159.1116]$&$159.11053$&$0.00047$&$[159.1096,159.1114]$\\
orbital period $P$ (days)&$45.15396$&$0.00022$&$[45.15350,45.15436]$&$45.15405$&$0.00039$&$[45.1532,45.1548]$\\
eccentricity $e$&$0.0162$&$0.0038$&$[0.009,0.024]$&$0.026$&$0.010$&$[0.007,0.046]$\\
$e\cos\omega$&$-0.0156$&$0.0038$&$[-0.023,-0.008]$&$-0.0114$&$0.0048$&$[-0.021,-0.002]$\\
$e\sin\omega$&$-0.0012$&$0.0044$&$[-0.010,0.007]$&$-0.022$&$0.011$&$[-0.045,-0.002]$\\
mass ($M_\oplus$)&$6.7$&$2.8$&$[1.39,12.22]$&$3.5$&$2.1$&$[0.1,7.3]$\\
radius ($R_\oplus$)&$6.83$&$0.13$&$[6.57,7.08]$&$6.83$&$0.13$&$[6.57,7.08]$\\
density ($\mathrm{g/cm^3}$)&$0.115$&$0.048$&$[0.02,0.21]$&$0.060$&$0.037$&$[0.00,0.13]$\\
\textit{\textbf{(Kepler-51c)}}\\
mass ratio $m$ ($M_\oplus/M_\odot$)&$6.34$&$0.40$&$[5.5,7.1]$&$5.89$&$0.83$&$[4.4,7.5]$\\
time of inferior conjunction (BKJD)&$210.0049$&$0.0025$&$[209.9999,210.0097]$&$210.0078$&$0.0029$&$[210.002,210.013]$\\
orbital period $P$ (days)&$85.3147$&$0.0018$&$[85.3111,85.3182]$&$85.3139$&$0.0020$&$[85.3103,85.3180]$\\
eccentricity $e$&$0.0093$&$0.0077$&$[0.000,0.026]$&$0.063$&$0.020$&$[0.021,0.104]$\\
$e\cos\omega$&$0.0073$&$0.0077$&$[-0.003,0.026]$&$0.029$&$0.014$&$[0.001,0.054]$\\
$e\sin\omega$&$-0.00073$&$0.00575$&$[-0.012,0.013]$&$-0.054$&$0.019$&$[-0.094,-0.016]$\\
mass ($M_\oplus$)&$6.09$&$0.41$&$[5.3,6.9]$&$5.65$&$0.81$&$[4.1,7.2]$\\
radius ($R_\oplus$)&$6.4$&$1.4$&$[4.5,8.9]$&$6.4$&$1.4$&$[4.5,8.9]$\\
density ($\mathrm{g/cm^3}$)&$0.151$&$0.068$&$[0.03,0.28]$&$0.140$&$0.066$&$[0.02,0.27]$\\
\textit{\textbf{(Kepler-51d)}}\\
mass ratio $m$ ($M_\oplus/M_\odot$)&$6.9$&$1.1$&$[4.9,9.1]$&$5.9$&$1.3$&$[3.4,8.4]$\\
time of inferior conjunction (BKJD)&$212.0223$&$0.0017$&$[212.0194,212.0258]$&$212.02524$&$0.00081$&$[212.0237,212.0269]$\\
orbital period $P$ (days)&$130.1858$&$0.0018$&$[130.1824,130.1891]$&$130.1820$&$0.0024$&$[130.1776,130.1868]$\\
eccentricity $e$&$0.0061$&$0.0056$&$[0.000,0.018]$&$0.048$&$0.016$&$[0.013,0.080]$\\
$e\cos\omega$&$0.0027$&$0.0062$&$[-0.006,0.018]$&$0.020$&$0.011$&$[-0.001,0.041]$\\
$e\sin\omega$&$0.00099$&$0.00467$&$[-0.008,0.012]$&$-0.042$&$0.016$&$[-0.076,-0.012]$\\
mass ($M_\oplus$)&$6.6$&$1.0$&$[4.7,8.7]$&$5.6$&$1.2$&$[3.4,8.2]$\\
radius ($R_\oplus$)&$9.32$&$0.18$&$[8.98,9.66]$&$9.32$&$0.18$&$[8.98,9.66]$\\
density ($\mathrm{g/cm^3}$)&$0.0448$&$0.0071$&$[0.031,0.059]$&$0.0381$&$0.0085$&$[0.022,0.055]$\\
\textit{\textbf{(Kepler-51e)}}\\
mass ratio $m$ ($M_\oplus/M_\odot$)&$5.6$&$1.6$&$[2.6,8.5]$&$3.9$&$1.3$&$[1.8,6.5]$\\
time of inferior conjunction (BKJD)&$262.508$&$11.445$&$[240.75,283.64]$&$285.174$&$9.210$&$[266.89,302.55]$\\
orbital period $P$ (days)&$264.284$&$0.905$&$[262.5,265.9]$&$256.860$&$0.631$&$[255.8,258.2]$\\
eccentricity $e$&$0.020$&$0.015$&$[0.000,0.048]$&$0.080$&$0.032$&$[0.01,0.13]$\\
$e\cos\omega$&$-0.011$&$0.015$&$[-0.044,0.015]$&$0.023$&$0.019$&$[-0.010,0.060]$\\
$e\sin\omega$&$-0.0068$&$0.0148$&$[-0.040,0.020]$&$-0.074$&$0.032$&$[-0.13,0.00]$\\
mass ($M_\oplus$)&$5.4$&$1.5$&$[2.5,8.1]$&$3.8$&$1.3$&$[1.6,6.1]$\\
\enddata
\tablecomments{
The values and uncertainties shown in this table are derived assuming that Kepler-51e is near the $2:1$ resonance with Kepler-51d; other orbital periods are not excluded by the data, and the mass of Kepler-51e varies from $\sim 0.1\,M_\oplus$ to $\sim M_\mathrm{Jup}$ depending on the value of the correct period.
As indicated by the HDI (highest density interval, within which every point has a higher probability density than any point outside it), we do not find a very stringent lower bound on the mass of Kepler-51b. This upper limit is also sensitive to the period of Kepler-51e. Among the solutions we found, the ``outside $2:1$'' case shown here provides the largest upper limit and may be considered most conservative. The masses of Kepler-51c and d are more robustly determined and remain within $\sim 2\sigma$ regardless of the period of Kepler-51e. BKJD refers to $\mathrm{BJD}-2454833$.}
\end{deluxetable*}

\begin{figure*}
	\epsscale{1.1}
	\plotone{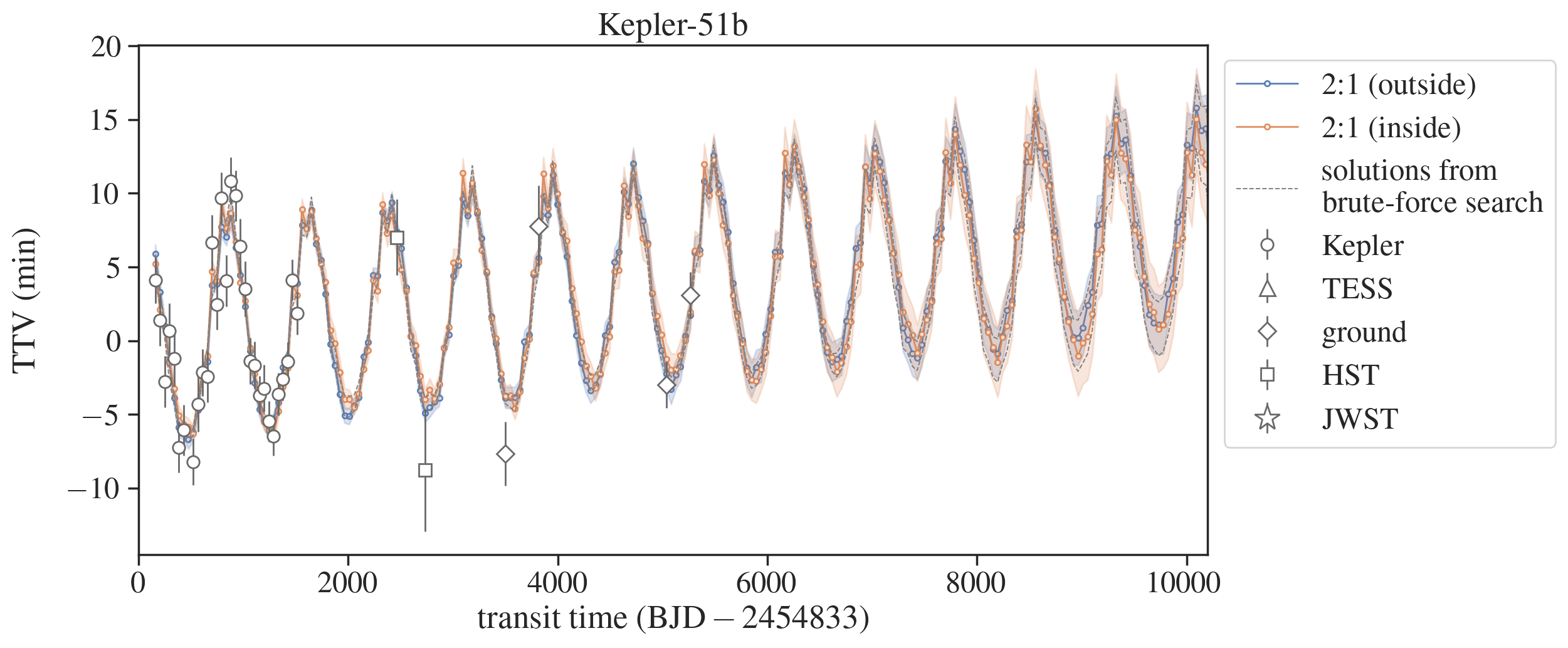}
    \plotone{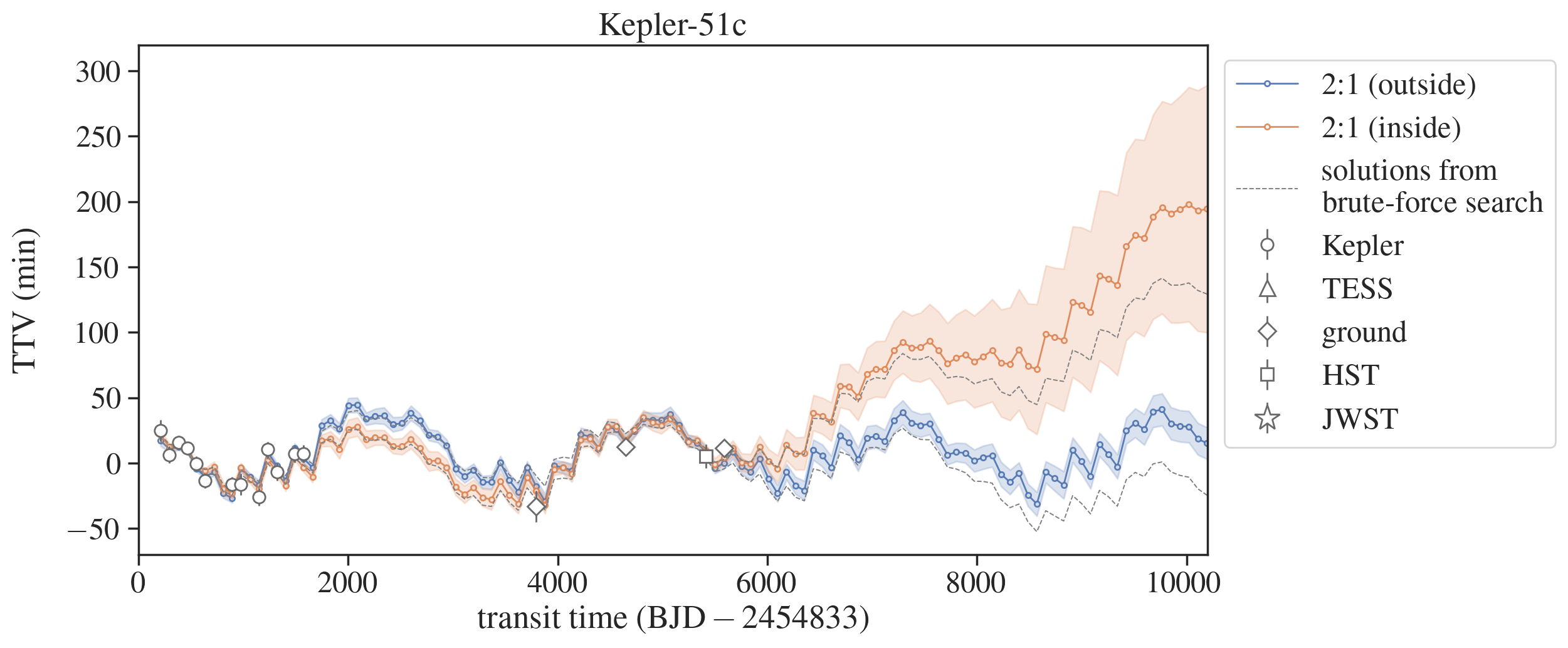}
    \plotone{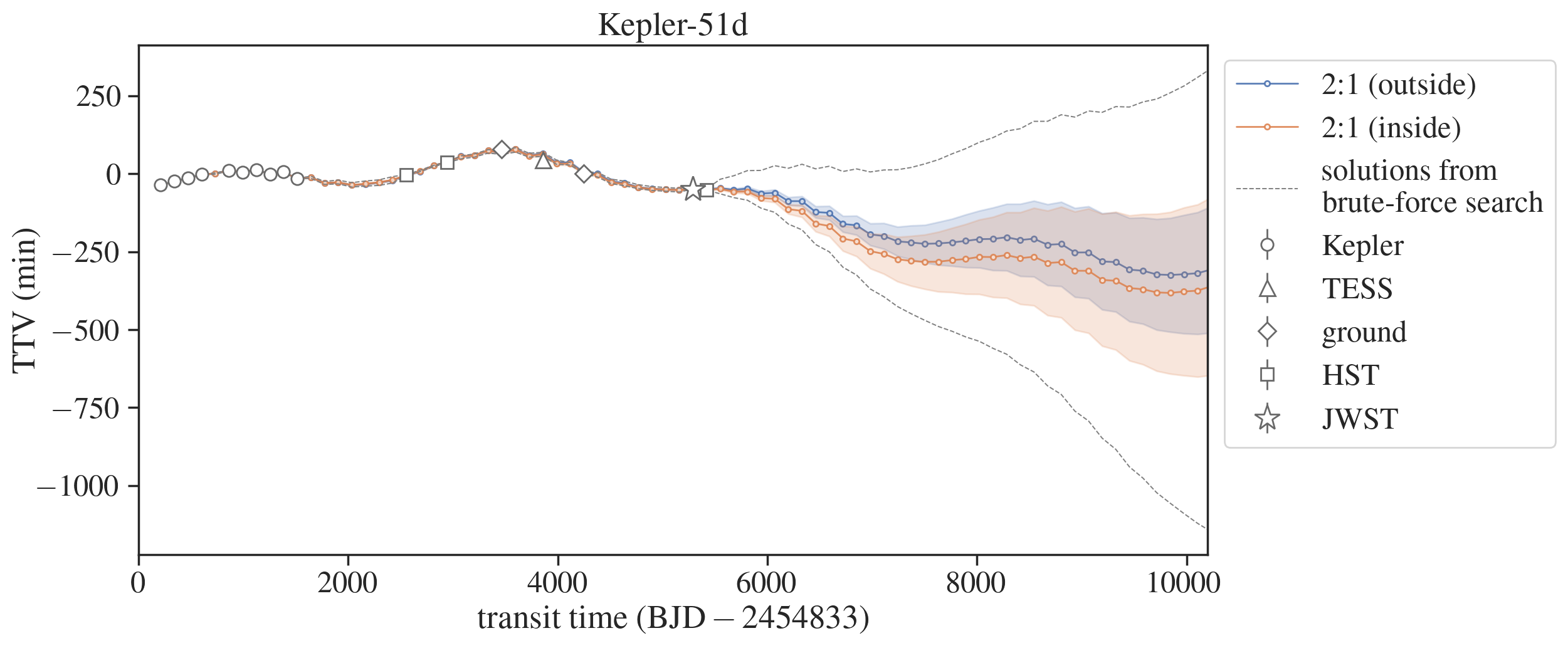}
	\caption{The TTVs of Kepler-51b, c, and d predicted up to 2036. 
    Here the TTVs in the vertical axes are shown with respect to the linear ephemerides given by $t_0(\mathrm{BJD}-2454833)=159.1068, 209.9946, 212.0493$ and $P(\mathrm{days})=45.155296,  85.316963, 130.175183$ for planet b, c, and d respectively; note that the appearance of the plot depends on these arbitrary choices.
    The 68\% intervals of the solutions from the brute-force search are shown with the gray dashed lines. The 68\% intervals for the $2:1$ solutions based on posterior samples are shown with blue (outside) and orange (inside) shades; these are not the only possible solutions, and they are meant to illustrate the prediction uncertainty within a solution with a given period ratio.}
	\label{fig:ttv_predict}
\end{figure*}

\section{Discussion}\label{sec:discussion}

\subsection{Prospects of Determining the Properties of Kepler-51e with Future TTVs}\label{ssec:4planet_future}

We showed that all observed TTVs, including the discrepant JWST data, can be accounted for with a wide range of masses and orbital periods for Kepler-51e. In other words, the properties of Kepler-51e are not well constrained with the available timing data. What observations could remedy this situation?

We used the solutions found from the brute-force search (Section~\ref{ssec:4planet_brute}) as well as the posterior samples for the $2:1$ solution (Section~\ref{ssec:4planet_2to1}) to predict future TTVs of the three transiting planets. The blue and orange lines and shaded areas in Figure~\ref{fig:ttv_predict} show the mean and standard deviation of the posterior models computed for the solutions outside and inside $2:1$, respectively. 
As we discussed in Section~\ref{ssec:4planet_brute}, these solutions are not the only ones allowed by the data, but here we show them as an example to illustrate the prediction uncertainty associated with a solution with a given period for planet e.
In contrast, the gray dashed lines show the same interval (one standard deviation around the mean) for all the solutions in Figure~\ref{fig:p4_m4_e4} found from the brute-force search, which involve a wide range of orbital periods for planet e. This interval does not possess a statistically well-defined interpretation, but here we show it to illustrate the degree of scatter of all possible solutions. These predicted transit times are given in Table~\ref{tab:ttvprediction} in Appendix~\ref{sec:app_corner}.

As exemplified by the blue and orange curves for planet c and d, the high-probability regions differ for each solution. At the same time, there is overlap in the prediction ranges at a given epoch: the areas shaded in orange and blue overlap with, or fall between, the two gray dashed lines.
Therefore, a single timing measurement or just a few will not be able to completely single out a specific solution. Instead, continuous measurements of transit times that constrain the shape of the TTV curve will be necessary to distinguish different solutions. These follow-up observations will be particularly informative for planet c and d, for which predictions made by different solutions exhibit a larger dispersion than for b. However, we also emphasize the importance of monitoring planet b to continuously verify the accuracy of the four-planet model, considering how the presence of planet e was revealed.

It is currently challenging to measure the masses of the Kepler-51 planets using the radial velocity method for multiple reasons: the host star Kepler-51 is faint ($G\approx 14.7$) and exhibits a strong stellar activity, and its planets have relatively low masses and long orbital periods.
For the $2:1$ solution, for example, the radial velocity semi-amplitude induced by Kepler-51e ($m\approx 5\,M_\oplus$ and $P\approx 260\,\mathrm{days}$) is only $\approx 0.5\,\mathrm{m/s}$. 
The amplitudes induced by the inner transiting planets are $\lesssim 1\,\mathrm{m/s}$.

\subsection{Implications of the Updated Masses of the Inner Three Transiting Planets}

Despite the uncertainties in the properties of the fourth planet, all the solutions found from our brute-force search imply $\lesssim 10\,M_\oplus$ for the inner three transiting planets (Section~\ref{ssec:4planet_brute}, \ref{ssec:4planet_profile}).
More quantitatively, the masses of Kepler-51c and d were found to be relatively insensitive to the assumed properties of the fourth planet 
and remain consistent with the previous values within $\sim 2\sigma$. A firm lower bound on the mass was not found for Kepler-51b, and the upper limit turned out to depend on the assumed period of Kepler-51e. The largest upper limit of $\lesssim 12(16)\,M_\oplus$ at the 95(99.7)\% level was found for the solutions in which Kepler-51e is slightly outside the $2:1$ mean-motion resonance with Kepler-51d. 
The three planets' densities therefore remain low, as discussed in Section~\ref{sec:intro}, although the larger upper limit for planet b's mass may help alleviate some of the theoretical difficulties for understanding its origins and atmospheric loss history. 

\subsection{Peas-in-a-pod Pattern Extending Beyond Kepler-51d?}


It is not well understood what determines the orbital period distribution of the outermost transiting planets in the Kepler sample. \citet{2022AJ....164...72M} argued that the outermost transiting planets have shorter periods than expected purely from geometric and detection biases, and that this could result from a drop-off in planet occurrence (i.e., physical truncation), a decrease in the typical planet radius, an increase in the period ratios between the adjacent planets, an increase in the mutual orbital inclination (i.e., change in architecture), or a combination of all of the above in the outer parts of planetary systems. 
In any of these cases, the system architecture beyond the transiting planets provides important clues for the formation and dynamical evolution of compact multi-planet systems. 

Our analysis demonstrates that long-term TTV monitoring can provide useful insights for testing the various scenarios as described above. For Kepler-51, we show that the system is not physically truncated at the outermost transiting planet by identifying another longer-period planet. We also find a solution that implies similar masses and period ratios for all four planets, i.e., the ``peas-in-a-pod'' architecture extending beyond transiting planets \citep{2017ApJ...849L..33M, 2018AJ....155...48W}. If this turns out to be a generic feature of Kepler multi-transiting systems, then it might be that an increase in the mutual inclination is responsible for the ``truncation'' of the transiting systems. At this point, this is just one possible solution for Kepler-51, but our analysis suggests that this scenario is testable with more observations. To probe mutual orbital inclinations, searches for transit duration variations (TDVs) in addition to TTVs will also be informative, as we will discuss next. 

\subsection{Is the Whole System Indeed Coplanar?}\label{ssec:disc_tdv}

Throughout the paper, we assumed that the orbits of the Kepler-51 planets are coplanar. 
The fact that Kepler-51b, c, and d are all transiting supports this assumption at least for these three. The assumption may also be supported from a dynamical point of view: a misalignment in the orbits necessarily induces nodal precession of the orbital planes, which would have easily caused the grazing orbit of Kepler-51c to rotate out of our line of sight to the star \edit1{(see also Section~\ref{ssec:3planet_photod}).}
That said, this argument is less constraining for the orbit of Kepler-51e, which affects Kepler-51c's orbit less than Kepler-51b and d. Indeed, a larger mutual inclination is consistent with the lack of its transits, although we note that a misalignment as small as $1^\circ$ is sufficient for this to happen. 

If the orbit of Kepler-51e is misaligned, Kepler-51d may exhibit TDVs associated with a slow drift of the transit impact parameter due to nodal precession of its orbit \citep{2002ApJ...564.1019M}. We searched for TDVs of Kepler-51d using the precise transit durations from JWST and Kepler separated by 10~years. We did not use the transits from the other facilities because they include gaps during transits, in particular around the ingress and egress. 
We evaluated the interval $T$ between the halfway points of ingress and egress as an average of the ``total'' and ``full'' durations \citep[Equations~14 and 15 of][]{2011exop.book...55W} using the posterior samples obtained from transit fits in Section~\ref{ssec:obs_jwst} and Section~\ref{ssec:obs_kepler}. 
The mean and standard deviation of the posterior distribution for the duration difference $\Delta T$ (JWST minus Kepler) were found to be $0.9\pm1.7\,\mathrm{min}$ for the Kepler long-cadence data, and $0.1\pm2.2\,\mathrm{min}$ for the short-cadence data. Thus, there is no evidence of TDVs \edit1{$\dot{T}$} at a rate higher than a few $0.1\,\mathrm{min/yr}$.
On the other hand, nodal precession due to planet e should induce the TDVs of planet d at a rate of 
\begin{align}
\label{eq:tdv}
    \dot{T}\sim 1\,\mathrm{min/yr}\left({b_{\rm d}\over\sqrt{1-b_{\rm d}^2}}\right)\left(P_{\rm d} \over P_{\rm e}\right)^2\left(m_{\rm e}/M_\star \over M_\oplus/M_\odot\right)\,\Delta i,
\end{align}
where $\Delta i$ \edit1{(radians)} denotes the mutual inclination between the orbits of planets d and e \citep{2021MNRAS.505.1293S}.
This reduces to $\dot{T} \sim 0.1\Delta i\,\mathrm{min/yr}$ adopting $b_{\rm d}\sim 0.1$, $P_{\rm d}/P_{\rm e}\sim 1/2$, and $\mue\sim 5\muunit$ (i.e., $2:1$ solution). Thus the lack of clear TDVs for Kepler-51d 
\edit1{does not provide useful constraints on $\Delta i$}
at the moment: $\Delta i$ of order unity is not excluded. 
We note, though, that a dynamical analysis also incorporating the transit durations/shapes of Kepler-51c may improve the constraint on $\Delta i$, because its grazing transits may serve as a more sensitive probe of nodal precession (note the $1/\sqrt{1-b^2}$ dependence in Eq.~\ref{eq:tdv}). We leave such an analysis for future work.

\subsection{Are Four Planets Enough?}

\edit1{
While we find no clear flaw in the four-planet model based on the available timing data, this does not necessarily rule out the possibility of a fifth planet in the system. There may indeed be theoretical motivations for investigating such a hypothesis. For example, the period ratio of Kepler-51b and c deviates more from $1:2$ ($(P_{\rm c}/P_{\rm b})/2 - 1 = -0.055$) than is typically observed in other compact multi-transiting systems \citep[e.g.,][]{2014ApJ...790..146F, 2024arXiv240606885D}. This might be more naturally explained by the presence of another planet between b and c, forming a chain of three-body generalized Laplace resonances as seen in the TRAPPIST-1 system \citep[e.g.,][]{2021PSJ.....2....1A}.
Although we found that a single planet located in between b and c does not appear to provide a satisfactory explanation for the entire TTV data (see Figure~\ref{fig:p4_m4_e4}), such solutions could work better in the presence of a fifth planet in the system, perhaps outside of Kepler-51d as in our four-planet solutions so that it also affects the TTVs of Kepler-51d. Given the vast parameter space and the challenges in distinguishing between solutions even within the four-planet model, we leave further exploration of these possibilities for future work.
}

\section{Summary and Conclusion}\label{sec:summary}

The Kepler-51 system is unlike any other discovered to date, with a young ($\lesssim 1\,\mathrm{Gyr}$) Sun-like star hosting three low-mass transiting planets all with mean densities $\sim 0.1\,\mathrm{g/cm^3}$. The masses of these planets are derived via TTVs, and thus understanding the dynamical architecture of the entire system is critical for accurate mass measurements. 

In this paper, we presented the discovery of a fourth planet in the Kepler-51 system based on an extensive transit timing data set spanning over 14~years, which we compiled by reanalyzing the available Kepler/TESS photometry and by adding new measurements from JWST, HST, and ground-based facilities (Section~\ref{sec:obs}). Our key findings are summarized as follows:
\begin{itemize}
\item The new transit time of the outermost transiting planet, Kepler-51d, from JWST Cycle 1 observations significantly deviates from the prediction based on TTV models that consider the three known transiting planets alone (Section~\ref{sec:3planet}),
which we confirmed with ground-based and follow-up HST observations.
This demonstrates the presence of a fourth planet in the system,\footnote{This is the first incidental planet discovery made by JWST.} whose transits have not been identified.
\item The fourth planet, Kepler-51e, located beyond Kepler-51d's orbit with a wide range of masses and orbital periods can account for all the observed TTVs including the discrepancy found in the JWST data (Section~\ref{sec:4planet}). In other words, the property of Kepler-51e is not well determined by the available TTVs. It could either be a low-mass ($\lesssim 10\,M_\oplus$) planet on a near-circular orbit around a certain low-order mean-motion resonance with Kepler-51d, or a more massive planet on a distant, eccentric orbit. 
\item Despite this uncertainty, all the coplanar four-planet solutions found from our brute-force search imply $\lesssim 10\,M_\oplus$ for the inner three transiting planets (Section~\ref{ssec:4planet_brute}, \ref{ssec:4planet_profile}): thus their densities remain low. However, the increase in the mass upper limit for planet b will need to be considered carefully in the quantitative discussion. 
\item One of the best solutions found from the brute-force search implies that Kepler-51e is around the $2:1$ mean-motion resonance with Kepler-51d. This solution, unlike others, implies that Kepler-51e has a near-circular orbit and has a mass of $\sim 5\,M_\oplus$ consistent with those of the inner three transiting planets. Thus the near-circular, flat, ``peas-in-a-pod'' type architecture as seen in many compact multi-planet systems may extend to $\approx 260\,\mathrm{days}$ in this system. This is also the solution that allows for the largest mass for Kepler-51b and is least fine-tuned in terms of the orbital phase of Kepler-51e.
\end{itemize}

Long-term monitoring of Kepler systems for dynamical signatures of additional planets is proving essential for uncovering their full architecture and gaining the insights needed to understand the formation and evolution of compact multi-planet systems.
Future observations of the Kepler-51 planets are vital both for robustly testing the various possible solutions for Kepler-51e and for improving the mass constraints on the three transiting planets, while also searching for any additional long-period TTVs and TDVs. Regardless of whether it is possible to uniquely determine Kepler-51e's properties from the timing data alone, it is important to keep observing transits of the Kepler-51 planets so that their ephemerides are kept up-to-date to enable follow-up transit studies of this planet. The two hour deviation of Kepler-51d's transit time from the prediction of the three-planet model during the JWST observation was fortunately not large enough for us to miss the transit. However, without more timing measurements for Kepler-51d, the prediction uncertainty grows to even larger values in the next few years, jeopardizing future transit investigations. 
We provide future predicted transit times of all three transiting planets (Table~\ref{tab:ttvprediction}) to facilitate such follow-up observations.

\acknowledgments

This work is based on observations made with the NASA/ESA/CSA James Webb Space Telescope. The data were obtained from the Mikulski Archive for Space Telescopes at the Space Telescope Science Institute, which is operated by the Association of Universities for Research in Astronomy, Inc., under NASA contract NAS 5-03127 for JWST. These observations are associated with program GO-2571. Support for program GO-2571 was provided by NASA through a grant from the Space Telescope Science Institute, which is operated by the Association of Universities for Research in Astronomy, Inc., under NASA contract NAS 5-03127.

This research is based on observations made with the NASA/ESA Hubble Space Telescope obtained from the Space Telescope Science Institute, which is operated by the Association of Universities for Research in Astronomy, Inc., under NASA contract NAS 5–26555. These observations are associated with program: DD Program 17585.

This paper made use of data collected by the Kepler and TESS missions. Funding for these mission is provided by the NASA Science Mission directorate. This research made use of Lightkurve, a Python package for Kepler and TESS data analysis (Lightkurve Collaboration, 2018).

This work has made use of data from the European Space Agency (ESA) mission {\it Gaia} (\url{https://www.cosmos.esa.int/gaia}), processed by the {\it Gaia} Data Processing and Analysis Consortium (DPAC, \url{https://www.cosmos.esa.int/web/gaia/dpac/consortium}). Funding for the DPAC has been provided by national institutions, in particular, the institutions participating in the {\it Gaia} Multilateral Agreement.

This publication makes use of data products from the Two Micron All Sky Survey, which is a joint project of the University of Massachusetts and the Infrared Processing and Analysis Center/California Institute of Technology, funded by the National Aeronautics and Space Administration and the National Science Foundation.

This work is partly based on observations made with the MuSCAT2 instrument, developed by ABC, at Telescopio Carlos S\'{a}nchez operated on the island of Tenerife by the IAC in the Spanish Observatorio del Teide.

This work is partly based on observations made with the MuSCAT3 instrument, developed by the Astrobiology Center and under financial supports by JSPS KAKENHI (JP18H05439) and JST PRESTO (JPMJPR1775), at Faulkes Telescope North on Maui, HI, operated by the Las Cumbres Observatory.

The research leading to these results has received funding from the ARC
grant for Concerted Research Actions, financed by the Wallonia-Brussels Federation. TRAPPIST is funded by the Belgian Fund for Scientific Research (Fond National de la Recherche Scientifique, FNRS) under the grant PDR T.0120.21. TRAPPIST-North is a project funded by the University of Liege (Belgium), in collaboration with Cadi Ayyad University of Marrakech (Morocco). 
MG is F.R.S.-FNRS Research Director and EJ is F.R.S.-FNRS Senior Research Associate.
The postdoctoral fellowship of KB is funded by F.R.S.-FNRS grant T.0109.20 and by the Francqui Foundation.

This work is partly based on observations from the Apache Point Observatory 3.5-meter telescope, owned and operated by the Astrophysical Research Consortium. 
We acknowledge support from NSF grants AST 1907622, 1909506, 1909682, 1910954 and the Research Corporation. 

The Center for Exoplanets and Habitable Worlds is supported by Penn State and the Eberly College of Science.

Computations for this research were performed on the Pennsylvania State University’s Institute for Computational and Data Sciences Advanced CyberInfrastructure (ICDS-ACI), including the CyberLAMP cluster supported by NSF grant MRI-1626251.  This content is solely the responsibility of the authors and does not necessarily represent the views of the Institute for Computational and Data Sciences.

Part of this research was carried out at the Jet Propulsion Laboratory, California Institute of Technology, under a contract with the National Aeronautics and Space Administration (80NM0018D0004).

We acknowledge financial support from the Agencia Estatal de Investigaci\'on of the Ministerio de Ciencia e Innovaci\'on MCIN/AEI/10.13039/501100011033 and the ERDF ``A way of making Europe'' through project PID2021-125627OB-C32, and from the Centre of Excellence ``Severo Ochoa'' award to the Instituto de Astrofisica de Canarias.

Work by KM was supported by JSPS KAKENHI grant No.~21H04998.
E. E-B. acknowledges financial support from the European Union and the State Agency of Investigation of the Spanish Ministry of Science and Innovation (MICINN) under the grant PRE2020-093107 of the Pre-Doc Program for the Training of Doctors (FPI-SO) through FSE funds.
CIC acknowledges support by NASA Headquarters through an appointment to the NASA Postdoctoral Program at the Goddard Space Flight Center, administered by ORAU through a contract with NASA. 
KAC acknowledges support from the TESS mission via subaward s3449 from MIT.
This work is partly supported by JSPS KAKENHI Grant Numbers JP24H00017 and JP24K00689, and JSPS Bilateral Program Number JPJSBP120249910.



\vspace{5mm}
\facilities{JWST, HST(WFC3), Kepler, TESS, ARC 3.5m (ARCTIC), TCS (MuSCAT2), LCO 0.4m (SBIG), LCO 2.0m (MuSCAT3)}

\software{
corner \citep{corner}, JAX \citep{jax2018github}, NumPyro \citep{bingham2018pyro, phan2019composable}, AstroImageJ \citep{2017AJ....153...77C}, TAPIR \citep{Jensen:2013}, jaxstar \citep{jaxstar}}

\vspace{5mm}
Some of the data presented in this article were obtained from the Mikulski Archive for Space Telescopes (MAST) at the Space Telescope Science Institute. The specific observations analyzed can be accessed via \dataset[doi: 10.17909/kxr3-7m59]{https://doi.org/10.17909/kxr3-7m59} (Transits of Kepler-51c and Kepler-51d) and \dataset[doi: 10.17909/t002-ht63]{https://doi.org/10.17909/t002-ht63} (JWST transit of Kepler-51d). 

\bibliography{references,references_masuda}
\bibliographystyle{aasjournal}

\appendix

\section{Corner Plots}\label{sec:app_corner}

\subsection{Three-planet Model}

\begin{figure*}
	\epsscale{1.1}
	\plotone{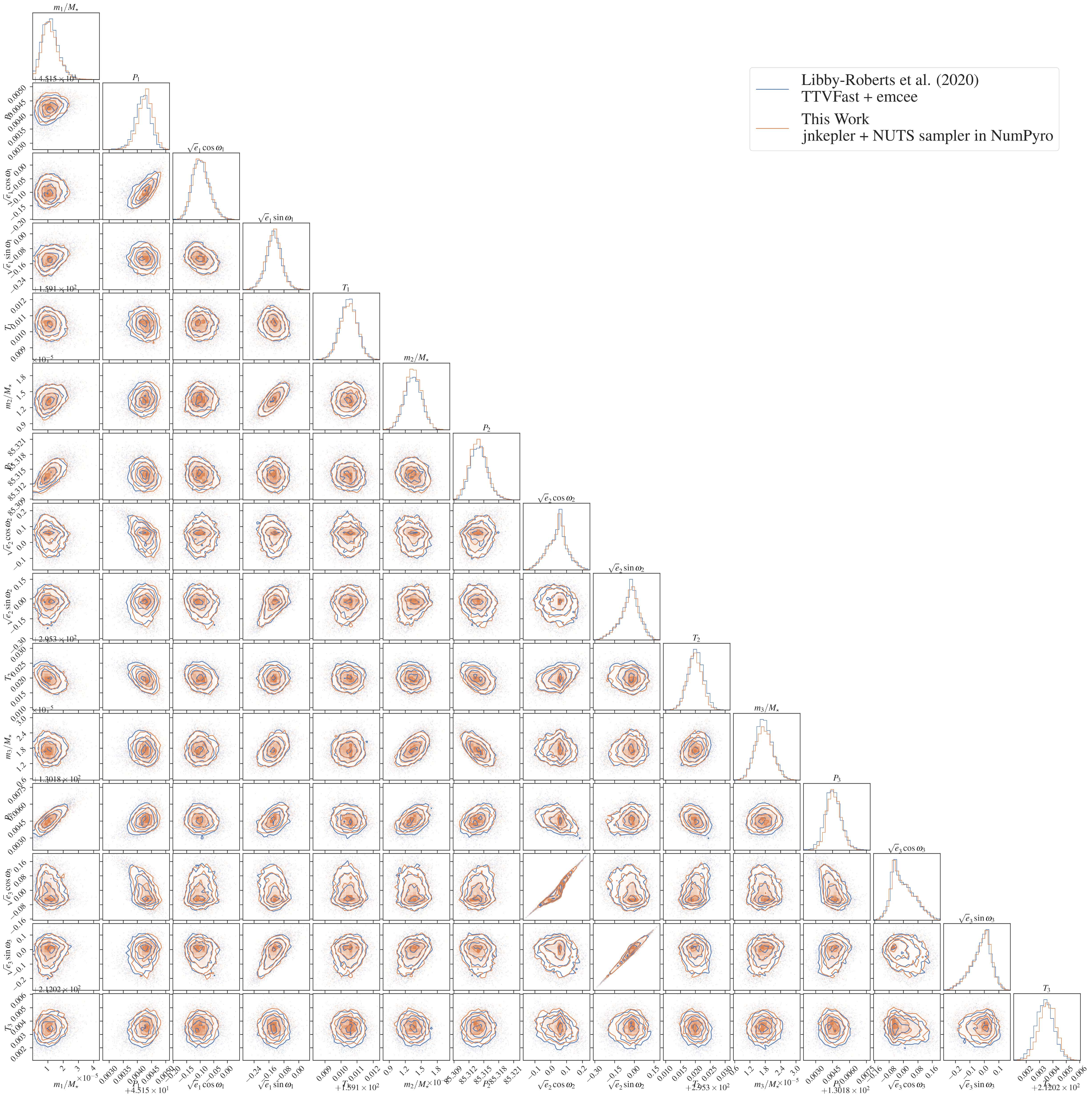}
	\caption{Corner plot \citep{corner} for the posterior samples from the three-planet fit to the TTV data in \citet{libbyroberts2020}.}
	\label{fig:corner_3planets}
\end{figure*}

In Section~\ref{sec:3planet}, we reproduced the TTV analysis in \citet{libbyroberts2020} using the code described in Section~\ref{sec:model}. Figure~\ref{fig:corner_3planets} compares the distributions of the posterior samples from the two analyses.

\subsection{Four-planet Model Assuming $P_{\rm e}/P_{\rm d} \approx 2$ in Section~\ref{ssec:4planet_2to1}.}

\begin{figure*}
	\epsscale{1.1}
	\plotone{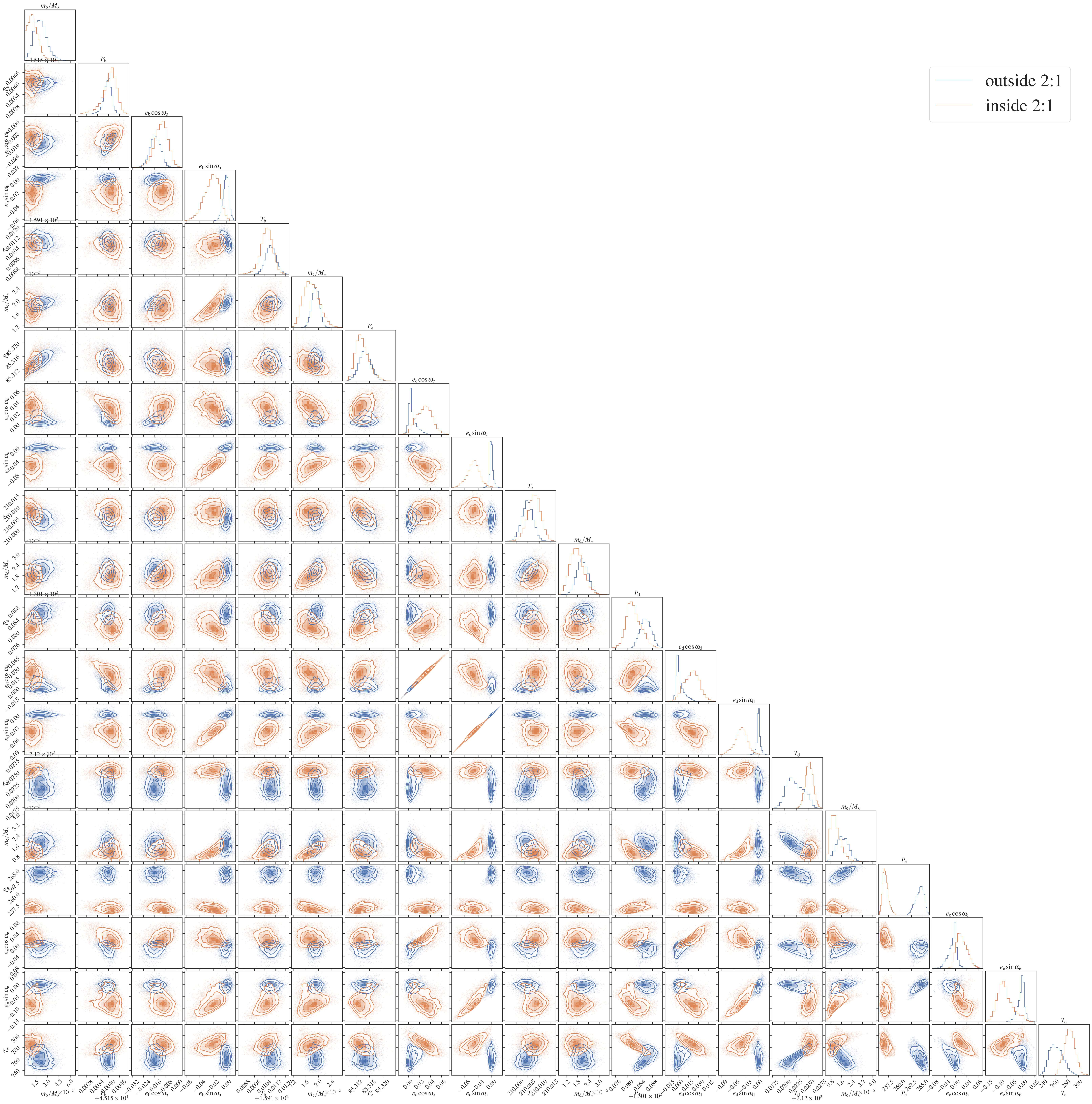}
	\caption{Corner plot \citep{corner} for the posterior samples from the four-planet fit to the entire TTV data in Table~\ref{tab:transit_times} assuming $P_\mathrm{e}\approx 2P_\mathrm{d}$.}
	\label{fig:corner_2to1}
\end{figure*}

In Section~\ref{ssec:4planet_2to1}, we performed an HMC sampling from the joint posterior PDF for the system parameters, assuming that Kepler-51e is around the $2:1$ resonance with Kepler-51d. Figure~\ref{fig:corner_2to1} shows the corresponding corner plot, in which the blue and orange contours show samples for the priors assuming $P_\mathrm{e}/P_\mathrm{d}>2$ and $P_\mathrm{e}/P_\mathrm{d}<2$, respectively.

\section{Prediction of Future Transit Times}\label{sec:ttvprediction}

\startlongtable
\begin{deluxetable*}{cccccc}
\tablecaption{Future predicted transit times of Kepler-51b, c, and d based on the $2:1$ solutions and all solutions from the brute-force search.\label{tab:ttvprediction}}
\tablehead{
\multicolumn{2}{c}{outside $2:1$}  & \multicolumn{2}{c}{inside $2:1$} & \multicolumn{2}{c}{all solutions}\\
\colhead{mean (BJD)} & \colhead{SD (days)} & \colhead{mean (BJD)} & \colhead{SD (days)} & \colhead{mean (BJD)} & \colhead{SD (days)}
}
\startdata
\textit{\textbf{(Kepler-51b)}}\\
$2460591.36365$&$0.00088$&$2460591.36346$&$0.00091$&$2460591.36313$&$0.00043$ \\
$2460636.51763$&$0.00084$&$2460636.51745$&$0.00085$&$2460636.51743$&$0.00034$ \\
$2460681.67237$&$0.00082$&$2460681.67235$&$0.00086$&$2460681.67222$&$0.00035$ \\
$2460726.82830$&$0.00082$&$2460726.82760$&$0.00091$&$2460726.82779$&$0.00050$ \\
$2460771.98367$&$0.00081$&$2460771.98341$&$0.00087$&$2460771.98346$&$0.00045$ \\
$2460817.14042$&$0.00083$&$2460817.13958$&$0.00098$&$2460817.13977$&$0.00066$ \\
$2460862.29691$&$0.00083$&$2460862.29649$&$0.00088$&$2460862.29660$&$0.00054$ \\
$2460907.45488$&$0.00087$&$2460907.4547$&$0.0011$&$2460907.45399$&$0.00073$ \\
$2460952.61018$&$0.00089$&$2460952.60994$&$0.00096$&$2460952.61010$&$0.00054$ \\
$2460997.76918$&$0.00094$&$2460997.7700$&$0.0011$&$2460997.76872$&$0.00057$ \\
$2461042.92390$&$0.00095$&$2461042.9239$&$0.0011$&$2461042.92368$&$0.00054$ \\
$2461088.0806$&$0.0010$&$2461088.0810$&$0.0011$&$2461088.08083$&$0.00043$ \\
$2461133.2351$&$0.0010$&$2461133.2354$&$0.0011$&$2461133.23504$&$0.00049$ \\
$2461178.3896$&$0.0010$&$2461178.3892$&$0.0011$&$2461178.38945$&$0.00060$ \\
$2461223.5432$&$0.0010$&$2461223.5434$&$0.0011$&$2461223.54304$&$0.00053$ \\
$2461268.6966$&$0.0010$&$2461268.6967$&$0.0011$&$2461268.69631$&$0.00057$ \\
$2461313.85053$&$0.00098$&$2461313.8509$&$0.0010$&$2461313.85034$&$0.00042$ \\
$2461359.00414$&$0.00098$&$2461359.0045$&$0.0010$&$2461359.00387$&$0.00040$ \\
$2461404.15842$&$0.00095$&$2461404.15857$&$0.00097$&$2461404.15838$&$0.00037$ \\
$2461449.31329$&$0.00094$&$2461449.31341$&$0.00099$&$2461449.31310$&$0.00040$ \\
$2461494.46873$&$0.00094$&$2461494.4681$&$0.0010$&$2461494.46829$&$0.00056$ \\
$2461539.62414$&$0.00093$&$2461539.62377$&$0.00100$&$2461539.62394$&$0.00057$ \\
$2461584.78109$&$0.00095$&$2461584.7799$&$0.0011$&$2461584.78022$&$0.00083$ \\
$2461629.93727$&$0.00096$&$2461629.9363$&$0.0011$&$2461629.93668$&$0.00076$ \\
$2461675.0951$&$0.0010$&$2461675.0943$&$0.0013$&$2461675.09392$&$0.00096$ \\
$2461720.2507$&$0.0010$&$2461720.2497$&$0.0012$&$2461720.25010$&$0.00083$ \\
$2461765.4095$&$0.0011$&$2461765.4096$&$0.0013$&$2461765.40845$&$0.00085$ \\
$2461810.5640$&$0.0011$&$2461810.5634$&$0.0013$&$2461810.56335$&$0.00089$ \\
$2461855.7210$&$0.0011$&$2461855.7208$&$0.0013$&$2461855.72068$&$0.00066$ \\
$2461900.8756$&$0.0011$&$2461900.8753$&$0.0013$&$2461900.87493$&$0.00092$ \\
$2461946.0299$&$0.0012$&$2461946.0292$&$0.0014$&$2461946.02938$&$0.00098$ \\
$2461991.1835$&$0.0012$&$2461991.1835$&$0.0013$&$2461991.18304$&$0.00093$ \\
$2462036.3372$&$0.0012$&$2462036.3372$&$0.0013$&$2462036.3366$&$0.0010$ \\
$2462081.4909$&$0.0011$&$2462081.4915$&$0.0012$&$2462081.49052$&$0.00082$ \\
$2462126.6443$&$0.0011$&$2462126.6450$&$0.0012$&$2462126.64405$&$0.00082$ \\
$2462171.7991$&$0.0011$&$2462171.7997$&$0.0012$&$2462171.79895$&$0.00068$ \\
$2462216.9537$&$0.0011$&$2462216.9545$&$0.0012$&$2462216.95363$&$0.00068$ \\
$2462262.1088$&$0.0011$&$2462262.1090$&$0.0011$&$2462262.10874$&$0.00068$ \\
$2462307.2647$&$0.0011$&$2462307.2651$&$0.0011$&$2462307.26482$&$0.00068$ \\
$2462352.4216$&$0.0011$&$2462352.4212$&$0.0012$&$2462352.42108$&$0.00079$ \\
$2462397.5774$&$0.0011$&$2462397.5773$&$0.0012$&$2462397.57734$&$0.00072$ \\
$2462442.7356$&$0.0011$&$2462442.7354$&$0.0013$&$2462442.73487$&$0.00087$ \\
$2462487.8914$&$0.0011$&$2462487.8909$&$0.0013$&$2462487.89108$&$0.00078$ \\
$2462533.0498$&$0.0012$&$2462533.0502$&$0.0014$&$2462533.04912$&$0.00081$ \\
$2462578.2046$&$0.0012$&$2462578.2042$&$0.0014$&$2462578.20421$&$0.00086$ \\
$2462623.3619$&$0.0013$&$2462623.3617$&$0.0015$&$2462623.36162$&$0.00075$ \\
$2462668.5162$&$0.0013$&$2462668.5156$&$0.0015$&$2462668.51551$&$0.00097$ \\
$2462713.6706$&$0.0013$&$2462713.6695$&$0.0015$&$2462713.6700$&$0.0011$ \\
$2462758.8244$&$0.0013$&$2462758.8239$&$0.0015$&$2462758.8237$&$0.0011$ \\
$2462803.9779$&$0.0013$&$2462803.9772$&$0.0015$&$2462803.9769$&$0.0013$ \\
$2462849.1313$&$0.0013$&$2462849.1313$&$0.0014$&$2462849.1307$&$0.0011$ \\
$2462894.2849$&$0.0013$&$2462894.2848$&$0.0014$&$2462894.2842$&$0.0010$ \\
$2462939.4396$&$0.0012$&$2462939.4395$&$0.0014$&$2462939.43893$&$0.00097$ \\
$2462984.5939$&$0.0012$&$2462984.5941$&$0.0014$&$2462984.59342$&$0.00096$ \\
$2463029.7490$&$0.0012$&$2463029.7487$&$0.0014$&$2463029.7486$&$0.0010$ \\
$2463074.9052$&$0.0012$&$2463074.9051$&$0.0014$&$2463074.9047$&$0.0010$ \\
$2463120.0617$&$0.0012$&$2463120.0610$&$0.0015$&$2463120.0608$&$0.0012$ \\
$2463165.2174$&$0.0012$&$2463165.2172$&$0.0014$&$2463165.2171$&$0.0011$ \\
$2463210.3760$&$0.0012$&$2463210.3758$&$0.0016$&$2463210.3749$&$0.0012$ \\
$2463255.5316$&$0.0013$&$2463255.5314$&$0.0015$&$2463255.5312$&$0.0011$ \\
$2463300.6898$&$0.0013$&$2463300.6907$&$0.0016$&$2463300.6892$&$0.0011$ \\
$2463345.8452$&$0.0013$&$2463345.8452$&$0.0016$&$2463345.8448$&$0.0010$ \\
$2463391.0025$&$0.0014$&$2463391.0029$&$0.0016$&$2463391.00235$&$0.00092$ \\
$2463436.1564$&$0.0014$&$2463436.1566$&$0.0016$&$2463436.1561$&$0.0010$ \\
$2463481.3114$&$0.0014$&$2463481.3110$&$0.0016$&$2463481.3112$&$0.0011$ \\
$2463526.4653$&$0.0014$&$2463526.4653$&$0.0016$&$2463526.4648$&$0.0011$ \\
$2463571.6184$&$0.0014$&$2463571.6181$&$0.0016$&$2463571.6178$&$0.0012$ \\
$2463616.7722$&$0.0014$&$2463616.7724$&$0.0016$&$2463616.7718$&$0.0010$ \\
$2463661.9258$&$0.0014$&$2463661.9259$&$0.0016$&$2463661.9253$&$0.0011$ \\
$2463707.0801$&$0.0013$&$2463707.0800$&$0.0016$&$2463707.0797$&$0.0010$ \\
$2463752.2346$&$0.0013$&$2463752.2345$&$0.0016$&$2463752.2342$&$0.0011$ \\
$2463797.3898$&$0.0013$&$2463797.3891$&$0.0016$&$2463797.3893$&$0.0012$ \\
$2463842.5456$&$0.0013$&$2463842.5450$&$0.0016$&$2463842.5450$&$0.0013$ \\
$2463887.7020$&$0.0013$&$2463887.7006$&$0.0017$&$2463887.7010$&$0.0015$ \\
$2463932.8579$&$0.0013$&$2463932.8569$&$0.0016$&$2463932.8573$&$0.0014$ \\
$2463978.0164$&$0.0014$&$2463978.0151$&$0.0019$&$2463978.0148$&$0.0016$ \\
$2464023.1717$&$0.0014$&$2464023.1706$&$0.0018$&$2464023.1709$&$0.0015$ \\
$2464068.3301$&$0.0014$&$2464068.3300$&$0.0018$&$2464068.3290$&$0.0015$ \\
$2464113.4855$&$0.0015$&$2464113.4848$&$0.0019$&$2464113.4846$&$0.0015$ \\
$2464158.6426$&$0.0015$&$2464158.6426$&$0.0018$&$2464158.6421$&$0.0013$ \\
$2464203.7966$&$0.0015$&$2464203.7964$&$0.0019$&$2464203.7960$&$0.0015$ \\
$2464248.9521$&$0.0015$&$2464248.9514$&$0.0019$&$2464248.9514$&$0.0015$ \\
$2464294.1057$&$0.0015$&$2464294.1057$&$0.0019$&$2464294.1050$&$0.0014$ \\
$2464339.2587$&$0.0015$&$2464339.2586$&$0.0019$&$2464339.2581$&$0.0015$ \\
$2464384.4130$&$0.0015$&$2464384.4134$&$0.0018$&$2464384.4125$&$0.0013$ \\
$2464429.5665$&$0.0015$&$2464429.5669$&$0.0018$&$2464429.5659$&$0.0013$ \\
$2464474.7204$&$0.0015$&$2464474.7209$&$0.0018$&$2464474.7202$&$0.0012$ \\
$2464519.8753$&$0.0014$&$2464519.8758$&$0.0018$&$2464519.8750$&$0.0012$ \\
$2464565.0305$&$0.0014$&$2464565.0303$&$0.0018$&$2464565.0301$&$0.0013$ \\
$2464610.1858$&$0.0014$&$2464610.1858$&$0.0018$&$2464610.1856$&$0.0013$ \\
$2464655.3426$&$0.0015$&$2464655.3417$&$0.0018$&$2464655.3419$&$0.0015$ \\
$2464700.4986$&$0.0015$&$2464700.4980$&$0.0018$&$2464700.4982$&$0.0015$ \\
$2464745.6565$&$0.0015$&$2464745.6556$&$0.0020$&$2464745.6553$&$0.0016$ \\
$2464790.8122$&$0.0015$&$2464790.8112$&$0.0019$&$2464790.8116$&$0.0016$ \\
\textit{\textbf{(Kepler-51c)}}\\
$2460588.5952$&$0.0047$&$2460588.5966$&$0.0050$&$2460588.5946$&$0.0041$ \\
$2460673.9091$&$0.0049$&$2460673.9127$&$0.0055$&$2460673.9094$&$0.0051$ \\
$2460759.2335$&$0.0046$&$2460759.2385$&$0.0062$&$2460759.2321$&$0.0066$ \\
$2460844.5395$&$0.0049$&$2460844.5477$&$0.0067$&$2460844.5417$&$0.0078$ \\
$2460929.8488$&$0.0052$&$2460929.8606$&$0.0074$&$2460929.8537$&$0.0090$ \\
$2461015.1778$&$0.0047$&$2461015.1903$&$0.0085$&$2461015.180$&$0.011$ \\
$2461100.4873$&$0.0053$&$2461100.5026$&$0.0088$&$2461100.492$&$0.012$ \\
$2461185.8012$&$0.0060$&$2461185.8196$&$0.0091$&$2461185.808$&$0.012$ \\
$2461271.1403$&$0.0061$&$2461271.1579$&$0.0099$&$2461271.143$&$0.013$ \\
$2461356.4540$&$0.0067$&$2461356.473$&$0.010$&$2461356.459$&$0.014$ \\
$2461441.7643$&$0.0071$&$2461441.787$&$0.011$&$2461441.773$&$0.015$ \\
$2461527.0987$&$0.0067$&$2461527.123$&$0.012$&$2461527.105$&$0.016$ \\
$2461612.4119$&$0.0066$&$2461612.439$&$0.012$&$2461612.421$&$0.016$ \\
$2461697.7198$&$0.0066$&$2461697.751$&$0.013$&$2461697.733$&$0.017$ \\
$2461783.0487$&$0.0064$&$2461783.080$&$0.014$&$2461783.060$&$0.018$ \\
$2461868.3664$&$0.0062$&$2461868.400$&$0.015$&$2461868.379$&$0.019$ \\
$2461953.6803$&$0.0065$&$2461953.716$&$0.015$&$2461953.694$&$0.019$ \\
$2462039.0083$&$0.0070$&$2462039.043$&$0.016$&$2462039.020$&$0.019$ \\
$2462124.3296$&$0.0071$&$2462124.364$&$0.017$&$2462124.341$&$0.020$ \\
$2462209.6408$&$0.0071$&$2462209.678$&$0.017$&$2462209.654$&$0.020$ \\
$2462294.9562$&$0.0068$&$2462294.995$&$0.018$&$2462294.970$&$0.021$ \\
$2462380.2742$&$0.0062$&$2462380.315$&$0.019$&$2462380.288$&$0.022$ \\
$2462465.5829$&$0.0059$&$2462465.627$&$0.020$&$2462465.599$&$0.023$ \\
$2462550.8915$&$0.0056$&$2462550.937$&$0.021$&$2462550.909$&$0.024$ \\
$2462636.2099$&$0.0053$&$2462636.257$&$0.022$&$2462636.226$&$0.025$ \\
$2462721.5261$&$0.0057$&$2462721.575$&$0.023$&$2462721.541$&$0.025$ \\
$2462806.8387$&$0.0065$&$2462806.888$&$0.024$&$2462806.854$&$0.026$ \\
$2462892.1574$&$0.0069$&$2462892.207$&$0.025$&$2462892.172$&$0.027$ \\
$2462977.4755$&$0.0069$&$2462977.527$&$0.026$&$2462977.489$&$0.028$ \\
$2463062.7821$&$0.0071$&$2463062.837$&$0.027$&$2463062.798$&$0.029$ \\
$2463148.0949$&$0.0071$&$2463148.153$&$0.029$&$2463148.112$&$0.030$ \\
$2463233.4171$&$0.0063$&$2463233.478$&$0.031$&$2463233.432$&$0.031$ \\
$2463318.7224$&$0.0066$&$2463318.786$&$0.032$&$2463318.741$&$0.032$ \\
$2463404.0345$&$0.0070$&$2463404.101$&$0.033$&$2463404.054$&$0.034$ \\
$2463489.3688$&$0.0063$&$2463489.436$&$0.035$&$2463489.384$&$0.035$ \\
$2463574.6821$&$0.0073$&$2463574.751$&$0.035$&$2463574.699$&$0.036$ \\
$2463659.9953$&$0.0081$&$2463660.067$&$0.036$&$2463660.014$&$0.037$ \\
$2463745.3316$&$0.0073$&$2463745.403$&$0.038$&$2463745.346$&$0.039$ \\
$2463830.6423$&$0.0074$&$2463830.718$&$0.039$&$2463830.660$&$0.040$ \\
$2463915.9514$&$0.0075$&$2463916.032$&$0.040$&$2463915.972$&$0.041$ \\
$2464001.2859$&$0.0067$&$2464001.368$&$0.042$&$2464001.304$&$0.043$ \\
$2464086.5977$&$0.0066$&$2464086.683$&$0.044$&$2464086.619$&$0.044$ \\
$2464171.9080$&$0.0069$&$2464171.996$&$0.044$&$2464171.931$&$0.045$ \\
$2464257.2445$&$0.0073$&$2464257.333$&$0.047$&$2464257.263$&$0.046$ \\
$2464342.5655$&$0.0082$&$2464342.656$&$0.048$&$2464342.585$&$0.046$ \\
$2464427.8787$&$0.0088$&$2464427.971$&$0.048$&$2464427.900$&$0.047$ \\
$2464513.2055$&$0.0091$&$2464513.299$&$0.050$&$2464513.225$&$0.048$ \\
$2464598.5235$&$0.0092$&$2464598.620$&$0.052$&$2464598.544$&$0.049$ \\
$2464683.8329$&$0.0092$&$2464683.934$&$0.054$&$2464683.856$&$0.050$ \\
$2464769.1486$&$0.0088$&$2464769.252$&$0.055$&$2464769.172$&$0.051$ \\
\textit{\textbf{(Kepler-51d)}}\\
$2460642.5493$&$0.0043$&$2460642.5429$&$0.0044$&$2460642.556$&$0.033$ \\
$2460772.7134$&$0.0058$&$2460772.7033$&$0.0062$&$2460772.723$&$0.042$ \\
$2460902.8905$&$0.0074$&$2460902.8765$&$0.0084$&$2460902.899$&$0.052$ \\
$2461033.0471$&$0.0091$&$2461033.028$&$0.011$&$2461033.058$&$0.062$ \\
$2461163.223$&$0.011$&$2461163.199$&$0.015$&$2461163.231$&$0.073$ \\
$2461293.374$&$0.013$&$2461293.345$&$0.019$&$2461293.385$&$0.084$ \\
$2461423.548$&$0.016$&$2461423.514$&$0.024$&$2461423.555$&$0.095$ \\
$2461553.698$&$0.018$&$2461553.660$&$0.029$&$2461553.71$&$0.11$ \\
$2461683.871$&$0.021$&$2461683.828$&$0.035$&$2461683.88$&$0.12$ \\
$2461814.027$&$0.024$&$2461813.981$&$0.041$&$2461814.03$&$0.13$ \\
$2461944.199$&$0.029$&$2461944.149$&$0.047$&$2461944.20$&$0.14$ \\
$2462074.363$&$0.033$&$2462074.311$&$0.053$&$2462074.37$&$0.15$ \\
$2462204.536$&$0.038$&$2462204.482$&$0.059$&$2462204.54$&$0.16$ \\
$2462334.709$&$0.042$&$2462334.654$&$0.065$&$2462334.71$&$0.17$ \\
$2462464.886$&$0.048$&$2462464.829$&$0.071$&$2462464.88$&$0.19$ \\
$2462595.064$&$0.053$&$2462595.007$&$0.077$&$2462595.06$&$0.20$ \\
$2462725.243$&$0.059$&$2462725.183$&$0.084$&$2462725.23$&$0.21$ \\
$2462855.422$&$0.064$&$2462855.362$&$0.090$&$2462855.41$&$0.22$ \\
$2462985.599$&$0.070$&$2462985.536$&$0.097$&$2462985.58$&$0.23$ \\
$2463115.778$&$0.075$&$2463115.71$&$0.10$&$2463115.76$&$0.25$ \\
$2463245.948$&$0.081$&$2463245.88$&$0.11$&$2463245.92$&$0.26$ \\
$2463376.126$&$0.085$&$2463376.06$&$0.12$&$2463376.10$&$0.28$ \\
$2463506.288$&$0.091$&$2463506.22$&$0.13$&$2463506.26$&$0.29$ \\
$2463636.465$&$0.095$&$2463636.39$&$0.13$&$2463636.43$&$0.31$ \\
$2463766.62$&$0.10$&$2463766.55$&$0.14$&$2463766.59$&$0.33$ \\
$2463896.80$&$0.10$&$2463896.72$&$0.15$&$2463896.76$&$0.35$ \\
$2464026.95$&$0.11$&$2464026.88$&$0.16$&$2464026.91$&$0.36$ \\
$2464157.13$&$0.11$&$2464157.05$&$0.17$&$2464157.08$&$0.38$ \\
$2464287.29$&$0.12$&$2464287.21$&$0.18$&$2464287.24$&$0.40$ \\
$2464417.46$&$0.12$&$2464417.38$&$0.18$&$2464417.40$&$0.42$ \\
$2464547.63$&$0.13$&$2464547.54$&$0.19$&$2464547.57$&$0.44$ \\
$2464677.80$&$0.13$&$2464677.72$&$0.20$&$2464677.74$&$0.46$ \\
$2464807.98$&$0.13$&$2464807.90$&$0.21$&$2464807.91$&$0.48$ \\
\enddata
\end{deluxetable*}

Table~\ref{tab:ttvprediction} lists the future predicted transit times of Kepler-51b, c, and d discussed in Section~\ref{ssec:4planet_future}. The predictions are based on the solutions with $P_{\rm e}/P_{\rm d} \approx 2$ (first four columns), and for all the solutions from the brute-force search (last two columns). For the former, the mean and standard deviation of the posterior models obtained in Section~\ref{ssec:4planet_2to1} are shown. For the latter, the same statistics are shown for timing models computed for all the solutions found from the brute-force search in Section~\ref{ssec:4planet_brute}.

\section{Three-Planet Photodynamical Modeling of the Kepler Data}\label{sec:app_photod}

We performed a so-called photodynamical modeling of the Kepler light curves, in which model transit light curves are computed using the planets' orbits from numerical integration. Here we consider only three planets (Kepler-51b, c, d) that are known to transit. 

\subsection{The Data}

We used the normalized and detrended flux obtained in Section~\ref{ssec:obs_kepler}. We used short-cadence data whenever available, and excluded from the analysis the transits with more than 20\% of the data points missing in the one-day window, as well as the last transits of Kepler-51b and Kepler-51d that occurred simultaneously around $\mathrm{BJD} = 2456346.8$.

\subsection{The Flux Model}

Our model computes the relative flux loss $F$ at a given time due to all three transiting planets, taking into account the mutual gravitational interactions between the planets when solving their motion. We assume that the planets are fully optically thick spheres and that the stellar surface brightness profile is described by the quadratic limb-darkening law, $\propto 1 - u_1(1-\mu) - u_2(1-\mu)^2$ with $\mu$ being the sky-projected distance from the star's center normalized by the stellar radius. 

The model parameters are:
\begin{itemize}
    \item planet-to-star mass ratio $m$,
    \item planet-to-star radius ratio $r$,
    \item osculating orbital elements at the start of integration, here chosen to be $\mathrm{BJD}=2454833+155$,
    \item mean density of the star $\rho_\star$,
    \item limb-darkening coefficients $q_1=(u_1+u_2)^2$ and $q_2=u_1/2(u_1+u_2)$.
\end{itemize} 
The first three quantities are defined for each planet and are specified by the subscript b, c, and d when necessary. 
We use the same set of osculating orbital elements as given in Section~\ref{sec:model}, though in practice we use the transit impact parameter $b$ instead of the inclination $i$, computing the latter from the former and other parameters via Equation 7 of \citet{2011exop.book...55W}. Because the relative flux data alone do not constrain the absolute dimensions of the system, we fix the stellar mass to be $1\,M_\odot$ and fit $R_\star$ and $\rho_\star$ interchangeably in the dynamical modeling.
Instead of $u_1$ and $u_2$, we use $q_1$ and $q_2$ that can be sampled from the bounded prior following \citet{2013MNRAS.435.2152K}.

Given the mass ratios and initial osculating orbital elements, we numerically compute the transit times of the planets as well as their positions and velocities at the transit centers in the same way as described in Section~\ref{sec:model}. The positions of the planets at each time are then calculated assuming that the in-transit motion is linear. The flux loss is computed using the solution vector as defined in \citet{2020AJ....159..123A}, where we use {\tt exoplanet-ops[jax]} that provides {\tt JAX} implementations of {\tt exoplanet} \citep{exoplanet}. For long-cadence data, the flux is upsampled by a factor of 11 and averaged to take into account finite integration time \citep{2010MNRAS.408.1758K}. We checked the flux calculation by comparing the code output (with planetary mass ratios set to be $10^{-12}$) with flux computed using {\tt exoplanet} \citep{exoplanet} for several sets of the common orbital parameters. The relative flux difference without finite time integration was found to be $\sim 10^{-6}$ at most, and the errors due to finite time integration was confirmed to be at most $\sim 20\%$ of the assigned flux errors, whose medians are $2.9\times10^{-4}$ and $1.6\times10^{-3}$ for the long- and short-cadence data, respectively.

\subsection{Likelihood}

\newcommand{\pphys}{\theta_{\rm phys}}
\newcommand{\pnoise}{\theta_{\rm noise}}
\newcommand{\lc}{\mathrm{long}}
\renewcommand{\sc}{\mathrm{short}}
\newcommand{\dt}{\mathrm{double}}
\newcommand{\fbump}{\delta F}

The above model provides the flux loss for the long and short cadence data, $F_\lc$ and $F_\sc$, for given times and a set of physical model parameters $\theta_{\rm phys}=\{m, r, P, e, \omega, b, \Omega, T, q_1, q_2, R_\star\}$. The flux model was then used to define the likelihood $\mathcal{L}_\lc$ and $\mathcal{L}_\sc$ for the long-cadence data $f_\lc$ and the short-cadence data excluding the simultaneous transit of Kepler-51b and Kepler-51d $f_\sc$, respectively:
\begin{align}
    \notag
    &\mathcal{L}_\lc(\pphys, \pnoise, \mu_\lc)\\
    \label{eq:loglike_long}
&\equiv \mathcal{N}(f_\lc; F_\lc (\pphys)+\mu_\lc, \Sigma(\pnoise))\\
    \notag
    &\mathcal{L}_\sc(\pphys, \pnoise, \mu_\sc)\\
    \label{eq:loglike_short}
&\equiv \mathcal{N}(f_\sc; F_\sc (\pphys)+\mu_\sc, \Sigma(\pnoise))
\end{align}
where $f$ and $F$ are treated as vectors, and $\mathcal{N}(x; \mu, \Sigma)$ is a multivariate normal distribution with the mean $\mu$ and covariance matrix $\Sigma$. We adopt the same covariance matrix given by Equation~\ref{eq:kernel}, which is a the sum of the Mat\'ern-3/2 covariance and the white-noise term, and so $\theta_{\rm noise}$ consists of $\alpha$, $\rho$, and $\sigma_{\rm jit}$ defined separately for the long- and short-cadence data.

\subsection{Inference}

To obtain a reasonable initial guess for the full $N$-body photodynamical fitting, we first fitted the whole long-cadence light curve with an $N$-body transit model by minimizing chi-squared using {\tt jaxopt.ScipyBoundedMinimize}. Here we fixed the mass ratios and orbital parameters except for the inclination to the values derived from the TTV fit, but optimized the impact parameters, radius ratios, limb-darkening coefficients, and mean stellar density. This way we obtained a set of system parameters for the $N$-body light curve model that fits the long cadence data. 

We then sampled from the joint posterior distribution for all the parameters $\theta = \{\pphys, \pnoise, \mu\}$ conditioned on the flux data $f$:
\begin{equation}
    p(\theta | f) \propto \mathcal{L}(f |\theta) \, \pi (\theta) 
\end{equation}
adopting the log-likelihood 
\begin{align}
    \ln \mathcal{L}(\theta) = \ln \mathcal{L}_\lc(\theta) + \ln \mathcal{L}_\sc(\theta)
\end{align}
and the prior PDF $\pi$ separable for the parameters. We adopted either of a uniform, log-uniform, or normal distribution for the prior of each parameter as summarized in Table~\ref{tab:photod}). 
Here we fixed $\Omega_{\rm d}$ to be zero without loss of generality, and restricted $|\Omega_{\rm b}|$ and $|\Omega_{\rm c}|$ to be less than one radian. Care should be taken about the range of impact parameters; usually it is chosen to be positive, but in multi-transiting systems both signs (i.e., $i<90^\circ$ and $i>90^\circ$) need to be considered if the difference matters. 
Here we assume $b_{\rm c}>0$, again without loss of generality, and allow $b_{\rm b}$ and $b_{\rm d}$ to take both positive and negative values. 
We were able to sample $b_{\rm b}$ and $b_{\rm d}$ well from such a prior, as the joint posterior PDF happened to have large enough density at $(b_{\rm b}, b_{\rm d})\approx (0,0)$ in this problem.

The sampling was performed using the No-U-Turn Sampler \citep{DUANE1987216, 2017arXiv170102434B} as implemented in {\tt NumPyro} \citep{bingham2018pyro, phan2019composable}. We initialized the Markov chains at the parameter values derived in the first paragraph of this subsection, and ran four chains for 1,000 tuning steps and for 2,000 steps to draw 8,000 posterior samples in total, setting the target acceptance probability to be 0.95 and the maximum tree depth to be 11. We found $\hat{R}=1.00$--1.05 and estimated $n_{\rm eff}=100$--$1000$ except for $b_{\rm b}$, for which $\hat{R}=1.08$ and $n_{\rm eff}\approx 80$. The poorer convergence is a result of the bimodal nature of the parameter's posterior distribution, and we did not attempt to improve the statistics through extended sampling.

\subsection{Results}

The summary statistics based on the posterior samples are shown in Table~\ref{tab:photod}. The resulting posterior models appear to be almost identical to those in Figure~\ref{fig:keptransit1}--\ref{fig:keptransit3} and thus are not shown. The residuals from the maximum likelihood model (i.e., $f-F-\mu$ in Eqs.~\ref{eq:loglike_long}--\ref{eq:loglike_short}) divided by the assigned flux errors (i.e., $\sigma$) were found to be consistent with zero-mean normal distributions with the standard deviations of 1.1 and 0.98 for the long- and short-cadence data, respectively, except for a small number of outliers. The chi-squared values are $\approx 1465$ for the 1458 long-cadence data points and $\approx 22800$ for the 24208 short-cadence data points. When the mean GP prediction was subtracted from these residuals, the normalized standard deviations were 0.90 and 0.97. These are consistent with small inferred jitters $\sigma_{\rm jit}$. 

The photodynamical modeling constrains the mutual orbital inclinations, apparently driven by planet c's grazing orbit, which makes its transits highly sensitive to small variations in orbital inclination. 
The inclinations of the orbits of Kepler-51b and Kepler-51c relative to that of Kepler-51d are inferred to be $0$--$13\,\mathrm{deg}$ and $0$--$3\,\mathrm{deg}$ (95\% highest density intervals), respectively. 
Despite this difference, the resulting planetary masses and eccentricities remain to be consistent with those from TTV-only fitting (cf.~Section~\ref{sec:3planet}), and so do the transit-time predictions after the Kepler observations. Therefore, we conclude that the three-planet model is incompatible with the JWST timing measurement, even when accounting for the possibility of non-zero mutual orbital inclinations.

\newcommand{\logu}{\mathcal{U}_\mathrm{log}}

\begin{deluxetable*}{l@{\hspace{.1cm}}cc@{\hspace{.1cm}}c}[!ht]
\tablecaption{Parameters of the Kepler-51 system from three-planet photodynamical modeling in Appendix~\ref{sec:app_photod}. Note that this model fails to reproduce the JWST transit time, as does the three-planet TTV-only model in Section~\ref{sec:3planet}.\label{tab:photod}}
\tablehead{
\colhead{} & \colhead{mean} & \colhead{$95\%$ HDI} & \colhead{Prior} 
}
\startdata
\textbf{Physical Parameters}\\
\textit{(host star)}\\
mean density $\rho_\star$ ($\rho_\odot$) 
& $1.818$ & [$1.572$, $2.118$] & \\
radius $R_\star/M_\star^{1/3}$ ($R_\odot/M_\odot^{1/3}$) & $0.820$ & [$0.777$, $0.857$]& $\mathcal{N}(1,1)$, $R_\star/M_\star^{1/3}>0$ \\
limb-darkening coefficients $q_1$ & $0.296$ & [$0.198$, $0.401$] & $\mathcal{U}(0,1)$\\
limb-darkening coefficients $q_2$ & $0.400$ & [$0.265$, $0.556$] & $\mathcal{U}(0,1)$
\vspace{0.1cm}\\
\textit{(Kepler-51b)}\\
mass ratio $m$ ($M_\oplus/M_\odot$) & $2.060$ & [$0.001$, $4.897$] & $\mathcal{U}(0,5\times10^{-4})$\\
radius ratio $r$ & $0.073$ & [$0.072$, $0.073$] & $\mathcal{U}(0,0.2)$\\
mean density $\rho_{\rm p}$ ($\mathrm{g\,cm^{-3}}$) & $0.041$ & [$0.000$, $0.096$] \\
time of inferior conjunction (BKJD) & $159.110$ & [$159.109$, $159.111$] & $\mathcal{U}(T_0-0.1,T_0+0.1)$\\
orbital period (days) & $45.154$ & [$45.153$, $45.155$] & $\mathcal{U}(P_0-0.5,P_0+0.5)$\\
eccentricity $e$ & $0.042$ & [$0.014$, $0.079$] & $\mathcal{U}(0, 0.3)$\\
$e\cos\omega$ & $-0.022$ & [$-0.039$, $-0.006$] & $\omega \sim \mathcal{U}(0,2\pi)$\\
$e\sin\omega$ & $-0.034$ & [$-0.072$, $-0.003$] & $\omega \sim \mathcal{U}(0,2\pi)$\\
impact parameter $b$ & $0.043$ & [$-0.300$, $0.315$] &  $\mathcal{U}(-1.1,1.1)$\\
longitude of ascending node $\Omega$ (deg) & $-0.625$ & [$-13.976$, $12.200$] & $\mathcal{U}(-1\,\mathrm{rad},1\,\mathrm{rad})$
\vspace{0.1cm}\\
\textit{(Kepler-51c)}\\
mass ratio $m$ ($M_\oplus/M_\odot$) & $4.080$ & [$3.062$, $5.153$] & $\mathcal{U}(0,5\times10^{-4})$\\
radius ratio $r$ & $0.060$ & [$0.048$, $0.077$] & $\mathcal{U}(0,0.2)$\\
mean density $\rho_{\rm p}$ ($\mathrm{g\,cm^{-3}}$) & $0.158$ & [$0.053$, $0.269$]\\
time of inferior conjunction (BKJD) & $210.009$ & [$210.003$, $210.016$] & $\mathcal{U}(T_0-0.1,T_0+0.1)$\\
orbital period (days) & $85.312$ & [$85.308$, $85.316$] & $\mathcal{U}(P_0-0.5,P_0+0.5)$\\
eccentricity $e$ & $0.038$ & [$0.000$, $0.092$] & $\mathcal{U}(0, 0.3)$\\
$e\cos\omega$ & $0.016$ & [$-0.009$, $0.051$] &
$\omega \sim \mathcal{U}(0,2\pi)$\\
$e\sin\omega$ & $-0.031$ & [$-0.090$, $0.007$] &
$\omega \sim \mathcal{U}(0,2\pi)$ \\
impact parameter $b$ & $0.979$ & [$0.956$, $1.005$] &  $\mathcal{U}(0,1.1)$\\
longitude of ascending node $\Omega$ (deg) & $0.782$ & [$-2.149$, $3.660$] & $\mathcal{U}(-1\,\mathrm{rad},1\,\mathrm{rad})$
\vspace{0.1cm}\\
\textit{(Kepler-51d)}\\
mass ratio $m$ ($M_\oplus/M_\odot$) & $6.210$ & [$3.988$, $8.573$] & $\mathcal{U}(0,5\times10^{-4})$\\
radius ratio $r$ & $0.099$ & [$0.098$, $0.099$] & $\mathcal{U}(0,0.2)$\\
mean density $\rho_{\rm p}$ ($\mathrm{g\,cm^{-3}}$) & $0.050$ & [$0.034$, $0.067$] \\
time of inferior conjunction (BKJD) & $212.024$ & [$212.022$, $212.026$] & $\mathcal{U}(T_0-0.1,T_0+0.1)$\\
orbital period (days) & $130.184$ & [$130.182$, $130.186$] & $\mathcal{U}(P_0-0.5,T_0+0.5)$\\
eccentricity $e$ & $0.030$ & [$0.000$, $0.075$] & $\mathcal{U}(0, 0.3)$\\
$e\cos\omega$ & $0.010$ & [$-0.011$, $0.038$] & $\omega \sim \mathcal{U}(0,2\pi)$ \\
$e\sin\omega$ & $-0.025$ & [$-0.075$, $0.006$] &
$\omega \sim \mathcal{U}(0,2\pi)$\\
impact parameter $b$ & $0.012$ & [$-0.256$, $0.262$] & $\mathcal{U}(-1.1,1.1)$\\
longitude of ascending node $\Omega$ (deg) & \multicolumn{3}{c}{0 (fixed)}
\vspace{0.1cm}\\
\textbf{Noise Parameters}\\
mean flux $\mu \times 10^5$ (long cadence) & $1.196$ & [$-0.941$, $3.460$]& $\mathcal{N}(0, 10)$\\
$\ln \sigma_{\rm jit}$ (long cadence) & $-11.463$ & [$-13.953$, $-9.277$] & $\mathcal{U}(-14, -4)$\\
$\ln \alpha$ (long cadence) & $-8.812$ & [$-9.025$, $-8.607$] & $\mathcal{U}(-14, -4)$\\
$\ln \rho$ (long cadence) & $-3.577$ & [$-4.079$, $-3.089$] & $\mathcal{U}(-5, 1)$\\
mean flux $\mu \times 10^5$ (short cadence) & $-0.425$ & [$-3.199$, $2.433$] & $\mathcal{N}(0, 100)$\\
$\ln \sigma_{\rm jit}$ (short cadence) & $-11.939$ & [$-13.996$, $-9.907$] & $\mathcal{U}(-14, -4)$\\
$\ln \alpha$ (short cadence) & $-8.679$ & [$-8.909$, $-8.460$] & $\mathcal{U}(-14, -4)$\\
$\ln \rho$ (short cadence) & $-3.926$ & [$-4.395$, $-3.416$] & $\mathcal{U}(-5, 1)$
\enddata
\end{deluxetable*}

\end{document}